\documentclass[12pt]{article}

\usepackage{epsf}
\usepackage{rotate}
\usepackage{graphicx}

\newcommand{\insertfig}[2]{\mbox{\epsfxsize=#1cm 
\epsfbox{#2.eps}}}
\newcommand{\MS}{{$\overline{\mbox{MS}}$}}
\newcommand{\MSeq}{\overline{\rm MS}}
\newcommand{\CS}{{$\overline{\mbox{CS}}$}}
\newcommand{\CSeq}{\overline{\rm CS}}

\newcommand{\req}[1]{(\ref{#1})}
\newcommand{\VF}{\widetilde{v}}
\newcommand{\FF}{\widetilde{f}}
\newcommand{\PF}{\widetilde{P}}
\newcommand{\pF}{\widetilde{p}}

\renewcommand{\theequation}{\arabic{section}.\arabic{equation}}%

\begin{document}


\begin{titlepage}


\centerline{\large\bf Next-to-next-to-leading order prediction for }
\centerline{\large\bf the photon-to-pion transition form factor}

\vspace{15mm}

\renewcommand{\thefootnote}{\fnsymbol{footnote}}

\centerline{\bf B. Meli{\' c}$^{a}$\footnotemark[1], D. M\"uller$^{b}$,
K. Passek-Kumeri\v{c}ki$^{b}$\footnotemark[1]}

\vspace{10mm}

\centerline{\it $^a$ Institut f\"{u}r Physik,
        Universit\"{a}t Mainz}
\centerline{\it D-55099 Mainz, Germany}
 
\centerline{\it Institut f\"{u}r Theoretische Physik und Astrophysik,
     Universit\"{a}t W\"{u}rzburg}
\centerline{\it D-97074 W\"{u}rzburg, Germany} 

\vspace{5mm}

\centerline{\it $^b$Fachbereich Physik,
                Universit\"at  Wuppertal}
\centerline{\it D-42097 Wuppertal, Germany}

\footnotetext[1]{On leave of absence from the 
Rudjer Bo\v{s}kovi\'{c} Institute, Zagreb, Croatia.}

\renewcommand{\thefootnote}{\arabic{footnote}}

\vspace{15mm}


\centerline{\bf Abstract}

\vspace{0.5cm}

\noindent
We evaluate the next-to-next-to-leading order corrections to the
hard-scattering amplitude of the photon-to-pion transition form factor.
Our approach is based on the predictive power of the conformal operator
product expansion, which is valid for a vanishing $\beta$-function in
the so-called conformal scheme. The Wilson--coefficients appearing in
the non-forward kinematics are then entirely determined from those of
the polarized deep-inelastic scattering known to next-to-next-to-leading
order accuracy. We propose different schemes to include explicitly also the
conformal symmetry breaking term proportional to the $\beta$-function,
and discuss numerical predictions calculated in different kinematical
regions. It is demonstrated that the photon-to-pion transition form
factor can provide a fundamental testing ground for our QCD
understanding of exclusive reactions.

\vspace{1cm}


\vspace{0.5cm}


\end{titlepage}
 
\tableofcontents

\newpage

\section{Introduction}
\label{Sec-Intro}
\setcounter{equation}{0}

At a sufficiently large scale exclusive QCD reactions factorize into a
perturbative calculable partonic hard-scattering amplitude and universal
hadron distribution amplitudes \cite{BroLep80,BroLep81,EfrRad80}. The study
of such reactions offers the possibility to directly explore 
non-perturbative features of hadrons at the amplitude level, as well
as to test our understanding of the amplitude factorization. 
Unfortunately, exclusive reactions are still challenging to both 
experimentalists and theoreticians and the onset of the perturbative 
approach is controversially discussed in the literature.

The photon-to-pion transition form factor, appearing in the two-photon
amplitude of the process $\gamma^{\ast}(q_1)\; \gamma^{(\ast)}(q_2) \to
\pi^0(P)$ can serve for a thorough study of the mentioned problem.
In this process the partonic content of a meson is only probed by the
electromagnetic interaction. Furthermore, since we require that the
meson is produced at light-like distances, i.e., that at least one
photon is far off-shell, this process belongs to quite a large class of
two-photon processes calculable by means of the operator product
expansion (OPE) \cite{MueRobGeyDitHor94}. Deeply virtual Compton
scattering (DVCS), deeply inelastic lepton--hadron scattering (DIS) and
the production of various hadronic final states by photon--photon fusion
belong to this class of processes. Such processes can be described by a
general scattering amplitude given by the time-ordered product of
two-electromagnetic currents sandwiched between the hadronic states. For
a specific process, the generalized Bjorken kinematics at the light-cone
can be reduced to the corresponding kinematics, while the particular
hadron content of the process reflects itself in the non-perturbative
part of the amplitude. Hence, the generalized hard-scattering amplitude
enables us to relate predictions of different two-photon processes on
the partonic level.

In the leading twist-2 approximation, the light-cone OPE approach is
equivalent to the collinear factorization scheme
\cite{BroLep80,BroLep81,EfrRad80}. The transition form factor factorizes
as a convolution of the 
hard-scattering amplitude $T$ and the pion distribution amplitude 
$\phi$, with respect to the momentum fraction $x$:
\begin{eqnarray}
\label{RG-transfor}
F_{\gamma\pi}(\omega,Q) = f_{\pi} \;
	T(\omega,x,Q,\mu_f)\otimes\phi(x,\mu_f),
\qquad 	\otimes \equiv \int_0^1 dx,
\label{eq:Fconv}
\end{eqnarray}
 where 
\begin{eqnarray*}
Q^2=-\frac{(q_1-q_2)^2}{4},\qquad \omega= \frac{q_1^2 - q_2^2}{q_1^2+
q_2^2}.
\end{eqnarray*}
In the above, the resolution scale $Q^2$ is large and 
the asymmetry parameter $\omega$
is fixed%
, 
i.e., $|\omega| \le 1$, while $\mu_f$ 
represents the factorization scale. 
Due to Bose symmetry the transition form
factor is symmetric in $\omega$.
The perturbative expansion of the hard-scattering
amplitude reads 
\begin{eqnarray}
T(\omega,x,Q,\mu_f) & \!\!\!=\!\!\! & \frac{\sqrt{2}}{6 Q^2} \Bigg[
T^{(0)}(\omega,x) + \frac{\alpha_s(\mu_r)}{2\pi}
T^{(1)}\!\!\left(\!\omega,x,\frac{Q}{\mu_f}\right)
+\frac{\alpha^2_s(\mu_r)}{(2\pi)^2}
T^{(2)}\!\!\left(\!\omega,x,\frac{Q}{\mu_f},\frac{Q}{\mu_r}\right)
\nonumber\\
 &&\hspace{1.3cm} +  O\left(\alpha_s^3\right) +  
\big\{x \to 1-x \big\}\Bigg],
\label{eq:Texp}
\end{eqnarray}
where $\mu_r$ is the renormalization scale
and the leading-order (LO) contribution is given by
\begin{equation}
T^{(0)}(\omega,x) = \frac{1}{1-\omega(2x-1)} .
\label{eq:T0omega}
\end{equation}
The normalization of $T$ given above corresponds to $\phi(x,\mu_f)$
normalized to one and $f_\pi=131$ MeV. 
Note that a residual dependence on the renormalization scale $\mu_r$ 
appears in the truncated perturbative expansion of the hard-scattering 
amplitude. 
The
next-to-leading order (NLO) correction to the hard scattering amplitude
has been calculated in the modified minimal subtraction (\MS) scheme
\cite{AguCha81,Bra83,KadMikRad86}. In the next-to-next-to-leading order
(NNLO), only the contributions coming from the quark bubble insertions
have been evaluated \cite{GodKiv97,BelSch98,MelNizPas01}, again using
the \MS\ scheme. 
The pion distribution amplitude $\phi(x,\mu_f)$ is
intrinsically a non-perturbative quantity and 
cannot be determined from the perturbation theory. 
However, its evolution is governed by the
evolution equation
\begin{equation}
  \mu_f^2 \frac{d}{d \mu_f^2} \phi(x,\mu_f)   =
   V(x,u,\mu_f) \, \otimes \, \phi(u,\mu_f)
         \, ,
\label{eq:eveq}
\end{equation}
in which the evolution kernel has a perturbative expansion as 
\begin{equation}
V(x,y,\mu_f) = \frac{\alpha_s(\mu_f)}{2\pi} V^{(0)}(x,y) +
\frac{\alpha_s^2(\mu_f)}{(2\pi)^2}  V^{(1)}(x,y) +
O\left(\alpha_s^3\right) \, .
\label{eq:Vexp}
\end{equation}
The evolution kernel has been estimated to NLO accuracy using the \MS\
scheme \cite{Sar84,DitRad84,MikRad85} and the corresponding solution of
the evolution equation has been obtained \cite{MikRad86,Mue94,Mue95}.
The latter can be expressed in the form 
\begin{equation}
\phi(x,\mu_f|\mu_0)=\phi^{(0)}(x,\mu_f|\mu_0) 
+\frac{\alpha_s(\mu_f)}{2\pi} \phi^{(1)}(x,\mu_f|\mu_0)
+ O\left(\alpha_s^2\right) \, ,
\label{eq:phiexp}
\end{equation}
where the scale $\mu_0$ denotes some low scale at which the
non-perturbative input has been obtained. The solution \req{eq:phiexp}
satisfies the initial condition
$\phi(x,\mu_0|\mu_0)=\phi^{(0)}(x,\mu_0|\mu_0)$ and for $\mu_f \to
\infty$ takes the asymptotic form $\phi(x,\mu_f \to \infty |\mu_0)=6 x
(1-x)$. We stress that the evolution equation as defined by
\req{eq:eveq} and \req{eq:Vexp}
corresponds to the simplified scheme fixed by the
preference that the distribution amplitude $\phi$ should have no residual
dependence on the renormalization scale\footnote{Note that, in general,
such a residual dependence appears along with 
the evolution kernel depending on two scales:
\begin{displaymath}
V(x,y, \mu_f) = \frac{\alpha_s(\mu_r)}{2\pi} V^{(0)}(x,y) +
\frac{\alpha_s^2(\mu_r)}{(2\pi)^2}  \left( V^{(1)}(x,y) 
- \frac{\beta_0}{2} V^{(0)}(x,y) \ln
  \left(\frac{\mu_r^2}{\mu_f^2}\right) \right) +
O\left(\alpha_s^3\right) \, .
\end{displaymath}
}.

The photon-to-pion transition form factor has been measured at large
$Q^2$ by the CELLO \cite{Behetal91} and CLEO \cite{Sav97}
collaborations, where one photon is almost on-shell, while the second
one has a virtuality up to $9\ \mbox{GeV}^2.$ 
Different authors analyzed the data
\cite{KroRau96,RadRus96,CaoHuaMa96,RadRus96a,MusRad97,
      SteSchKim98,SteSchKim00, BakMikSte01} 
and it is often stated that the pion distribution amplitude
is close to its asymptotic shape; for previous work
see also \cite{BraHal94,JakKroRau96,Ong95}.
However, in this kinematics, the
shape of the distribution amplitude can be constrained only from the
scaling violation that arises from the evolution of the distribution
amplitude. In the small $ | \omega | $-region, perturbative QCD gives a
parameter free prediction of the photon-to-pion transition form factor
and in the intermediate region, one might extract a few lowest moments
of the distribution amplitude and confront them with non-perturbative
results, see Ref. \cite{DieKroVog01}. However, both high precision data
as well as a precise understanding of perturbative and non-perturbative
effects are necessary for such analysis.

Thus, the computation of both perturbative and power suppressed
contributions is an important task. In this way, we can gain insight
into the perturbative approach to exclusive processes. However,
calculations of exclusive amplitudes beyond the leading order are quite
cumbersome. Besides the photon-to-meson transition form factor and
similar exclusive two-photon processes, the perturbative
next-to-leading order predictions are known only for the pion form
factor
\cite{FieGupOttCha81,DitRad81,Dit82,Sar82,RadKha85,BraTse87,MelNizPas98}
and for the amplitude of charged meson pair production in two-photon
collision for the case of equal momenta sharing meson
distribution amplitude \cite{Niz87}. Fortunately, in the perturbative sector 
massless QCD is invariant under conformal transformation provided the
coupling has a fixed point, so that the $\beta$-function, the
renormalization group coefficient of the running coupling, vanishes.
In lowest order of $\alpha_s$, the conformal symmetry breaking part,
which is consequently proportional to $\beta/g = \beta_0 \alpha_s/(4\pi) +
O(\alpha_s^2)$, can be determined by calculating the abelian part of the
gluon self-energy proportional to the number of quarks $n_f$. Additional
subtleties may appear owing to the factorization procedure and they can
be resolved by a finite renormalization of the hard-scattering and the
distribution amplitude. Making use of conformal symmetry constraints,
together with the explicit calculation of terms proportional to the
$\beta$-function, offers a considerable simplification of the
perturbative calculation and, in our case, gives the possibility of
going beyond the NLO approximation.

Indeed, for the photon-to-pion transition form factor we can take
advantage of this symmetry and its predictive power \cite{Mue97a,Mue98}
by means of the conformal OPE (COPE) \cite{FerGriGat71a,FerGriGat72a},
in which the form of the Wilson--coefficients is constrained. The
normalization of these coefficients can be recovered in the forward
kinematics from the DIS results for the non-singlet coefficient function of
the polarized structure function $g_1$ known to NNLO \cite{ZijNee94}.
This field-theoretical approach has been explored
\cite{Mue94,Mue95,BelMue97a,BelMue98a,BelMue98c} and tested to NLO
\cite{BelMueSch98}, where the $\beta$-function is absent in 
the Wilson-coefficients. We emphasise that the ``conformal
symmetry breaking'' due to the factorization procedure 
in the \MS\ scheme and the restoration of conformal symmetry 
by finite renormalization are well understood at NLO. 
Further consistency checks are based on 
comparison with explicit results (e.g.\ hard-scattering amplitudes for
two-photon processes in the light-cone dominated region, the flavour
non-singlet kernel, quark bubble insertions in singlet kernels), and
with constraints coming from the $N=1$ super Yang-Mills theory
\cite{BelMue00b}.

In this paper we apply the COPE combining the NNLO result for the
non-singlet coefficient function of the polarized structure function
$g_1$ \cite{ZijNee94} with the explicit result for the $n_f$ (i.e.,
$\beta_0$) proportional NNLO contribution to the hard-scattering
amplitude of the photon-to-pion transition form factor
\cite{BelSch98,MelNizPas01}, both being evaluated in the \MS\ scheme, to
obtain a full NNLO result for the photon-to-pion transition form factor
in the so-called conformal factorization scheme. Alternatively, we
propose a scheme in which the hard-scattering amplitude can already be
constructed from a knowledge of the non-singlet coefficient function
of $g_1$ and the corresponding anomalous dimensions.

The paper is organized as follows. For the convenience of the reader, in
Section 2 we review the predictive power of conformal symmetry relevant
to the photon-to-pion transition form factor. We then propose two
treatments of terms proportional to the $\beta$-function and discuss the
remaining freedom in the choice of the factorization procedure. The
general structure of the hard-scattering amplitude in the \MS \ scheme,
and the NNLO term that is proportional to $\beta_0$, are analyzed in
Section 3. For the phenomenologically important case, in which one
photon is quasi-real, we evaluate conformal moments for the
hard-scattering amplitude and making use of the NNLO results for the
$g_1$ function we obtain the NNLO prediction for the photon-to-pion
transition form factor in the conformal factorization scheme. We extend
this procedure to other values of photon virtualities and present a
detailed investigation of the conformal partial wave decomposition of
the transition form factor in different $ \omega $ regions. Based on these
results, in Section 4 we analyze the size of the NLO and NNLO effects
for one quasi-real photon and in the small and intermediate $ | \omega | $
regions. Finally, a summary and conclusions are given in Section 5.
The appendices are devoted to technical details: the Feynman-Schwinger
representation of the hard-scattering amplitude, a consistency check at
NNLO between the $n_f$-proportional \MS \ results for the
hard-scattering amplitude of the photon-to-pion transition form factor
and the results for the non-singlet coefficient function of the DIS
polarized structure function $g_1$, evaluation of the conformal moments
of the hard scattering amplitude, the Taylor expansions in $\omega$, and
the prescription for reconstructing the hard-scattering amplitude in the
momentum fraction space from the known conformal moments.

\section{Outlining the conformal symmetry formalism}
\label{Sec-Conf}
\setcounter{equation}{0}

In the physical sector, massless QCD at the tree level is invariant under
conformal transformations, i.e., under space-time transformations
containing the Poincar{\' e} transformations, dilatation, and special
conformal transformations. The latter are composed of translation,
inversion, and translation again. Conformal symmetry implies an
improvement of the energy-momentum tensor, which then becomes traceless.
Owing to this symmetry, one has additional constraints for
field-theoretical quantities, e.g.\, for Green functions. This subject
was intensively studied in the 60th and 70th in the four-dimensional
field theory. In QCD, conformal symmetry is manifested in the Crewther
relation \cite{Cre72AdlCalGroJac72BroKat93Cre97} and in the solution for
the mixing problem of composite operators under renormalization
\cite{BroFriLepSac80, Mue94,BelMue98c}. The reduced matrix elements of
the conformal operators ${\cal O}$, sandwiched between the vacuum
$|\Omega\rangle$ and one pion $\langle \pi(P)| $ states, are pertinent
to the expansion of the distribution amplitude\footnote{ It is common in
the literature that the distribution amplitude is expanded in the form
\begin{displaymath}
\phi(x,\mu_f) = 6 \sum_{j=0}^{\infty}{}' 
 x (1-x) \, B_j(\mu_f)\, C\, _j^{3/2}(2 x -1)\, ,
\end{displaymath}
where $B_j$ ($j=2,4,\cdots$) essentially represent the non-perturbative
input. Comparing with our definition (\ref{Rel-DA-ConMom}), we have 
$\langle \pi(P)| {\cal O}_{jj}(\mu_f)|\Omega\rangle^{\rm red} 
= 6 N_j B_j(\mu_f)$, where
$\langle \pi(P)| {\cal O}_{00}(\mu_f)|\Omega\rangle^{\rm red}
= B_0=1$ is a renormalization group invariant quantity.}
\begin{eqnarray}
\label{Rel-DA-ConMom}
\phi(x,\mu_f)=\sum_{j=0}^{\infty}{}' \;
 \frac{x (1-x)}{N_j}  C_j^{3/2}(2 x -1)
\langle \pi(P)|  {\cal O}_{jj}(\mu_f) |\Omega\rangle^{\rm red}
\, ,\qquad
N_j = \frac{(j+1)(j+2)}{4(2j+3)}\, .
\end{eqnarray}
Here $C_j^{3/2}$ are the Gegenbauer polynomials with the index $\nu=3/2$ of
order $j$ and the sum runs over even $j$. In this representation, the
transition form factor reads
\begin{equation}
F_{\gamma \pi}(\omega,Q)=
f_{\pi} \,\sum_{j=0}^{\infty}{}' \; T_j(\omega,Q,\mu_f) \,
\langle \pi(P)|  {\cal O}_{jj}(\mu_f) |\Omega\rangle^{\rm red}
\, ,
\label{eq:Fpij}
\end{equation}
where $T_j$ denotes the $j$th conformal moment of the hard-scattering 
amplitude:  
\begin{eqnarray}
\label{Def-Har-Sca-Amp}
T_j(\omega,Q,\mu_f) &\!\!\!=\!\!\!&
\int_0^1 \; dx \; T(\omega,x,Q,\mu_f) \; \frac{x (1-x)}{N_j} \; 
C_j^{3/2}(2 x -1)
\\ &\!\!\! =\!\!\! &
\frac{\sqrt{2}}{3 Q^2}
\left[ T_j^{(0)}(\omega)
+ \frac{\alpha_s(\mu_r)}{2 \pi} T_j^{(1)}\left(\omega,\frac{Q}{\mu_f}\right)
+ \frac{\alpha_s^2(\mu_r)}{(2 \pi)^2}
  T_j^{(2)}\left(\omega,\frac{Q}{\mu_f},\frac{Q}{\mu_r}\right)
+ O(\alpha_s^3) \right]
\, .
\nonumber 
\end{eqnarray}

As reviewed in Section \ref{SubSec-RenPro}, the operator mixing problem
under renormalization beyond the one-loop level is solved by the
restoration of conformal symmetry. 
In Section \ref{SubSec-COPE} this allows us to employ 
conformal symmetry to the OPE of two
electromagnetic currents, and to fix the hard-scattering amplitude
(\ref{Def-Har-Sca-Amp}) in the conformal limit. Additionally, we include
$\beta$ proportional terms and in Section \ref{SubSec-Ambi} we discuss
the corresponding ambiguities of this procedure. The solution of the
renormalization group equation to NNLO is worked out in Section
\ref{SubSec-Evo}.

\subsection{Renormalization properties of conformal operators
and the conformal scheme}
\label{SubSec-RenPro}

Let us start with the constraints for the renormalization of composite
operators. In the flavour non-singlet and parity odd sector 
the twist-2 operators read \cite{Mak81,Ohr82,BalBra89}
\begin{equation}
\label{Def-treeCO-NS}
{\cal O}_{jl}
=
\bar{\psi}(x) \,
(n\cdot\gamma) \; \gamma_5 \: \,
 C^{3/2}_j\! \!
\left( \frac{n\cdot\stackrel{\leftrightarrow}{D}}{n\cdot\partial} \right) 
(i n\cdot \partial)^l\, \psi(y){\Big|_{x=y=0}}\,
\end{equation}
where $\partial \!= \stackrel{\rightarrow}{\partial}
\!\!+\!\! \stackrel{\leftarrow}{\partial}$, 
 $\stackrel{\leftrightarrow}{D}
= \stackrel{\rightarrow}{D} - \stackrel{\leftarrow}{D}$ 
and $n$ is a light-like vector that makes these operators symmetric and
traceless. The Gegenbauer polynomials $C^{3/2}_j$ arise from the
group-theoretical construction of the operators and they are of order
$j$, where this label is related to the conformal spin $j+1$, i.e., the
eigenvalue of the Casimir operator of the so-called collinear conformal
group. These operators have spin $l+1$ and canonical dimension
$l+3$. In other words, we have different infinite irreducible
representations of the conformal algebra, called towers, that are
characterized by the conformal spin $j+1$, while the members of each
representation are labeled by the spin $l+1$. The conformal operators
with $l=j$ are the lowest members of each conformal tower, and we can
climb the tower by acting with the generator of translation.

Employing Poincar\'e invariance, the general form of the renormalization
group equation for the operators introduced above reads
\begin{eqnarray}
\mu \frac{d}{d\mu}{\cal O}_{j l} =- \sum_{k=0}^j \gamma_{jk} {\cal O}_{kl}.
\label{eq:RGE}
\end{eqnarray}
In the conformal invariant theory operators of different conformal
towers do not mix under renormalization. Indeed, the anomalous dimension
matrix
\begin{eqnarray}
\gamma_{jk} =
\frac{\alpha_s}{2\pi}\delta_{jk}\gamma_{j}^{(0)} + 
\frac{\alpha_s^2}{(2\pi)^2}
\gamma_{jk}^{(1)} + \frac{\alpha_s^3}{( 2 \pi)^3} \gamma_{jk}^{(2)}
+ O(\alpha_s^4) \, , \quad \mbox{with} \quad \gamma_{j} \equiv 
\gamma_{jj} \, ,
\label{eq:gammajk}
\end{eqnarray} 
is diagonal at LO. 
This property is induced by conformal symmetry at tree level. The
fact that these operators will mix beyond LO even for the
vanishing $\beta$-function 
in the $\overline{\rm MS}$ scheme, has been considered as an
unexpected breakdown of conformal symmetry. Note that the appearance of the
anomalous dimension already requires a `redefinition' of the conformal
representation at tree level, i.e., the scaling dimensions of the operators
change.

The understanding of this subtlety is the key for the application of 
conformal symmetry in {\em all} orders of perturbation theory. It is
well known that, on the quantum level, conformal symmetry is broken owing to
the regularization of ultraviolet divergences, which shows up in
the trace anomaly of the (improved) energy-momentum tensor. This trace
anomaly is given as a linear combination of different renormalized
operators. In the dimensionally regularized theory with space-time
dimension $n=4-2\epsilon$, it reads 
\begin{eqnarray}
\label{def-tra-ano}
{\mit\Theta}_{\mu\mu} (x)
=  \frac{\beta_\epsilon}{2 g} \left(G^a_{\mu\nu}(x)\right)^2 +
\cdots, 
\end{eqnarray}
where the $\beta$-function in the regularized theory is defined as
\begin{eqnarray}
\beta_\epsilon\equiv \mu\frac{\partial g}{\partial\mu}  =
-\frac{4-n}{2}g + \beta\, \quad  \mbox{with}\quad
 \frac{\beta}{g}= \frac{\alpha_s}{4\pi} \beta_0 +
O(\alpha_s^2)
\end{eqnarray}
and $\beta_0 = (2/3) n_f - (11/3) C_A$.
Besides the square of the renormalized field strength tensor
$G^a_{\mu\nu}$ multiplied by the $\beta$-function, the trace anomaly
(\ref{def-tra-ano}) contains equations of motion and BRST-exact
operators. Therefore, it is sometimes believed that in the physical
sector of the theory the breaking of conformal symmetry is in general
proportional to the $\beta$-function. However, if one deals with
composite operators, the operator product and the
trace anomaly of these operators contain additional ultraviolet divergences. 
Being multiplied by the $(4-n)$ contribution in $\beta_\epsilon$, 
these UV divergences produce finite symmetry
breaking terms that are not proportional to the $\beta$-function. The
appearance of anomalous dimensions of composite operators can be
understood in this way, too.

A detailed analysis shows that the non-diagonality of the anomalous
dimension matrix observed in the $\overline{\rm MS}$ scheme at NLO
originates from such an effect of conformal symmetry breaking. It
already appears at LO in the Ward--identities of these operators with
respect to the special conformal transformation. The calculation of this
special conformal anomaly matrix $\hat{\gamma}^{c}(l) =
(\alpha_s/2\pi)\hat{\gamma}^{c(0)}(l) + O(\alpha_s^2) $ results in 
\begin{equation}
\hat{\gamma}^{c(0)}(l) = -\hat{b}(l) \hat{\gamma}^{(0)}+\hat{w}\, ,
\label{CWI-scBre2}
\end{equation}
where
\begin{eqnarray}
b_{jk}(l)&=&\left\{\begin{array}{c@{\quad}l} 
                         2(l+k+3) \delta_{jk}
                           - 2(2k+3) & 
                          j -k \ge 0 \mbox{ and even}\\
                             0     &   \mbox{otherwise} 
                          \end{array}\right. \, ,
\nonumber
\end{eqnarray}
and
\begin{eqnarray}
w_{jk}&=&C_F \left\{
  \begin{array}{c@{\quad}l} 
 \begin{array}{l}
     -4(2k+3)(j-k)(j+k+3) \\[0.1cm]\times
        \displaystyle
 \left[{A_{jk}-\psi(j+2)+\psi(1)\over (k+1)(k+2)} 
               +{2A_{jk}\over (j-k)(j+ k+3)}\right] 
    \end{array} 
         & 
       j -k > 0 \mbox{ and even}\\[0.2cm]
   0     &   \mbox{otherwise} 
                          \end{array}\right. \, , \qquad
\nonumber \\[0.2cm]
A_{jk}&=&
		 \psi\left({j+k+4\over 2}\right)-
\psi\left({j-k\over 2}\right)
       +2\psi\left(j-k\right)-\psi\left(j+2\right)-\psi(1)\, , 
\end{eqnarray}
with $\psi\left(z\right)= {d\over dz} \ln\Gamma(z)$ and $C_F=4/3$.
It induces off-diagonal entries in the anomalous dimension matrix 
\req{eq:gammajk}:
\begin{eqnarray}
\label{Def-gamND}
\gamma_{jk}^{\MSeq(1)} =
 - \frac{\left[\hat{\gamma}^{c(0)}+\beta_0 \hat{b},\gamma^{(0)} 
\right]_{jk}}
{a(j,k)}\quad \mbox{\ for\ }  j>k\, ,
\end{eqnarray}
where
$
a(j,k)=2(j-k)(j+k+3)
$.
The prediction (\ref{Def-gamND}) coincides with the moments of the
explicitly calculated evolution kernel at NLO \cite{Sar84,DitRad84,MikRad85}.
On the other hand, the diagonal part of the anomalous dimension 
matrix \req{eq:gammajk}
\begin{equation}
\gamma_{jj} \equiv \gamma_j=
\frac{\alpha_s}{2\pi}\gamma_{j}^{(0)} 
+ \frac{\alpha_s^2}{(2\pi)^2}
\gamma_{j}^{(1)} + \frac{\alpha_s^3}{(2\pi)^3}
\gamma_{j}^{(2)} + O(\alpha_s^4)
\label{eq:gammajj}
\end{equation}
coincides with the anomalous dimensions of the operators that are
restricted to the forward kinematics and are known as the moments of the
splitting kernels in DIS.

As we can see, there is a complete understanding of conformal symmetry
breaking in the $\overline{\rm MS}$ scheme. Now the questions arise:
\begin{itemize}
\item
Can we find a scheme at which conformal symmetry
holds true?
\item
Can we then use the predictive power of conformal symmetry?
\end{itemize}

The first question has a positive answer in the case that the
$\beta$-function has a fixed point. Instead of relying on the hypothetical
fixed point, we simply freeze the coupling by hand, which implies 
$\beta=0$. It is then possible to find a scheme in which the
renormalized conformal operators (\ref{Def-treeCO-NS}) form an
irreducible representation of the collinear conformal group, i.e. their
special conformal anomaly and anomalous dimension matrices are
simultaneously diagonal. The rotation from the $\overline{\rm MS}$ to
such a scheme, which we call the conformal subtraction (CS) scheme, is given
by the matrix $\hat{B}$ defined by
\begin{eqnarray}
\label{blefdt-1}
{\cal O}^{\rm CS} =\hat{B}^{-1} {\cal O}^{\MSeq}, \qquad
B_{jk} = \delta_{jk} +\frac{\alpha_s}{2\pi} B_{jk}^{(1)} + O(\alpha_s^2)\, .
\end{eqnarray}
The NLO term $B_{jk}^{(1)}$ is  
entirely determined by the special conformal anomaly
(\ref{CWI-scBre2}) and reads
\begin{eqnarray}
B^{(1)}_{jk} \: = \: -\theta(j>k) \; \frac{\gamma^{c(0)}_{jk}}{a(j,k)} 
\: = \: \theta(j>k) \: \left\{ \hat d \hat\gamma^{(0)} - \hat g \right\}_{jk}.
\label{eq:B1jk}
\end{eqnarray}
Here we introduce the notation $d_{jk} = b_{jk}/a(j,k)$ and
$g_{jk}=w_{jk}/a(j,k)$. In the case of a non-vanishing $\beta$-function, an
additional off-diagonal term appears and, thus, the complete anomalous
dimension in the CS scheme reads \cite{Mue94,Mue95}
\begin{eqnarray}
\label{AnoDim-CS-MS}
\gamma^{\rm CS}_{jk} =
\left\{	\hat{B}^{-1}\hat{\gamma}^{\MSeq} \hat{B} + 
	\hat{B}^{-1} \left[\mu \frac{d}{d \mu}\hat{B}\right]\right\}_{jk}
=\delta_{jk} \gamma_j + \theta(j>k)\frac{\beta}{g} \Delta_{jk}
\, .
\end{eqnarray}
The addenda of the anomalous dimension matrix (\ref{AnoDim-CS-MS}) is
known in the lowest order of $\alpha_s$ \cite{Mue98}
\begin{eqnarray}
\Delta_{jk}= \frac{\alpha_s}{2\pi}\Delta^{(0)}_{jk} +
\frac{\alpha_s^2}{(2\pi)^2}\Delta^{(1)}_{jk}+ 
O(\alpha_s^3)
\quad \mbox{with}\quad 
 \Delta^{(0)}_{jk} =B^{(1)}_{jk} -\left[\hat d,\hat \gamma^{(0)} \right]_{jk}\, 
. 
\label{eq:Djk}
\end{eqnarray}
Note that the diagonal part \req{eq:gammajj}, which corresponds to the
moments of the splitting kernels, also includes $\beta$ proportional
terms. The moments of the splitting kernels are completely known to
two-loop accuracy \cite{FloRosSac77,GonLopYnd79,CurFurPet80,FloLacKou81}
and the first thirteen entries at three-loop level are given in
\cite{RetVer00}. Their scheme dependence is fixed by the fact 
that they are evaluated in the \MS\ scheme. This implies that the
diagonal part of $\hat B$ is given by the identity matrix.

\subsection{Conformal operator product expansion}
\label{SubSec-COPE}

Let us now turn to the second question we have raised. As we have
discussed in the preceding subsection, one advantage of the conformal
scheme is that, up to off-diagonal terms proportional to $\beta$, the
anomalous dimensions are fixed by the DIS results, and, in the conformal
limit, they are partly known to NNLO order\footnote{In principle, we
then also know the Efremov-Radyushkin--Brodsky-Lepage (ER-BL) evolution
kernels, which can be obtained from the
Dokshitzer--Gribov-Lipatov--Altarelli-Parisi (DGLAP) kernels through an
integral transformation \cite{BelMue98b}.}. Furthermore, the class of
two-photon processes that are light-cone dominated, i.e., for $Q^2$
large, can be treated by means of OPE. That includes the evaluation of
the corresponding non-forward Wilson-coefficients. Under the assumption
that conformal symmetry holds true, these coefficients are fixed up to
normalization factors that coincide with the Wilson--coefficients
appearing in the deep inelastic scattering structure functions $F_1$ and
$g_1$ \cite{FerGriGat71a,FerGriGat72a}. Hence, in the conformal scheme,
taking the conformal limit in which conformal symmetry holds true, we
can use this predictive power of the COPE to avoid cumbersome
higher-order calculations. Indeed, the NLO coefficient functions for the
hadronic tensor in the general off-forward kinematics were predicted in
this way \cite{BelMue97a} and they coincide, after rotation to the
$\overline{\rm MS}$ scheme, with explicitly calculated ones
\cite{ManPilSteVanWei97,JiOsb97,JiOsb98}.

For the photon-to-pion transition form factor, the leading twist-2
result of the OPE is given by Eq.\ (\ref{eq:Fpij}), where the conformal
moments $T_j$ of the hard-scattering amplitude correspond to the
Wilson--coefficients $C_j$, which are conventionally normalized as 
\begin{equation}
T_j(\omega,Q,\mu) = \frac{\sqrt{2}}{3 Q^2 }
C_j(\omega|\alpha_s(\mu),Q/\mu) \, .
\label{eq:T2C}
\end{equation} 
As we have mentioned, in the formal conformal limit the Wilson-coefficients
are constrained in the CS scheme by the predictive power of the COPE: 
\begin{eqnarray*}
\alpha_s(\mu) \Rightarrow \alpha_s^\ast - 
\mbox{fixed\ \ implies the reduction\ \ }
C_j (\omega|\alpha_s(\mu),Q/\mu) \Rightarrow
\overline{C}_j(\omega|\alpha_s^\ast,Q/\mu)\, , 
\end{eqnarray*}
where $\overline{C}_j(\omega|\alpha_s^\ast,Q/\mu)$  is given by 
\cite{Mue97a} 
\begin{eqnarray}
\label{Def-OveC}
\overline{C}_j=
	c_j(\alpha_s^\ast) \left(
				\frac{\mu^2}{Q^2}
				\right)^{\frac{\gamma_j}{2}}
	\frac{2(2 \omega)^{j}}{(1+\eta \, \omega)^{j+1+\frac{\gamma_j}{2}}}
	B(j+1,j+2)
{_{2}F}_1\left(
{j+1+\frac{1}{2}\gamma_j, j+2+\frac{1}{2} \gamma_j
 \atop
2(j+2+\frac{1}{2}\gamma_j)}\Bigg|\frac{2\eta\, \omega}{1+\eta\, \omega}
\right)\, .
\nonumber \\
\end{eqnarray}
In the limit $\eta \rightarrow 1$, one obtains the result for the
production of a (pseudo)scalar meson by two virtual photons, while for
$\eta=0$ the forward case is reproduced\footnote{ The
Wilson--coefficients appearing in the deep inelastic scattering
structure functions $F_1$ and $g_1$ are derived in the usual DIS
operator basis, which differs in the normalization from the definition
for the basis of conformal operators. Hence, these Wilson--coefficients
differ slightly from the $\eta=0$ limit of \req{Def-OveC}.}. Note that
${C}_j(\omega)$ is an even (odd) function of $\omega$ for even (odd)
$j$, which is guaranteed by the linear transformation properties of the
hypergeometric functions ${_{2}F}_1$. The normalization $c_j(\alpha_s)$
coincides with the flavour non-singlet Wilson-coefficients of the
polarized structure function $g_1$ taken at $\mu=Q$. It is given as a
perturbative expansion
\begin{eqnarray}
c_j(\alpha_s)
=
c_j^{(0)} + \frac{\alpha_s}{2\pi} c_j^{(1)} +
\frac{\alpha^2_s}{(2\pi)^2} c_j^{(2)} + O(\alpha^3_s)\quad \mbox{with}\quad 
 c_j^{(0)} =1
\label{eq:cjI}
\end{eqnarray}
and is known to NNLO in the \MS\ scheme \cite{ZijNee94}. Strictly
speaking, this coincidence appears just at the hypothetical fixed point
$\alpha_s=\alpha^\ast_s$. However, since we know the forward anomalous
dimensions and the Wilson-coefficients perturbatively in the \MS\ scheme,
we can easily restore $\beta$ proportional terms in these quantities
beyond the conformal limit. 

Conformal symmetry breaking terms,
proportional to the $\beta$-function alter the COPE result
(\ref{Def-OveC}) in the full theory. Obviously, in the irreducible
conformal representation used, the $\beta$ term cannot be fixed from the
requirement of conformal invariance. Thus, the definition of the
conformal scheme 
\begin{eqnarray}
C^{{\rm CS}}\left(\alpha_s(\mu),Q/\mu\right) =
C^{{\MSeq}}\left(\alpha_s(\mu),Q/\mu\right)
 \hat{B}\left(\alpha_s(\mu)\right) \ \mbox{with} \ \
C^{{\rm CS}}\left(\alpha_s^\ast,Q/\mu\right) =
\overline{C}\left(\alpha_s^\ast,Q/\mu\right) \, ,
\end{eqnarray}
is ambiguous and, consequently, the $\hat{B}$-matrix is uniquely
defined only up to $\beta$ proportional terms that are off-diagonal. 
At the two-loop level, ${C}$  contains $\beta_0$
proportional terms appearing in both $c_j^{(2)}$ and $\gamma_j^{(2)}$.
Let us first set such terms to zero and, with the help
of this, single out all $\beta_0$ proportional terms in $C^{\rm CS}$: 
\begin{eqnarray}
\label{Def-CinCS}
C^{{\rm CS}} \!\!\!&=&\!\!\!
C^{{\MSeq}}  \hat{B} {\big|}_{\beta=0} + \frac{\beta}{g} \delta C
\\
\!\!\!&=&\!\!\! 
\overline{C}{\big|}_{\beta=0} +
\frac{\beta}{g} \delta C,
\qquad
\mbox{with}\quad
\delta C= \frac{g}{\beta}
\left( C^{{\MSeq}}\hat{B} -  C^{{\MSeq}}\hat{B}{|}_{\beta=0} \right)
\nonumber
\end{eqnarray}
and  $\alpha_s$ remains running.
At LO
a $\beta_0$ term does not appear in $C^{{\MSeq}}$, thus,
the perturbative expansion 
\begin{equation}
\delta C_j=\delta C_j^{(0)} + \frac{\alpha_s}{2\pi} \delta C^{(1)}_j 
+ O(\alpha_s^2)
\end{equation}
holds true. Note that since $ B^{(0)}_{jk}=\delta_{jk}$ \req{blefdt-1}
the LO coefficient $C^{\MSeq(0)} \equiv C^{(0)}$ is independent of the
scheme. Similarly to \req{Def-CinCS}, we can write the matrix $\hat{B}$
in the general form
\begin{eqnarray}
\hat{B} = \hat{B}{|}_{\beta=0} + 
\frac{\beta}{g}  \delta\hat{B}\quad \mbox{with}\quad
\delta\hat{B} = \delta\hat{B}^{(0)} 
  + \frac{\alpha_s}{2\pi} \delta\hat{B}^{(1)} + O(\alpha_s^2)
\, .
\label{bb}
\end{eqnarray} 
If we define $\hat{B}$ so that it contains no
$\beta_0$-term in order $\alpha_s$, i.e.,
if we take $\delta\hat{B}^{(0)}=0$ as in our definition
of the CS scheme (\ref{blefdt-1}-\ref{eq:B1jk}),
the coefficients $\delta C$ up to NNLO read
\begin{eqnarray}
\label{Tra-WilCoe-1}
\delta C^{(0)} =0, \qquad
\delta C^{(1)} = \frac{2}{\beta_0} 
\left( C^{{\MSeq}(2)} -  C^{{\MSeq}(2)}{|}_{\beta_0=0} \right) 
+ C^{(0)}
\delta\hat{B}^{(1)}. 
\end{eqnarray}
Since we have required that the diagonal entries of $\hat{B}$ should be one, 
the normalization coefficients $c_j$ coincide in the forward limit with the
Wilson-coefficients of $g_1$ calculated in the $\overline{\rm MS}$-scheme.

\subsection{Ambiguities in the definition of the conformal scheme
}
\label{SubSec-Ambi}

As we have discussed, the ambiguity left in the definition of the
conformal scheme in the full theory resides in the $\beta$ proportional
off-diagonal terms, i.e., in the choice of $\delta \hat B$ in
(\ref{bb}). Adopting the definitions \req{blefdt-1} and \req{bb}, we set
$\delta\hat{B}^{(0)}=0$ and in the following discuss different choices of
$\delta\hat{B}^{(1)}$, restricting ourselves to NNLO.

\subsubsection{Defining CS and $\overline{\mbox{CS}}$ schemes}

The naive choice is to set
\begin{eqnarray}
\label{naiveCS}
\delta \hat B^{(1)}=0,
\end{eqnarray}
which means that the $\beta_0$ proportional term in the NNLO
Wilson--coefficients, i.e., $\delta C^{(1)}$, is entirely evaluated in
the \MS\ scheme. Since the conformal symmetry breaking part appearing in
the COPE has to be proportional to the $\beta$-function, we can
calculate $\delta C^{(1)}$ by evaluating the contributions
proportional to the $n_f$ piece of $\beta_0$, i.e., from the one-loop
Feynman graphs with an additional quark-bubble insertion.

In this naive scheme (\ref{naiveCS}), which we denote by CS, both the
conformal operators and the Wilson--coefficients will mix under
renormalization to NLO accuracy owing to the running of the coupling. Let
us consider this in more detail. Since the transition form factor is
invariant under renormalization, i.e.,
\begin{eqnarray}
\mu \frac{d}{d \mu} F_{\gamma \pi}(\omega,Q) = 0,
\end{eqnarray}
these effects will compensate one another. Thus, the renormalization
group equation for the Wilson-coefficients reads
\begin{eqnarray}
\left[\mu\frac{\partial}{\partial \mu} + \beta \frac{\partial}{\partial g}
\right] C_j^{\rm CS} =
\left[\frac{\alpha_s(\mu)}{2\pi} \gamma_j^{(0)} +
\frac{\alpha_s^2(\mu)}{(2\pi)^2} \gamma_j^{(1)} 
\right]C_j^{\rm CS} + \frac{\beta}{g} 
\sum_{i=j+2}^\infty\!\!{}'   
\frac{\alpha_s(\mu)}{2\pi}C_i^{\rm CS} \Delta_{ij}^{(0)}
+O(\alpha_s^3),
\end{eqnarray}
where the addenda $\Delta_{ij}^{(0)}$ is defined in Eq.\ (\ref{eq:Djk}).
Since $C_j^{\rm CS}$ is conformally covariant to the NLO approximation, 
i.e., it contains no partial waves (\ref{Def-OveC}) with a conformal spin 
larger than $j+1$, the off-diagonal entries on the r.h.s.\ arise from the
explicit $\mu$ dependence of $C_j^{{\rm CS}(2)}$, which has been taken
from the \MS\ scheme.

In the alternative conformal scheme, denoted in the following by
$\overline{\rm CS}$, this intermediate mixing is avoided by the complete
diagonalization of the renormalization group equation. This can be
achieved by including an explicit $\mu$ dependence in the $\hat
B$-matrix, i.e., by taking
\begin{eqnarray}
\label{ImpCS}
 \delta B^{(1)}_{jk} = \ln\left(\frac{\mu^{\ast 2}}{\mu^2}\right)
\Delta_{jk}^{(0)} \, \theta(j>k) + \cdots
\, .
\end{eqnarray}
In the order we are considering, the matrix $\hat B$ now reads
\begin{eqnarray}
\label{def-B-Dia}
B_{jk} = \delta_{jk} + \frac{\alpha_s(\mu) }{2\pi} B_{jk}^{(1)} +
\frac{\alpha_s^2(\mu) }{(2\pi)^2} \left\{ B_{jk}^{(2)} +
\frac{\beta_0}{2} 
\left[ \ln\left(\frac{\mu^{\ast 2}}{\mu^2}\right) \, 
\Delta_{jk}^{(0)}\theta(j>k)
+ \cdots \right]
   \right\} + O(\alpha_s^3).
\end{eqnarray}
This choice introduces $\gamma^{\overline{\rm CS}}$ given by Eqs.
(\ref{AnoDim-CS-MS}-\ref{eq:Djk}) with $\Delta^{(0)}_{jk} \to
\Delta^{\overline{\rm CS}(0)}_{jk}=0$, i.e., LO addenda to the anomalous
dimension matrix vanishes in this scheme. However, for dimensional
reasons the choice \req{def-B-Dia} additionally introduces a new
residual scale dependence $\mu^\ast$. The meaning of this procedure is
obvious. We do not resum the remaining off-diagonal $\ln\mu$ terms through
the renormalization group equation, rather we include them in the
Wilson-coefficients, where they will be annulled. This is indeed on the
same footing with that we have already discussed in the $\beta=0$ case, 
where the
off-diagonal entries present in the \MS\ scheme have been removed by a finite,
however, $\mu$ independent renormalization. 
Consequently,  both  the operators and the Wilson--coefficients  
\begin{eqnarray}
{\cal O}^{\overline{\rm CS}} =\hat{B}^{-1} {\cal O}^{\MSeq}, \qquad
{C}^{\overline{\rm CS}} = {C}^{\MSeq} \hat{B},
\end{eqnarray}
where $\hat{B}$ is defined by Eq.\ (\ref{def-B-Dia}), now satisfy the
desired renormalization group equations in the $\alpha_s^2$ approximation: 
\begin{eqnarray}
\label{def-O-Dia}
\mu\frac{d}{d\mu}{\cal O}_{jl}^{\overline{\rm CS}}(\mu) &\!\!\! =\!\!\! & -
\left[
\frac{\alpha_s(\mu)}{2\pi} \gamma_{j}^{(0)} +
\frac{\alpha_s^2(\mu)}{(2\pi)^2} \gamma_{j}^{(1)} + O(\alpha_s^3)
 \right]{\cal O}_{jl}^{\overline{\rm CS}} (\mu) ,
\\
\label{RGE-Dia-WilCoe}
\left[\mu\frac{\partial}{\partial \mu} + \beta \frac{\partial}{\partial g}
\right] C_j^{\overline{\rm CS}}\left(\alpha_s(\mu), Q/\mu\right) 
&\!\!\! =\!\!\! &
\left[\frac{\alpha_s(\mu)}{2\pi} \gamma_j^{(0)} +
\frac{\alpha_s^2(\mu)}{(2\pi)^2} \gamma_j^{(1)} + O(\alpha_s^3)
\right]C_j^{\overline{\rm CS}}\left(\alpha_s(\mu), Q/\mu\right).
\end{eqnarray}
Note that the forward anomalous dimensions $\gamma_{j}$ remain
explicitly $\mu$ independent. However, both the off-diagonal piece of
the anomalous dimensions and the Wilson--coefficients now possess a residual
$\mu^\ast$ dependence at the orders $\alpha_s^3$ and $\alpha_s^2$,
respectively.

To restore the $\mu$ dependence of the Wilson-coefficient, we
perturbatively solve its renormalization group equation
(\ref{RGE-Dia-WilCoe}). Up to an integration constant $\delta
C^{\prime}$, its solution to two-loop accuracy can be expressed by the
Wilson--coefficient (\ref{Def-OveC}), appearing in the COPE,
\begin{eqnarray}
 C_j^{\overline{\rm CS}}\left(\alpha_s(\mu), Q/\mu\right) =
\overline{C}_j\left(\alpha_s(\mu), Q/\mu\right)  +
 \frac{\alpha_s^2}{(2\pi)^2} \frac{\beta_0}{2} \delta C^{\prime}\, ,
\label{eq:CjI}
\end{eqnarray}
which now depend on the running coupling:
\begin{eqnarray}
\label{Def-OveC-run}
\overline{C}_j=
c_j\left(\alpha_s(\mu),\frac{Q}{\mu}, \frac{\partial}{\partial \gamma_j}\right)
\left(\frac{\mu^2}{Q^2}\right)^{\frac{\gamma_j}{2}}
	\frac{2(2\omega)^{j} B(j+1,j+2)}{(1+\omega)^{j+1+\frac{\gamma_j}{2}}}
  {_{2}F}_1\left({j+1+\frac{1}{2}\gamma_j, j+2+\frac{1}{2}\gamma_j
 \atop
2(j+2+\frac{1}{2}\gamma_j)}\Bigg|\frac{2\omega}{1+\omega}\right)\, .
\nonumber \\
\end{eqnarray}
The running of $\alpha_s(\mu)$, appearing in the lowest approximation of both
$\gamma_j$ and $c_j$, is compensated by introducing 
a logarithmic change of the $\mu$ 
dependence in the normalization factors:  
\begin{eqnarray}
\label{Def-Coe-FulThe}
c_j\left(\alpha_s(\mu),\frac{Q}{\mu}, \frac{\partial}{\partial \gamma_j}\right)
&\!\!\! =\!\!\! &
c_j^{(0)} + \frac{\alpha_s(\mu)}{2\pi} c_j^{(1)} +
\frac{\alpha^2_s(\mu)}{(2\pi)^2}  c_j^{(2)} 
+\frac{\alpha_s(\mu)}{2\pi}
\frac{\beta_0}{2}\ln\left(\frac{Q^2}{\mu^2}\right) 
\nonumber \\ & &
\times
\left[ \frac{\alpha_s(\mu)}{2\pi}
 \left(c_j^{(1)}+ c_j^{(0)} \frac{\gamma_j^{(0)}}{4}
\ln\left(\frac{Q^2}{\mu^2}\right)
\right) + c_j^{(0)}  \gamma_j^{(0)}\frac{\partial}{\partial \gamma_j^{(0)}}
\right] + O(\alpha_s^3)\, . \qquad 
\end{eqnarray}
Equation (\ref{Def-OveC-run}) should be understood in the sense of a consequent
expansion with respect to $\alpha_s$ up to the order $\alpha_s^2$.

So far we have found a rather natural way to include the effects of the
running coupling in the structure of the COPE result, with the advantage
that the conformal operators do not mix under renormalization in NLO. It
remains to fix the integration constant $\delta C^\prime$, which
vanishes in the kinematical forward limit. We can identify it with the
non-covariant part calculated in the \MS\ scheme, in an analogous way
to our discussion in the case of the CS scheme. On the other hand, it is
rather appealing that the Wilson--coefficients $C_j\left(\alpha_s(\mu),
Q^2/\mu^2\right)$ contain only conformally covariant terms to NNLO. For
the scheme we call $\overline{\mbox{CS}}$, we adopt this
prescription, i.e., we put $\delta C^\prime = 0$. In the NNLO approximation,
we then have a partial wave decomposition of the transition form factor
with respect to the `good' quantum number -- conformal spin. This in
principle allows us to extract the conformal moments of the distribution
amplitude with a `well-defined' conformal spin for experimental data.

\subsubsection{Calculational prescriptions}

Let us comment on the renormalization scale dependence 
and give the calculational prescription for the schemes we have proposed.

First, we have introduced a naive recipe (CS scheme) which combines the
COPE result with that one explicitly calculated in the \MS\ scheme. Now
we extend our analysis by distinguishing between the renormalization
scale $\mu_r$ (the argument of $\alpha_s$ in the Wilson--coefficients),
and the factorization scale $\mu_f$. We require that the matrix elements
of conformal operators should depend only on the factorization scale $\mu_f$.
Thus, the scheme transformation now reads 
\begin{eqnarray}
\label{Def-CinCS2Sca}
C^{{\rm CS}}(\alpha_s(\mu_r),Q/\mu_f,Q/\mu_r) =
C^{{\MSeq}}(\alpha_s(\mu_r),Q/\mu_f,Q/\mu_r) \: \: \hat{B}(\alpha_s(\mu_f)).
\nonumber
\end{eqnarray}
Employing the scale--changing relation
\begin{equation}
 \alpha_s(\mu_f) = \alpha_s(\mu_r)
  \left[ 1 + \frac{\alpha_s(\mu_r)}{2 \pi} \frac{\beta_0}{2}
       \ln \left(\frac{\mu_f^2}{\mu_r^2}\right)  + O(\alpha_s^2)
  \right]
\, ,
\label{eq:scalech}
\end{equation}
we expand the rotation matrix
\begin{eqnarray}
\hat{B}(\alpha_s(\mu_f)) = \hat{B}(\alpha_s(\mu_r)) +
\frac{\alpha_s^2(\mu_r)}{(2 \pi)^2} \frac{\beta_0}{2}
       \ln \left(\frac{\mu_f^2}{\mu_r^2}\right)   \hat{B}^{(1)}
+ O(\alpha_s^3).
\end{eqnarray}
Hence, in this scheme, the Wilson--coefficients read to NNLO accuracy as
\begin{eqnarray}
\label{Def-WilCoeCS}
C^{{\rm CS}}
\!\!\!&=&\!\!\!
\overline{C}(\alpha_s(\mu_r),Q/\mu_f){\bigg|}_{\beta=0}
\\
&&+
\frac{\beta_0}{2} \frac{\alpha_s^2(\mu_r)}{(2 \pi)^2} \left[
 -C^{{\MSeq} (2)}_{\beta}(\alpha_s(\mu_r),Q/\mu_f,Q/\mu_r) +
\ln \left(\frac{\mu_f^2}{\mu_r^2}\right)  C^{(0)} \hat{B}^{(1)}
\right]
+ O(\alpha_s^3)
,
\nonumber
\end{eqnarray}
where $C^{{\MSeq} (2)}_{\beta}$ denotes the  $(-\beta_0/2)$ proportional 
contribution evaluated in the \MS\ scheme, 
while the structure of $\overline{C}$ is fixed
by (\ref{Def-OveC}-\ref{eq:cjI}).

Alternatively in the $\overline{\rm CS}$ scheme, we have employed
renormalization group invariance to incorporate the running of the
coupling into the generic structure of the COPE result and have used a
finite renormalization to preserve the structure of the COPE to NNLO
accuracy:
\begin{eqnarray}
 C_j^{\overline{\rm CS}}\left(\alpha_s(\mu), Q/\mu\right) =
\overline{C}_j\left(\alpha_s(\mu), Q/\mu\right),
\label{eq:CjIbCS}
\end{eqnarray}
where $\overline{C}_j$ is defined by Eqs.\ 
(\ref{Def-OveC-run}-\ref{Def-Coe-FulThe}). The form
of the Wilson--coefficients in which the distinction between the scales
$\mu_r$ and $\mu_f$ is introduced can be obtained analogously to
the previously discussed case of the CS scheme, and will be presented in
Section \ref{SubSec-CS-lim}.

In Section \ref{Sec-NumEst} we employ both of these schemes to estimate
the size of NNLO effects at a given input scale. The missing ingredient
for a consistent NNLO analysis including the evolution of distribution
amplitude, is the anomalous dimension matrix at three-loop level.
Whereas the first few diagonal entries have been calculated, the
off-diagonal part is unknown. It could be read off from the
$n_f$-proportional part of the special conformal anomaly matrix in the
two-loop approximation. Also, generally, the trace anomaly will affect
the COPE at NNLO accuracy, see Eq.\ (\ref{def-tra-ano}). We rather make
use of the freedom to rotate the conformal symmetry breaking piece from
the perturbative sector to the non-perturbative one, as has been done
in the $\overline{\rm CS}$ scheme, or the reverse. However, we expect that
the mixing effect in the $\overline{\rm CS}$ scheme will be negligibly
small and its detailed investigation is beyond the scope of this paper.

\subsection{Evolutional behaviour of conformal operators}
\label{SubSec-Evo}

We end this section with a short review of the evolutional behaviour of
the conformal operators from which that of the distribution amplitude
$\phi(x,\mu)$ can easily be established. For the convenience of the
reader, we repeat here the basic steps for solving the renormalization
group equation \req{eq:eveq} and present the results in a form convenient
for the phenomenological analysis \cite{Mue94,Mue95,Mue98}.

The renormalization group equation (\ref{eq:RGE}) is an inhomogeneous
first-order partial differential equation and after sandwiching the
conformal operators between the hadronic states of interest we obtain
the evolution equation for the reduced matrix elements:
\begin{eqnarray}
\mu \frac{d}{d\mu}\langle \pi(P)| {\cal O}_{jj}(\mu) |\Omega\rangle^{\rm red}
=
-\gamma_{jj}(\mu)
\langle \pi(P)|  {\cal O}_{jj}(\mu) |\Omega\rangle^{\rm red}
- \sum_{k=0}^{j-2} \gamma_{jk}(\mu)
\langle \pi(P)|  {\cal O}_{kk}(\mu) |\Omega\rangle^{\rm red}.
\end{eqnarray}
The solution can be achieved by the ansatz
\begin{eqnarray}
\label{DA-ConSpin-Exp-CS-fin}
 \langle \pi(P)|  {\cal O}_{jj}(\mu) |\Omega\rangle^{\rm red} =
\sum_{k=0}^{j}{^\prime} {\cal B}_{jk}(\mu,\mu_0)
\exp\left\{- \int_{\mu_0}^\mu \frac{d\mu^\prime}{\mu^\prime}
\gamma_k(\mu^\prime) \right\}
\langle \pi(P)|  {\cal O}_{kk}(\mu_0) |\Omega\rangle^{\rm red} \,  ,\
\end{eqnarray}
with the initial condition
\begin{eqnarray}
\label{IntConB}
{\cal B}_{jk}(\mu=\mu_0,\mu_0)=\delta_{jk} \, .
\end{eqnarray}
The recursive solution of this set of differential equations, starting
with the homogeneous one for $j=0$, has been written for an arbitrary
scheme in a compact form (see Ref.\ \cite{Mue94}):
\begin{eqnarray}
\label{sol-B-mixing}
{\cal B}_{jk}=
	\frac{\delta_{jk}}{\delta_{jk}-{\cal L}\gamma^{\rm ND}_{jk}}
	=\delta_{jk} +{\cal L}{\gamma}^{\rm ND}_{jk}+
	 {\cal L}\left(\gamma^{\rm ND}{\cal L}
 \hat{\gamma}^{\rm ND}\right)_{jk}+\cdots\, ,
\end{eqnarray}
where $\hat{\gamma}^{\rm ND}$ represents the triangular off-diagonal
matrix 
\begin{equation}
 \gamma^{\rm ND}_{jk}=
 \left\{
 \begin{array}{cc}
   \gamma_{jk} & \quad \mbox{for} \quad  j>k \\
    0  & \quad \mbox{otherwise}
 \end{array}
 \right.
 \, ,
\end{equation}
and the operator ${\cal L}$ is an integral operator whose action is
defined by
\begin{eqnarray}
\label{sol-B-mixing-L}
{\cal L}\gamma^{\rm ND}_{jk}=
	-\int_{\mu_0}^{\mu} \frac{d\mu^\prime}{\mu^\prime} \;
\gamma^{\rm ND}_{jk}(\mu^\prime)
\; \exp\left\{
-\int_{\mu^\prime}^{\mu}\frac{d\mu^{\prime\prime}}{\mu^{\prime\prime}}
\left[\gamma_{j}(\mu^{\prime\prime})-
	\gamma_{k}(\mu^{\prime\prime})\right] \right\}.
\end{eqnarray}
In the \MS\ or CS scheme, the anomalous dimensions (\ref{eq:gammajk}) 
depend only implicitly
on the scale $\mu$ via the running coupling $\alpha_s(\mu)$.
For the  $\beta$-function, we employ the expansion \cite{TarVlaZho80LarVer93}
\begin{eqnarray}
&&\!\!\!\!\! \frac{\beta}{g} =
\frac{\alpha_s(\mu)}{4 \pi} \beta_0 + \frac{\alpha_s^2(\mu)}{(4 \pi)^2}
\beta_1 +\frac{\alpha_s^3(\mu)}{(4 \pi)^3} \beta_2+ O(\alpha_s^4),
\\
&&\!\!\!\!\!
\beta_0 = \frac{2}{3} n_f -11,\quad
\beta_1 = \frac{38}{3} n_f - 102,
\quad \beta_2 = -\frac{325}{54} n_f^2 + \frac{5033}{18} n_f -
\frac{2857}{2}\, .
\nonumber
\end{eqnarray}
Since the off-diagonal entries of the anomalous dimensions give only
subleading log's, which will not be resummed, we expand the ${\cal B}$
matrix in powers of $\alpha_s$: 
\begin{eqnarray}
{\cal B}_{jk}(\mu,\mu_0)= \delta_{jk} +
\frac{\alpha_s(\mu)}{2\pi } {\cal B}_{jk}^{(1)}(\mu,\mu_0) +
\frac{\alpha_s^2(\mu)}{(2\pi)^2 } {\cal B}_{jk}^{(2)}(\mu,\mu_0) +
O(\alpha_s^3)\, .
\end{eqnarray}
Performing the integrations in Eq.\ (\ref{sol-B-mixing}), we obtain 
the desired results for
\begin{eqnarray}
{\cal B}_{jk}^{(1)}\!\!\! &=&\!\!\!  -R_{jk}(\mu,\mu_0|1) 
\frac{\gamma_{jk}^{{\rm ND} (1)}}{\beta_0}\, ,
\\
{\cal B}_{jk}^{(2)} \!\!\! &=&\!\!\!
\left[R_{jk}(\mu,\mu_0|1) - R_{jk}(\mu,\mu_0|2)\right]
\left[
\frac{\gamma^{(1)}_j-\gamma^{(1)}_k}{\beta_0}-
\frac{\beta_1}{2\beta_0}
\frac{\gamma^{(0)}_j-\gamma^{(0)}_k}{\beta_0}
\right]
\frac{\gamma^{{\rm ND}(1)}_{jk}}{\beta_0} 
\nonumber\\
&&+
R_{jk}(\mu,\mu_0|2)
\left[\frac{\beta_1}{2\beta_0}\frac{\gamma^{{\rm ND}(1)}_{jk}}{\beta_0} -
\frac{\gamma^{{\rm ND}(2)}_{jk}}{\beta_0}\right]+
\sum_{m=k+2}^{j-2}{}^{\!\!\!\!\!\!\prime}
 \frac{\gamma^{{\rm ND}(1)}_{jm}}{\beta_0}
\frac{R_{mk}(\mu,\mu_0|1) - R_{jm}(\mu,\mu_0|2)}
{1+ \frac{\gamma^{(0)}_j-2\gamma^{(0)}_m+ \gamma^{(0)}_k}{\beta_0}}
 \frac{\gamma^{{\rm ND}(1)}_{mk}}{\beta_0} \, , 
\nonumber
\end{eqnarray}
where
\begin{eqnarray}
R_{jk}(\mu,\mu_0|n) = \frac{\beta_0}
{n \beta_0 +\gamma^{(0)}_j -  \gamma^{(0)}_k}
\left[1-
\left(\frac{\alpha_s(\mu_0)}{\alpha_s(\mu)}\right)^{
\frac{n \beta_0 +\gamma^{(0)}_j -  \gamma^{(0)}_k}{\beta_0}}\right]
\, .
\end{eqnarray}
The leading log's, associated with the diagonal entries, are resummed, while
the subleading ones are expanded with respect to $\alpha_s$:
\begin{eqnarray}
e^{\left\{
	- \int_{\mu_0}^\mu \frac{d\mu^\prime}{\mu^\prime}\gamma_k(\mu^\prime)
\right\}} =
\left[
\frac{\alpha_s(\mu)}{\alpha_s(\mu_0)}
\right]^{-\frac{\gamma^{(0)}_k}{\beta_0}}
\left[
	1 + \frac{\alpha_s(\mu)}{2\pi} {\cal A}_k^{(1)}(\mu,\mu_0)
      + \frac{\alpha_s^2(\mu)}{(2\pi)^2} {\cal A}_k^{(2)}(\mu,\mu_0)
+O(\alpha_s^3)
\right]\, ,
\end{eqnarray}
where
\begin{eqnarray}
{\cal A}_k^{(1)}(\mu,\mu_0)\!\!\! &=&\!\!\!
 \left[1-\frac{\alpha_s(\mu_0)}{\alpha_s(\mu)}\right]
\left[
\frac{\beta_1}{2\beta_0}
\frac{\gamma_k^{(0)}}{\beta_0} -\frac{\gamma_k^{(1)}}{\beta_0}
\right]\, ,
\\
{\cal A}_k^{(2)}(\mu,\mu_0) \!\!\! &=&\!\!\! \frac{1}{2} \left[{\cal
A}_k^{(1)}(\mu,\mu_0)\right]^2 -
\left[1-\frac{\alpha_s^2(\mu_0)}{\alpha_s^2(\mu)}\right]
\left[\frac{\beta_1^2-\beta_2 \beta_0}{8\beta_0} \frac{\gamma_k^{(0)}}{\beta_0}
- \frac{\beta_1}{4\beta_0}\frac{\gamma_k^{(1)}}{\beta_0} +
\frac{\gamma_k^{(2)}}{2\beta_0}\right]\, .
\nonumber
\end{eqnarray}
To the considered order, the evolution of the matrix elements then reads 
\begin{eqnarray}
\label{Sol-EvoEqu-NNLO}
\langle \pi(P)|  {\cal O}_{jj}(\mu) |\Omega\rangle^{\rm red} \!\!\! &=&\!\!\!
\sum_{k=0}^j{}^\prime \left[
\frac{\alpha_s(\mu)}{\alpha_s(\mu_0)}\right]^{-\frac{\gamma^{(0)}_k}{\beta_0}}
\Bigg\{\delta_{jk}+ \frac{\alpha_s(\mu)}{2\pi}
\left[ \delta_{jk} {\cal A}_k^{(1)} +
{\cal B}_{jk}^{(1)} \right](\mu,\mu_0)
+
\frac{\alpha_s^2(\mu)}{(2\pi)^2} 
\\
&&\hspace{0cm}\times\left[ \delta_{jk} {\cal A}_k^{(2)}+
 {\cal B}_{jk}^{(1)} {\cal A}_k^{(1)} +
{\cal B}_{jk}^{(2)}
\right](\mu,\mu_0)
+ O(\alpha_s^3)
\Bigg\} \langle \pi(P)|  {\cal O}_{kk}(\mu_0) |\Omega\rangle^{\rm red}\, .
\nonumber
\end{eqnarray}
The off-diagonal entries are known only at NLO and are given in the
\MS\ scheme in Eq.\ (\ref{Def-gamND}). In the CS scheme, they are proportional to
$\beta_0$, as given in Eq.\ (\ref{AnoDim-CS-MS}), whereas in 
the $\overline{\rm CS}$ scheme they are equal to zero by  definition. 
Therefore, the mixing
of operators is an $\alpha_s^2$ suppressed effect:
\begin{eqnarray}
\label{Def-B-barCS}
&&\!\!\!\! {\cal B}_{jk}^{(1)} = 0,
\\
&&\!\!\!\! {\cal B}_{jk}^{(2)} = \frac{\Delta^{(0)}_{jk}}{2}
\frac{\gamma_j^{(0)}-\gamma_k^{(0)}}
{\beta_0+\gamma_j^{(0)}-\gamma_k^{(0)}} 
\left[
\frac{\beta_0+\gamma_j^{(0)}-\gamma_k^{(0)}}{\beta_0}
\ln\left(\frac{\mu^2}{{\mu^\ast}^2} \right) R_{jk}(\mu,\mu_0|2)  \right.
\nonumber\\
&&\hspace{4.7cm} +
 \left. \ln\left(\frac{\mu^2}{\mu_0^2} \right)
\left( \frac{R_{jk}(\mu,\mu_0|2)}
{1-\frac{\alpha_s(\mu_0)}{\alpha_s(\mu)}} -1 \right)  \right] -
\frac{1}{2} R_{jk}(\mu,\mu_0|2)  \Delta^{\overline{\rm CS}(1)}_{jk}\, ,
\nonumber
\end{eqnarray}
where $\Delta^{(0)}_{jk}$ is defined in Eq.\ (\ref{eq:Djk}). Here we
have taken into account the explicit $\mu$--dependence in the anomalous
dimensions, induced by the transformation (\ref{def-B-Dia}). The
addenda
$\Delta^{\overline{\rm CS}(1)}_{jk}=
2\gamma^{\overline{\rm CS}(2)}_{jk}/\beta_0$, 
which is presently unknown,
has to be evaluated
at $\mu=\mu^\ast$.
Note that the $\alpha_s$--power counting remains correct as long as
the scales $\mu^\ast,\ \mu,$ and  $\mu_0$ are of the order of
$Q \gg \Lambda_{\rm QCD}$. The auxiliary scale can now be set, e.g.,
to $\mu^\ast=\mu$.

Let us remark that the evolution of the distribution amplitude can be
formally obtained by resummation of the conformal partial waves 
given in Eq. (\ref{Rel-DA-ConMom}).
Taking into account the evolution of the reduced matrix element in Eq.\
(\ref{Sol-EvoEqu-NNLO}), one finds the eigenfunctions of the
evolution equation, expanded with respect to the Gegenbauer polynomials
\cite{Mue94}.

\section{Hard-scattering amplitude to NNLO accuracy}
\label{Sec-NNLOcalc}
\setcounter{equation}{0}

In the preceding section we have outlined the structure of the conformal
predictions in the conformal momentum space and in this one we analyze
the structure of the NNLO results in the momentum fraction
representation. In Section \ref{SubSec-beta0} we derive a convolution
representation of the NNLO term proportional to $\beta_0$ and also give
the general structure of the hard-scattering amplitude in the \MS\
scheme up to the NNLO order. Furthermore, in Section \ref{SubSec-Lim} we
consider the phenomenologically important case of the asymmetry parameter
$ | \omega | $ equal to 1 and in Section \ref{SubSec-CS-lim} we then present the
NNLO result for the conformal moments in the CS and \CS\ schemes.
In Section \ref{SubSec-SmallOmega} we analogously present the results
at small and the intermediate values of $ | \omega | $.

\subsection{$\beta_0$-proportional NNLO terms and the general
structure of the NNLO results in the \MS\ scheme}
\label{SubSec-beta0}

First, we consider the term proportional to $n_f$, i.e., $\beta_0$,
appearing in the NNLO calculation of the two-photon hard-scattering
amplitude (\ref{eq:Texp}). For the case of general Bjorken kinematics,
the result has been given in Ref.\ \cite{BelSch98} and is easily
restricted to the kinematics of a particular process, i.e., in our case, 
to the kinematics of the photon-to-pion transition form factor 
(see Appendix \ref{App-ConChe}
for the definitions of generalized Bjorken kinematics). In the special
case $ | \omega | =1$ (one photon on-shell), these results coincide with the
results from Ref. \cite{MelNizPas01}.

The authors of Ref.\ \cite{BelSch98} have presented the regularized results 
in terms of hypergeometric functions ${_2\!F_1}$. It is instructive to
rewrite it as a convolution of the amplitude
\begin{eqnarray}
\label{Def-T0-eps}
T^{(0)}(\omega,x|\epsilon) =
\frac{1}{(1-\omega(2x-1))^{1+\epsilon}}
\, ,
\end{eqnarray}
with the kernels
\begin{eqnarray}
\label{Def-va}
v^{a}(x,y|\epsilon)&\!\!\!=\!\!\!& 
\theta(y-x)\left(\frac{x}{y}\right)^{1+\epsilon} + \left\{x \to \bar{x} \atop
y \to \bar{y}  \right\},
\\
\label{Def-vb}
v^{b}(x,y|\epsilon)&\!\!\!=\!\!\!& 
\theta(y-x)
\left(\frac{x}{y}\right)^{1+\epsilon} \frac{1}{y-x}
 + \left\{x \to \bar{x} \atop
y \to \bar{y}  \right\},
\\
\label{Def-g}
g(x,y|\epsilon,\sigma)&\!\!\!=\!\!\!&  \theta(y-x)
\frac{1}{y} \left(1-\frac{x}{y}
\right)^{-1+\epsilon+\sigma}
{\rm B}\!\left(\frac{x}{y},1+\sigma,-\epsilon-\sigma\right)+
\left\{x \to \bar{x} \atop
y \to \bar{y}  \right\},
\end{eqnarray}
where ${\rm B}(x,a,b)= \int_0^x dy\, y^{a-1} (1-y)^{b-1}$ is the
incomplete Beta--function. Here $\sigma$ and $\epsilon$ are the
dimensional regularization parameters ($n=4-2 \sigma [\epsilon]$)
associated with the quark bubble insertion and the overall loop,
respectively. The kernels $\left[v^{a}(x,y|\epsilon)\right]_+$ and
$\left[v^{b}(x,y|\epsilon)\right]_+$ are diagonal with respect to the
Gegenbauer--polynomials $C_j^{3/2 + \epsilon}(2x-1)$ and they are
regularized with the usual $[\cdots]_+$ prescription
\begin{eqnarray}
\left[v(x,y)\right]_+ = v(x,y) -\delta(x-y) \int_0^1dz\, v(z,y)\, .
\end{eqnarray}
Their eigenvalues are
\begin{eqnarray}
v_j^a= \frac{1 + \epsilon}{(1 + j + \epsilon)(2 + j  + \epsilon)} -
\frac{1}{2 + \epsilon},
\qquad
v_j^b= 2 \psi(2+\epsilon)- 2\psi(2+\epsilon+j).
\label{eq:vnab}
\end{eqnarray}

The $g$ kernel is not diagonal with respect to the Gegenbauer--polynomials
and is responsible for the apparent breaking of conformal
symmetry in the $\overline{\rm MS}$-scheme. Its expansion
\begin{eqnarray}
g(x,y|\epsilon,\sigma)= g(x,y) +  g^\prime(x,y)\, \epsilon +
 \dot{g}(x,y)\, \sigma    +
 O\left(\epsilon^2,\sigma^2,\epsilon\sigma\right)
\nonumber
\end{eqnarray}
 reads
\begin{eqnarray}
g(x,y)&\!\!\!=\!\!\!&  - \frac{ \theta(y-x)}{y-x}
\ln\left(1-\frac{x}{y}\right) +
\left\{x \to \bar{x} \atop
y \to \bar{y}  \right\},
\nonumber\\
g^\prime(x,y)&\!\!\!=\!\!\!& - \frac{ \theta(y-x)}{y-x}
 \frac{1}{2} \ln^2\left(1-\frac{x}{y}\right) +
\left\{x \to \bar{x} \atop
y \to \bar{y}  \right\},
\label{eq:g} \\
\dot{g}(x,y)&\!\!\!=\!\!\!&  \frac{ \theta(y-x)}{y-x}
\left[{\rm Li}_2\left(1-\frac{x}{y}\right) - {\rm Li}_2(1) \right]
+
\left\{x \to \bar{x} \atop
y \to \bar{y}  \right\} + g^\prime(x,y).
\nonumber
\end{eqnarray}
There is a similar expansion of the $v^i$ kernels ($i=a,b$) 
\begin{eqnarray}
v^i (x,y|\epsilon)= v^i(x,y) +
 \dot{v}^i(x,y)\, \epsilon +  \frac{1}{2} \ddot{v}^i(x,y)\, \epsilon^2   +
 O\left(\epsilon^3\right),
\nonumber
\end{eqnarray}
where
\begin{eqnarray}
\label{Def-dotKer}
v^i(x,y) &\!\!\!=\!\!\!& \theta(y-x) f^i(x,y)  +
\left\{x \to \bar{x} \atop y \to \bar{y}  \right\},
\nonumber\\
\dot{v}^i(x,y) &\!\!\!=\!\!\!& \theta(y-x) f^i(x,y) \ln\left(\frac{x}{y}\right)
+ \left\{x \to \bar{x} \atop y \to \bar{y}  \right\},
\\
\ddot{v}^i(x,y) &\!\!\!=\!\!\!& \theta(y-x) f^i(x,y) 
\ln^2\left(\frac{x}{y} \right)+
\left\{x \to \bar{x} \atop y \to \bar{y}  \right\},
\nonumber
\end{eqnarray}
and the functions $f^i$  can be read off from Eqs.\ (\ref{Def-va}),
(\ref{Def-vb}).

The LO kernel of Eq. \req{eq:Vexp} 
is expressed in terms of the $v^i$ kernels introduced above:
\begin{eqnarray}
\label{Def-V-LO}
V^{(0)}(x,y) = C_F \left[v(x,y)\right]_+, \qquad
v(x,y)= v^{a}(x,y) + v^{b}(x,y).
\end{eqnarray}
For the NLO kernel we use the colour decomposition
\begin{eqnarray}
V^{(1)}(x,y) = C_F\left[C_F v_F(x,y) -\frac{\beta_0}{2} v_\beta(x,y) -
\left(C_F-\frac{C_A}{2}\right)v_G(x,y) \right]_+ .
\end{eqnarray}
In the following we particularly need the $\beta_0$-proportional kernel
\begin{eqnarray}
\label{Def-EvoKer-Beta}
v_\beta(x,y) = \dot{v}(x,y) + \frac{5}{3} v(x,y) + v^a(x,y).
\end{eqnarray}

The unrenormalized NLO and NNLO corrections 
to the hard-scattering amplitude are of the form
\begin{eqnarray}
\label{Str-tT1}
\widehat{T}^{(1)}(\omega,x) &\!\!\!=\!\!\!& C_F 
\widehat{T}_F^{(1)}(\omega,x) \, ,
\\
\label{Str-tT2}
\widehat{T}^{(2)}(\omega,x) &\!\!\!=\!\!\!& C_F
\left[C_F \widehat{T}_F^{(2)}(\omega,x)-\frac{\beta_0}{2} 
\widehat{T}_\beta^{(2)}(\omega,x)  -
\left(C_F -\frac{C_A}{2}\right) \widehat{T}_G^{(2)}(\omega,x)
\right].
\nonumber
\end{eqnarray}
Employing the integral representation of the hypergeometric
functions, one can express the regularized results of Ref. \cite{BelSch98} 
\begin{eqnarray}
\widehat{T}^{(1)}_{F}(\omega,x)&\!\!\!=\!\!\!&
{\cal M}(\omega, x | \epsilon, 0) \, ,
 \\
\widehat{T}^{(2)}_\beta(\omega,x)&\!\!\!=\!\!\!&  -3
\frac{\Gamma(\epsilon)\Gamma(2 - \epsilon)^2}{\Gamma(4 - 2\epsilon)} 
{\cal M}(\omega, x | \epsilon, \epsilon) 
\end{eqnarray}
in terms of  convolutions
\begin{eqnarray}
\lefteqn{{\cal M}(\omega, x | \epsilon, \sigma)} 
\nonumber \\ &=\!\!\!&
\frac{\Gamma(\epsilon+\sigma)\Gamma(2-\epsilon)\Gamma(1-\epsilon-\sigma)}
{\Gamma(3-2\epsilon-\sigma)\Gamma(1 + \sigma)}
\left(\frac{4\pi \mu^2}{Q^2}\right)^{\epsilon+\sigma}
\int_0^1 dy\, T^{(0)}(\omega,y|\epsilon+\sigma)
\nonumber\\ & & \times
\; \Bigg\{
-\frac{(1-\epsilon)(1-4\epsilon-3\sigma)}{1+\sigma} 
\, \delta(x-y) 
+ \frac{(1-\epsilon) [2-2\epsilon(1-\epsilon)+\sigma(1-\sigma)]}{1+\sigma}
\, v^{a}(y,x|\sigma)
\nonumber\\ & & \quad
+ \frac{(1-\epsilon)[2-\epsilon+
2\epsilon(\epsilon + \sigma)]-\epsilon\sigma(\epsilon+\sigma)}{1 - \epsilon}
\left(\left[v^{b}(y,x|\sigma)\right]_+ - (\epsilon +\sigma)
\left[g(y,x|\epsilon,\sigma)\right]_+ 
\right)
\Bigg\}. \quad
\end{eqnarray}

The results given above contain UV and collinear singularities,
which are removed by renormalization (introduces the scale $\mu_r$)
and factorization (at the scale $\mu_f$) of collinear singularities. 
The renormalization procedure in the $\overline{\rm MS}$
scheme (for details see Ref.\
\cite{MelNizPas01}) induces the following general structure of NLO and 
NNLO corrections present in expansion (\ref{eq:Texp}):
\begin{eqnarray}
\label{Str-T1}
T^{(1)}(\omega,x,Q/\mu_f) &\!\!\!=\!\!\!& C_F T_F^{(1)}(\omega,x)
+ \ln\left(\frac{Q^2}{\mu_f^2}\right)
\left[T^{(0)} \otimes V^{(0)}\right](\omega,x)
\, ,
\\
\label{Str-T2}
T^{(2)}(\omega,x,Q/\mu_f,Q/\mu_r) &\!\!\!=\!\!\!& C_F
\left[C_F T_F^{(2)}-\frac{\beta_0}{2} T_\beta^{(2)}  -
\left(C_F -\frac{C_A}{2}\right) T_G^{(2)}
\right](\omega,x)
+\ln\left(\frac{Q^2}{\mu_f^2}\right)
\\
&&\times \left\{T^{(0)} \otimes  V^{(1)}
+\left[ C_F T^{(1)}_F  + \frac{1}{2}
\ln\left(\frac{Q^2}{\mu_f^2}\right)
T^{(0)} \otimes V^{(0)}\right]\otimes  V^{(0)} \right\}(\omega,x)
\nonumber \\
&&  + \frac{\beta_0}{2} \ln\left(\frac{Q^2}{\mu_r^2}\right)
T^{(1)}(\omega,x,Q/\mu_f)-\frac{\beta_0}{4} \ln^2\left(\frac{Q^2}{\mu_f^2}
\right)
\left[T^{(0)} \otimes  V^{(0)}\right](\omega,x),
\nonumber
\end{eqnarray}
where
\begin{eqnarray}
\label{Def-T1F}
\hspace{-0.5cm}
T^{(1)}_{F}(\omega,x)&\!\!\!=\!\!\!&  T^{(0)}(\omega,y)\otimes
\left\{ {\cal T}^{(1)}_{F}(y,x) + {\rm LN}(\omega,y) [v(y,x)]_+\right\} ,
 \\
\label{Def-T2F}
T^{(2)}_F(\omega,x)&\!\!\!=\!\!\!&  T^{(0)}(\omega,y)\otimes
\Bigg\{{\cal T}^{(2)}_{F}(y,x) + {\rm LN}(\omega,y)\left([v_F(y,x)]_+ +
{\cal T}^{(1)}_{F}\otimes[v]_+(y,x)\right)
\\
&&\hspace{4.4cm}+\frac{1}{2} {\rm LN}^2(\omega,y) [v]_+ \otimes [v]_+ (y,x)
\Bigg\},
 \nonumber\\
\label{Def-T2b}
T^{(2)}_\beta(\omega,x)&\!\!\!=\!\!\!& T^{(0)}(\omega,y)\otimes
\Bigg\{{\cal T}^{(2)}_{\beta}(y,x) + {\rm LN}(\omega,y)
\left([v_\beta]_+ - {\cal T}^{(1)}_{F}\right)(y,x)
\\
&&\hspace{4.4cm}
-\frac{1}{2} {\rm LN}^2(\omega,y) [v(y,x)]_+ 
\Bigg\},
\nonumber \\
\label{Def-T2G}
T^{(2)}_G(\omega,x)&\!\!\!=\!\!\!& T^{(0)}(\omega,y)\otimes
\Bigg\{{\cal T}^{(2)}_{G}(y,x) + {\rm LN}(\omega,y) [v_G(y,x)]_+  
\Bigg\},
\end{eqnarray}
while ${\rm LN}(\omega,x)= \ln[1+\omega-2x\omega]$ and
$T^{(0)}(\omega,x)$ is given by Eq.\ (\ref{eq:T0omega}).
For a detailed discussion of the appearance of the ${\rm LN}$ terms, see
Appendix \ref{App-MSstr}.

The explicit calculation provides us with the kernels
\begin{eqnarray}
\label{Def-C1F}
{\cal T}^{(1)}_{F}(x,y)&\!\!\!=\!\!\!&
\left[-\frac{3}{2} v^{b}  + g \right]_+\!\!(x,y) - \frac{3}{2} \delta(x-y),
\\
\label{Def-C2b}
{\cal T}^{(2)}_{\beta}(x,y)&\!\!\!=\!\!\!&
\Bigg[ \frac{29}{12} v^{a} + \dot{v}^{a}
- \frac{209}{36} v  -\frac{7}{3} \dot{v} - \frac{1}{4} \ddot{v}+ 
\frac{19}{6} g
+ \dot{g} \Bigg]_+\!\!(x, y)  - 3  \delta(x-y).
\end{eqnarray}
As we see, these $\cal T$-kernels are built up of $v$ and $g$ kernels
appearing in the evolution kernel, too. 
While the $g$ and $\dot{v}$ kernels appear at
NLO, the $\dot{g}$ and $\ddot{v}$ ones show up for the first time at NNLO. 
For the
missing two entries, namely, ${\cal T}^{(2)}_{F}$ and ${\cal T}^{(2)}_{G}$, we
expect a
similar structure, but with additional and unknown building blocks that
are related to the $g$-kernel.

Making use of the fact that both the photon-to-pion transition form
factor and the forward Compton scattering belong to the class of
light-cone dominated two-photon processes, which can be described by a
general scattering amplitude, we have performed a consistency check
between the previously presented results for the hard-scattering
amplitude of the photon-to-pion transition form factor (known up to
$\beta_0$-proportional NNLO terms) and the corresponding results for the
non-singlet coefficient function of the DIS polarized structure function
$g_1$ \cite{ZijNee94}. The procedure is presented in detail in Appendix
\ref{App-ConChe}.

A few comments regarding the ${\rm LN}(\omega,x)$ terms are in order. 
In NLO, we observe
that the ${\rm LN}(\omega,x)$ term match the $\ln$-term indicated in Eq.\
(\ref{Str-T1}), i.e., we can absorb it in Eq.\ (\ref{Def-T1F}) 
by an appropriate choice of the scale:
\begin{eqnarray}
\mu_i^2 \to
\widetilde{\mu}_i^2 = \mu_i^2 \left(1+\omega-2x\omega \right)^{-1}
\mbox{\ with\ } i=\{f,r\}. 
\end{eqnarray}
The explicit NNLO result for the $\beta$ proportional terms satisfies
the same rule for the scale redefinition, which indicates a general
property of the hard-scattering amplitude evaluated in the
$\overline{\rm{MS}}$ scheme. This is shown in Appendix \ref{App-MSstr}.
The terms proportional to ${\rm LN}(\omega,x)$ are vanishing in the
limit $ | \omega | \to  0$ and for $ | \omega | \to  1$ provide a logarithmic
enhancement in the end--point region. However, a resummation of such
terms through an appropriate scale setting is misleading, since other
logarithmically enhanced terms also appear. For instance, at NLO we have
\begin{eqnarray}
\frac{\ln(1-y)}{1-y} \otimes [v(y,x)]_+ =
\frac{\ln^2(1 - x)+2 \ln(1 - x)}{1 - x} + O\left(\ln(1-x)\right),
\end{eqnarray}
while the contribution of ${\cal T}^{(1)}_{F}(x,y)$ is [see Eq.\
(\ref{Def-C1F})]
\begin{eqnarray}
\frac{1}{1-y} \otimes {\cal T}^{(1)}_{F}(y,x) = - \frac{\ln^2(1 - x) + 
3 \ln(1 - x) +9}{2(1-x)}
+ O\left(\ln(1-x)\right),
\end{eqnarray}
so that only a partial cancellation of the $\ln^2(1 - x)/(1-x)$ term appears.

\subsection{Limit $ | \omega | \to  1$ and corresponding conformal moments
 in the \MS\ scheme}
\label{SubSec-Lim}

Of special interest is the limit $ | \omega | \to  1$, since different
(pseudo-scalar) meson-to-photon transition form factors are measured in
this kinematical region. We can trivially perform this limit in Eqs.\
(\ref{eq:T0omega}), (\ref{Def-T1F}), and 
(\ref{Def-T2b}) and after convolution we present the result
in the form of Ref.\ \cite{MelNizPas01}:
\begin{eqnarray}
T^{(0)}(x) &\!\!\!\! = \!\!\!\!& \frac{1}{2(1-x)} \, ,
\label{eq:T0o1}\\
T^{(1)}_{F}(x) &\!\!\!\! = \!\!\!\!&
\frac{1}{2(1-x)} \left[-\frac{9}{2} 
- \frac{1-x}{2x} \ln(1 - x) 
+ \frac{1}{2} \ln^2(1 - x) \right] \, ,
\label{eq:T1o1}\\
T^{(2)}_{\beta}(x) &\!\!\!\! = \!\!\!\!& \frac{1}{2(1-x)}\Bigg[
-\frac{457}{48} - \left(\frac{47}{36} - \frac{1}{4 x}\right) \ln(1 - x) + 
\left(\frac{13}{12} - \frac{1}{4 x}\right) \ln^2(1 - x)
\label{eq:T2bo1}\\
&&\hspace{2cm}- \frac{1}{6} \ln^3(1 - x) - 
  \frac{7}{3} {\rm Li}_2(x) + \frac{1}{2} {\rm Li}_3(x)  - {\rm S}_{12}(x)
\Bigg] \, ,
\nonumber
\end{eqnarray}
where the polylogarithms are defined by
\begin{eqnarray}
{\rm Li}_n(x) &\!\!\!\! = \!\!\!\!& \int_0^x \!dy\, 
\frac{{\rm Li}_{n-1}(y)}{y},
\quad\mbox{\ with\ } {\rm Li}_{1}(x)=-\ln(1-x) \, ,
\nonumber\\
{\rm S}_{12}(x) &\!\!\!\! = \!\!\!\!&
\frac{1}{2} \int_0^x \!dy\, \frac{\ln^2(1-y)}{y}
\\
&\!\!\!\! = \!\!\!\!&
\frac{1}{2} \ln^2(1 - x) \ln(x) +
\ln(1 - x)  {\rm Li}_2(1-x)-  {\rm Li}_3(1 - x) + \zeta(3) \, .
\nonumber
\end{eqnarray}
Here we introduce for the special case $ | \omega | =1$ a more convenient 
notation in which
the $\ln(2)$ terms arising from the ${\rm LN}(\omega=1,x)$ functions 
present in  Eqs.\ (\ref{Def-T1F}-\ref{Def-T2b}) are
absorbed in $\ln(2 Q^2/\mu_{f,(r)}^2) =  \ln(-q_1^2/\mu_{f,(r)}^2) $.  
We do not list here the terms  proportional to these log's but
refer to Ref.\ \cite{MelNizPas01} for their explicit expressions.

The conformal moments of the individual terms we encounter
in Eqs.\ (\ref{eq:T0o1}-\ref{eq:T2bo1})
are given in Appendix \ref{App-ConfMom}.
Here we list the conformal moments of 
$T^{(0)}(x)$, $T_F^{(1)}(x)$,
and $T_{\beta}^{(2)}(x)$ calculated in the \MS\ scheme:  
\begin{eqnarray}
T^{(0)}_{j}&\!\!\!\! = \!\!\!\!&
\frac{2j + 3}{(j + 1)(j + 2)} \, ,
\label{eq:T0MS}
\\[0.4cm]
\label{eq:T1MS}
T^{(1)}_{F,j}&\!\!\!\! = \!\!\!\!&
\frac{2j + 3}{(j + 1)(j + 2)}
\Bigg[
-\frac{9}{2} + \frac{3- 4S_{1}(j + 1)}{2(j + 1) (j + 2)}+
\frac{1}{(j + 1)^2(j + 2)^2}  +  2 S_{1}^2(j + 1)
 \Bigg] \, ,
\\[0.4cm]
T^{(2)}_{\beta,j}&\!\!\!\! = \!\!\!\!&
\frac{2j + 3}{(1 + j)(2 + j)} \Bigg[ 
 - \frac{457}{48}  
 -\frac{1}{2} \zeta(3) - \frac{7}{3} \zeta(2) 
 + \frac{1}{(j + 1)(j + 2)} \Bigg(
   \frac{115}{36} - \frac{(-1)^j}{2} \zeta(2)
\nonumber\\[0.2cm]
&& -  \frac{1}{3} S_1(j + 1) - 2 S_{1}^2(j + 1)
   - (-1)^j S_{-2}(j + 1) \Bigg)
   - \frac{1 - 8S_{1}(j + 1)}{2(j + 1)^2 (j + 2)^2}
\label{eq:Tj2betaMS}
\\[0.2cm]
&&- \frac{5}{2(j + 1)^3 (j + 2)^3}
   + \frac{19}{9} S_{1}(j + 1) + \frac{10}{3} S^2_{1}(j + 1) 
   + \frac{4}{3}  S^3_{1}(j + 1) 
    + \frac{2}{3} S_{3}(j + 1)  
\Bigg] \, .
\nonumber
\end{eqnarray}
After inspection of Eqs.\ (\ref{Str-T1}) and (\ref{Str-T2})
one notes that
the conformal moments of the terms proportional
to $\ln(2 Q^2/\mu_{f,(r)}^2)$ can be conveniently expressed
using the conformal moments of the kernels $v$ and $v_{\beta}$. 
For the definition of conformal moments of the kernels,
we refer to Appendix \ref{App-ConfMom}, Eq.\ \req{eq:tvlk}.
The conformal moments of the diagonal kernel  $[v(x,y)]_+$, given
in Eq.\ (\ref{Def-V-LO}), are denoted by $v_{j} \equiv v_{jj}$:
\begin{eqnarray}
\label{Def-ConMomv}
v_j = - 2S_{1}(j + 1) +\frac{3}{2} + \frac{1}{(1 + j)(2 + j)} 
\, .
\end{eqnarray}
The conformal moments of $T^{(0)} \otimes [v]_+$, determined using
\req{eq:FxkconvD}, are given by $T^{0}_j \; v_j$. On the other hand, the
kernel $[v_{\beta}]_+$ in Eq.\ (\ref{Def-EvoKer-Beta}) is non-diagonal,
and as shown in \req{eq:FxkconvND}, both the diagonal $v_{\beta,j}
\equiv v_{\beta,jj}$ as well as the non-diagonal $v_{\beta,kj}$ ($k > j$
and $k-j$ even) conformal moments contribute to the conformal moments of
$T^{(0)} \otimes [v_{\beta}]_+$, i.e. one obtains $\sum_{k>j} T^{0}_k \;
v_{\beta,kj}$. After performing the convolution and making use of the
results for the conformal moments summarized in Appendix
\ref{App-ConfMom}, one can express the conformal moment of $T^{(0)}
\otimes [v_{\beta}]_+$ in a form $T^{0}_j \; v_{\beta,j}^\Sigma$, where
we introduce the ``effective'' conformal moment of the $v_{\beta}$
kernel amounting to
\begin{equation}
v_{\beta,j}^{\Sigma} =
 \frac{5}{3} v_j + \zeta(2) - \frac{9}{4} - \frac{1}{(1 + j)^2 (2 + j)^2}
\, .
\label{eq:vbetasum}
\end{equation}

Finally, we summarize our $\overline{\rm MS}$ results.
The LO contribution is given in Eq.\ (\ref{eq:T0MS}), the NLO
contribution takes the form
\begin{equation}
 T_j^{(1)} = T_{F,j}^{(1)} + \ln\left( \frac{2 Q^2}{\mu_f^2} \right) \,
\frac{2 j + 3}{(j+1)(j+2)} \, v_j 
\label{eq:Tj1}
\, ,
\end{equation}
while the $(-\beta_0/2)$ proportional NNLO term 
is given by
\begin{eqnarray}
  T_{\beta,j}^{(2)}
+ \ln\left( \frac{2 Q^2}{\mu_f^2} \right) \,
\frac{2 j + 3}{(j+1)(j+2)} \left[ \, v_{\beta,j}^\Sigma 
+ \frac{1}{2} \ln\left( \frac{2 Q^2}{\mu_f^2} \right) \,
 \, v_{j} \right]
- \ln\left( \frac{2 Q^2}{\mu_r^2} \right) \, T_j^{(1)}
\, .
\label{eq:TjbetaMS}
\end{eqnarray}
Note that owing to the fact that $v_{\beta}$ is non-diagonal,
even the lowest partial wave, i.e., $j=0$, of the NNLO correction
depends on the factorization scale as well as the renormalization one.

\subsection{NNLO result in the CS and \CS\  schemes in the limit  
$ | \omega | \to  1$}
\label{SubSec-CS-lim}

Let us now turn to the conformal schemes CS and \CS. 
We make a distinction between the renormalization and 
factorization scale. 
Consequently, the argument of
the coupling in the Wilson--coefficients depends on $\mu_r$ and, 
as discussed in Section \ref{SubSec-Ambi}, 
we require that the matrix elements of conformal operators depend only on the
scale $\mu_f$. We use the COPE, where for $ | \omega | =1$ the Wilson--coefficients
(\ref{Def-OveC}) simplify to
\begin{eqnarray}
\label{Def-OveC-1}
\overline{C}_j=
c_j(\alpha_s(\mu_r)) \left(\frac{\mu^2_f}{2 Q^2}\right)^{\frac{\gamma_j}{2}}
\frac{\Gamma(j + 1) \Gamma(j + 2) \Gamma(2 j + 4 + \gamma_j)}
{\Gamma(j + 2 + \gamma_j/2)\Gamma(j + 3 + \gamma_j/2)\Gamma(2 j + 3)} .
\end{eqnarray}
The anomalous dimensions are given by
\begin{eqnarray}
\label{eq:gammaj}
\gamma_j &\!\!\! =\!\!\! &
 \frac{\alpha_s(\mu_r)}{2 \pi} \gamma_j^{(0)}
+ \frac{\alpha_s^2(\mu_r)}{( 2 \pi)^2} \gamma_j^{(1)}
+ O(\alpha_s^3)
  \\ 
         &\!\!\! =\!\!\! &
-2 \left\{ 
\frac{\alpha_s(\mu_r)}{2 \pi} C_F \, v_j
+  \frac{\alpha_s(\mu_r)}{(2 \pi)^2}  C_F
\left[ C_F \, v_{F,j}
- \frac{\beta_0}{2}  v_{\beta,j}
-\left(C_F - \frac{1}{2} C_A \right) v_{G,j}
\right] + O(\alpha_s^3) \right\}
, \qquad
\nonumber
\end{eqnarray}
where $v_j$, $v_{F,j}$, $v_{G,j}$, and $v_{\beta,j}$
are the diagonal conformal moments of the evolution kernels 
$[v(x,y)]_+$, $[v_F(x,y)]_+$, $[v_G(x,y)]_+$ and 
$[v_{\beta}(x,y)]_+$, respectively,
and they coincide with the moments of the DGLAP kernels.  
The LO moments are given by Eq.\ \req{Def-ConMomv},
while other entries can be found in Refs.\ \cite{FloRosSac77,
GonLopYnd79,CurFurPet80,FloLacKou81}.
Analogously, we decompose the normalization factor 
\begin{eqnarray}
c_j = 1 + \frac{\alpha_s}{2\pi} C_F \, c_j^{(1)}
+ \left(\frac{\alpha_s}{2\pi}\right)^2
C_F \left[C_F  \, c_{F,j}^{(2)} -\frac{\beta_0}{2}\, c_{\beta,j}^{(2)} -
\left(C_F-\frac{1}{2} C_A\right) c_{G,j}^{(2)}
 \right]+ O\left(\alpha_s^3\right).
\label{eq:cj}
\end{eqnarray}
Its NLO contribution reads
\begin{eqnarray}
c_j^{(1)} &\!\!\!\! = \!\!\!\!&
S^2_{1}(1 + j) + 
\frac{3}{2} S_{1}(j + 2) - \frac{9}{2}  +
\frac{3-2S_{1}(j)}{2(j + 1)(j + 2)} -   S_{2}(j + 1),
\label{eq:cj1}
\end{eqnarray}
while the NNLO contributions can be determined from
the Mellin moments of the coefficient functions calculated from the
$\alpha_s^2$ corrections to the polarized structure function $g_1$
 \cite{ZijNee94}. 
A consistency check of the $\beta_0$ proportional part of these results 
is given in Appendix \ref{App-ConChe}.

As discussed in Section \ref{SubSec-Ambi}, the Wilson--coefficients in
the CS scheme are obtained in a straightforward manner by means of Eq.\
(\ref{Def-WilCoeCS}), where $C^{{\MSeq}
(2)}_{\beta,j}(\alpha_s(\mu_r),Q/\mu_f,Q/\mu_r)$ is given by the
expression (\ref{eq:TjbetaMS}). Taking into account the proper
normalization, i.e. identifying $T^{(i)}$ with $C^{(i)}$ by
\req{eq:T2C}, the expansion of $\overline{C}_j$, defined by Eq.\
(\ref{Def-OveC-1}), leads to the complete NNLO result for the
hard-scattering amplitude:
\begin{eqnarray}
T^{\rm CS}_j(Q,\mu_f) =
\frac{\sqrt{2}}{3 Q^2}
\left[T_j^{(0)} 
+ \frac{\alpha_s(\mu_r)}{2 \pi} T_j^{{\rm CS}(1)}(Q/\mu_f)
+ \frac{\alpha_s^2(\mu_r)}{(2 \pi)^2}
  T_j^{{\rm CS}(2)}(Q/\mu_f,Q/\mu_r)
+ O(\alpha_s^3) \right],
\label{eq:TjCS}
\end{eqnarray}
where 
\begin{eqnarray}
T_j^{{\rm CS}(1)} &\!\!\! = &\!\!\! C_F \left(T_{F,j}^{{\rm CS}(1)} 
   + \frac{ 2 j + 3}{(j+1)(j+2)}
 \ln \left( \frac{2 Q^2}{\mu_f^2} \right) v_j \right)\, ,
\label{eq:Tj1CS}
 \\[0.3cm]
T_j^{{\rm CS}(2)} &\!\!\! = &\!\!\! C_F \left\{ 
   C_F \, T_{F,j}^{{\rm CS}(2)} 
  -\frac{\beta_0}{2} T_{\beta,j}^{(2)}
  -\left( C_F - \frac{1}{2} C_A \right) \, T_{G,j}^{{\rm CS}(2)}
+ \ln \left( \frac{2 Q^2}{\mu_f^2} \right) \frac{2 j + 3}{(j+1)(j+2)}
   \right. \nonumber \\ & & \left.
\qquad  \times    
     \Bigg[\, C_F \Bigg( v_{F,j} +c_j^{(1)} v_j + v_j^2 
             \left( S_1(j+1)+S_1(j+2)-2 S_1(2 j + 3) \right) 
\right.  \nonumber \\ & & \left.  
     \hspace{4cm}+ \frac{v_j^2}{2} \ln\left(\frac{2 Q^2}{\mu_f^2}\right)\Bigg)
-\frac{\beta_0}{2} \, v_{\beta,j}^{{\rm CS}, \Sigma}
     -\left( C_F - \frac{1}{2} C_A \right)\,  v_{G,j} 
    \Bigg]
    \right.  \nonumber \\ & & \left.
\qquad+\frac{\beta_0}{2}\ln \left( \frac{2 Q^2}{\mu_r^2} \right)
T_j^{{\rm CS}(1)}(Q/\mu_f)
 -  \frac{\beta_0}{4} \ln^2 \left( \frac{2 Q^2}{\mu_f^2} \right)
      \frac{ 2 j + 3}{(j+1)(j+2)} \, v_j
 \right\}\, ,
\label{eq:Tj2CS}
\end{eqnarray}
and
\begin{eqnarray}
T^{{\rm CS}(1)}_{F,j} &\!\!\! = \!\!\!&
\frac{2j+3}{(j+1)(j+2)}\left\{
c_j^{(1)} + v_j 
\left[S_{1}(j + 2) + S_{1}(j + 1) -2 S_{1}(2 j + 3) \right]
\right\}\, 
\label{eq:Tj1FCS}
\\
T^{{\rm CS}(2)}_{F,j} &\!\!\! = \!\!\!&
\frac{2j+3}{(j+1)(j+2)}\Big\{
c^{(2)}_{F,j} + \left(c^{(1)}_j v_j + v_{F,j} \right) 
\left[S_{1}(j + 1) + S_{1}(j + 2) - 2 S_{1}(2 j + 3)\right]
\nonumber\\
&&\hspace{3cm}+ \frac{v^2_j}{2}\bigg(
  \left[S_{1}(j + 1)+ S_{1}(j + 2) - 2 S_{1}(2 j + 3)\right]^2
\label{eq:Tj2FCS}
\\
&&\hspace{3cm} +
   S_{2}(j + 1) + S_{2}(j + 2) - 4 S_{2}(2 j + 3) + 2 \zeta(2)\bigg)
\Big\},
\nonumber\\
T^{{\rm CS}(2)}_{G,j} &\!\!\!\! = \!\!\!\!&
\frac{2j+3}{(j+1)(j+2)}\Big\{
c^{(2)}_{G,j} +  v_{G,j}  
\left[S_{1}(j + 1) + S_{1}(j + 2) - 2 S_{1}(2 j + 3)\right]\Big\}
\, ,
\label{eq:Tj2GCS}
\end{eqnarray}
while 
$T_{\beta,j}^{(2)}$
corresponds to the \MS\ result given by Eq.\ \req{eq:Tj2betaMS}.
Note that, in accordance with the renormalization group invariance,
the off-diagonal part of the anomalous dimension matrix 
proportional to $\beta_0$ has been changed and, consequently, also the
conformal moments
\begin{eqnarray}
v_{\beta,j}^\Sigma \Rightarrow v_{\beta,j}^{{\rm CS}, \Sigma} =
v_{\beta,j}^\Sigma + \frac{(j+1)(j+2)}{2j+3}
 \left(T_{F,j}^{{\rm CS}(1)}- T_{F,j}^{(1)}  \right) .
\label{Def-vCSbeta}
\end{eqnarray}

Finally, we present the result for the $\overline{\rm CS}$ scheme, in
which the conformal covariance of the partial wave decomposition is
preserved. The modification concerns only the terms proportional to
$\beta_0$ in which the off-diagonal entries in Eq.\ (\ref{eq:Tj2CS}) are 
removed by making the following replacements:
\begin{equation}
\begin{array}{lclcl}
 v^{{\rm CS}, \Sigma}_{\beta,j} &\Rightarrow& v_{\beta,j} \, ,
& &
\\[0.4cm]
 T^{(2)}_{\beta,j} &\Rightarrow& T^{\CSeq(2)}_{\beta,j} 
  &\!\!\!=\!\!\!&
\displaystyle \frac{2j+3}{(j+1)(j+2)}\Big\{
c^{(2)}_{\beta,j} +  v_{\beta,j}  
\left[S_{1}(j + 1) + S_{1}(j + 2) - 2 S_{1}(2 j + 3)\right] \Big\}
\\[0.4cm]
 & & & &
\displaystyle 
+ \ln(2)\left[ T_{F,j}^{{\rm CS}(1)} 
               + T_{j}^{(0)}  \frac{1}{2} \ln(2) \,  v_j \right] 
\, .
\end{array}
\label{eq:betaCSbar}
\end{equation}
The $\ln(2)$ terms appear here artificially from the absorption of such
terms into the factorization and renormalization log's, i.e.,
$\ln(Q^2/\mu_i^2) \to \ln(2Q^2/\mu_i^2) - \ln(2)$, see Eq.\
(\ref{Def-Coe-FulThe}). All other expressions in Eqs.\ (\ref{eq:Tj1CS}
-- \ref{eq:Tj2GCS}) remain unchanged, e.g., 
\begin{eqnarray}
T_j^{\overline{\rm CS}(1)} = T_j^{{\rm CS}(1)} ,
\qquad T^{\overline{\rm CS}(2)}_{F,j}=T^{{\rm CS}(2)}_{F,j} ,
\qquad
T^{\overline{\rm CS}(2)}_{G,j}=T^{{\rm CS}(2)}_{G,j} .
\end{eqnarray}	

\subsection{NNLO predictions for small and intermediate values of $ | \omega | $}
\label{SubSec-SmallOmega}

Based on numerical observations on the small and intermediate $ | \omega | $-
behaviour of the transition form factor predicted by perturbation
theory, interesting phenomenological aspects have been pointed out in
Ref.\ \cite{DieKroVog01}. Unfortunately, for these configurations both
photons are virtual, cf.\ Eq.\ (\ref{RG-transfor}), and thus the
statistics is rather low. Therefore, no measurements has been done yet,
although they could be possible at the existing $e^+ e^-$ machines of
the Babar, Belle and CLEO experiments. In the following we want to add some
comments on the pion transition form factor in the small and
intermediate $ | \omega | $ regions and to give predictions at NNLO.

From the representation (\ref{Def-OveC-run}) it follows that the $j$th
conformal moments for $|\omega| < 1$ are suppressed by $\omega^j$. In
addition, the hypergeometric functions appearing in the
Wilson--coefficients are sharply peaked at $|\omega| =1$ owing to a
logarithmical enhancement caused by the $\ln[(1-\omega)/(1+\omega)]$ term.
For fixed $|\omega| < 1$, one finds for growing $j$ an increasing
suppression of the hypergeometrical functions, in addition
to the power-like suppression due to $\omega^j$. To study this behaviour in
more detail, we employ the integral
representation for the hypergeometrical functions: 
\begin{eqnarray}
&&\hspace{-0.5cm}
\frac{2^{j+1+\frac{\gamma_j}{2}}}{(1+\omega)^{j+1+\frac{\gamma_j}{2}}}
 {_{2}F}_1\left({j+1+\frac{1}{2}\gamma_j, j+2+\frac{1}{2}\gamma_j
 \atop
2(j+2+\frac{1}{2}\gamma_j)}\Bigg|\frac{2\omega}{1+\omega}\right)
\\
&&\hspace{3cm}
=\frac{\Gamma(4 + 2j +  \gamma_j)}{\Gamma(2 + j + \gamma_j/2)^2}
\int_0^\infty ds\, s\, e^{-\frac{s^2}{2}} e^{\frac{1 + j + \gamma_j/2}{2}
\left\{-s^2 + \ln(1-e^{-s^2})-\ln(1-\omega^2 e^{-s^2}) \right\}}\, .
\nonumber
\end{eqnarray}
To evaluate this integral for large $j$, we rely on the saddle point
approximation, which is valid as long as the condition \begin{eqnarray}
\left(j+1+\frac{\gamma_j}{2}\right)\sqrt{1-\omega^2} > 1
\end{eqnarray}
is satisfied. To clearly illustrate the suppression we mentioned above, 
we write the Wilson--coefficients in the form
\begin{eqnarray}
\overline{C}_j = c_j\left(\!
\alpha_s(\mu),Q/\mu,\frac{\partial}{\partial \gamma}\!\right) \left(
				\frac{\mu^2}{2Q^2}\right)^{\frac{\gamma_j}{2}} 
\frac{\Gamma(j + 1) \Gamma(j + 2) \Gamma(2 j + 4 + \gamma_j) \omega^j}
{\Gamma(j + 2 + \gamma_j/2)\Gamma(j + 3 + \gamma_j/2)\Gamma(2 j + 3)}
 E_j(\omega|\gamma_j).
\end{eqnarray}
Consequently, we have the following normalization for the function
$E_j(\omega=1|\gamma_j)=1$, cf.\ Eq.\ (\ref{Def-OveC-1}). For $|\omega|<
1$, the approximation 
\begin{eqnarray}
E_j(\omega|\gamma_j) \approx 
\frac{\sqrt{\pi}\sqrt{2\left(j+1+\frac{\gamma_j}{2}\right)
\sqrt{1 - \omega^2}}}
{2\sqrt{1 + \sqrt{1 -\omega^2}} }\,
\exp{\left\{-\left(j+1+\frac{\gamma_j}{2}\right)
 \ln\left(1+\sqrt{1-\omega^2}\right)\right\}}
\end{eqnarray}
shows an exponential decrease for
$(j+1)
\ln(1+\sqrt{1-\omega^2}) > \sqrt{2}$ . The value of
$E_j(\omega|\gamma_j)$ is then smaller than $1/2$. Note that perturbative
corrections due to the anomalous dimensions, which are positive and grow
logarithmically with $j$, give a logarithmical enhancement of this
behaviour. In the case of rather small $ | \omega | $, the suppression factor is
proportional to $ 2^{-j-1} \sqrt{j+1} $, which affects even the lowest
partial wave $j = 2$. The suppression is already larger than 80\% for
given $j\ge 6$ as long as the inequality $1-|\omega| > 4/(j+1)^2$ is
satisfied. Increasing $|\omega|$ will then abruptly increase the value of
$E_j$ to reach the $E_j =1$ limit. 
To finish this general discussion, we
estimate the partial wave that will be suppressed by a factor $
{\cal E}=E_j(\omega) \le 1/e$ depending on $\omega$: 
\begin{eqnarray}
j+1\simeq
\left[\frac{-W_{-1}\left(-\frac{4 {\cal E}^2 (1 + \sqrt{1 -
\omega^2})\ln(1+\sqrt{1-\omega^2}}{\pi\sqrt{1 - \omega^2}}\right)}
{2 \ln(1+\sqrt{1-\omega^2})}\right]
\sim
\left[\frac{-\ln\left(4 {\cal E}^2/\pi\right)}
{4 (1-|\omega|)}\right]\, 
\end{eqnarray}
where $W_{-1}(-x)$ is the product log function that is real valued in
the region $0\le x\le 1/e$.

To estimate the  contribution of the first few non-vanishing
partial waves, we first consider the conformal moments 
 $\langle \pi(P)|  {\cal O}_{jj}(\mu_f) |\Omega\rangle^{\rm red} 
=6 N_j B_j$.  The distribution amplitude vanishes at the end--points
\cite{BroLep80} and from this behaviour 
it follows that $ N_j B_j$ 
vanishes at $j\to \infty$: 
\begin{eqnarray}
\label{Res-EndPoiBeh}
\phi(x) \sim (1-x)^\epsilon \ \mbox{for}\ x\to 1
\quad \Rightarrow \quad 6 N_j B_j \sim  j^{-\epsilon} \ \mbox{for}\ 
j\to\infty \quad \mbox{with}\quad \epsilon >0.
\end{eqnarray}
We want to add that different non-perturbative estimates, based on a
lattice calculation, sum rules or a model calculation give quite different
values of $6 N_2 B_2$ at a scale  $Q \le 1$ GeV, varying from $\sim
-1$ to $\sim +1$. Here the lower bound stems from a preliminary lattice
calculation \cite{DebPieDouSac00}, while the upper one arises from sum rule
estimates \cite{CheZhi84,BraFil89} and is also compatible with previous
lattice calculations (see \cite{DebPieDouSac00} for references). There
are other estimates that favour a rather small value of $B_2$. This
suggests that the absolute size of the lowest few conformal moments $6
N_j B_j$ are of the order one or even smaller. In the following
estimates we consider them of the order one, which serves us as an upper
bound for the contribution of the $j$th partial wave to the transition
form factor.

In the small $ | \omega | $ region, i.e., $|\omega| < 0.4$, the lowest partial
wave contributes essentially. In LO the relative contribution of the
second and the forth partial wave with respect to the first one is for
$ | \omega | =0.2 (0.4)$ about $0.08(2.3)\%$ and $0.004 (0.05)\%$,
respectively. The $\omega^2$ term of the zeroth partial wave varies in
the same order as the relative correction to the second partial wave,
which is in addition suppressed by a relative factor of 2/3. Thus, in the
small $ | \omega | $ region perturbative QCD provides us an (almost) parameter
free, factorization scheme independent prediction:
\begin{eqnarray}
F_{\gamma\pi}(\omega,Q) \simeq \frac{\sqrt{2}f_\pi}{3 Q^2}
c_0(\alpha_s(\mu_r),Q/\mu_r) \left(1+ \frac{\omega^2}{5}  + O(\omega^4)\right)
\quad \mbox{for}\quad |\omega| <  0.4\, .
\end{eqnarray}
The phenomenological consequences are obvious, since this prediction is
 practically independent of $\omega$
and its logarithmical $Q^2$ dependence is governed 
only by the running of the coupling.

For intermediate values of $ | \omega | $, defined as $0.4 \le |\omega|< 0.8$,
the second partial wave contributes between 2\% and 13\%, while the
fourth one is at least more than five times suppressed with respect to
the second one. Increasing $ | \omega | $ to the value $0.95$, the relative
contribution of the second and fourth partial wave grows to $25\%$ and
$10\%$, respectively, while the sixth (eighth) partial wave contributes
at the $4\% (2\%)$ level. It is illustrative to compare these numbers
with the suppression arising in the limit $ | \omega | \to  1$ in which the
contribution amounts to $39\%,\ 24\%,\ 18\%,\ 14\%$ for $j=2,\ 4,\ 6,\
8$ partial waves, respectively.

As we have realized,only the first two non-vanishing partial waves 
are essential for an intermediate value of $ | \omega | $. 
It could, therefore, be justified to 
employ the Taylor expansion of the hypergeometric functions at
$\omega=0$ and hence, the transition form factor reads in the $\overline{\rm
CS}$ scheme
\begin{eqnarray}
\label{Def-Ome2Zer-Exp}
F_{\gamma\pi}(\omega,Q) &\!\!\! \simeq \!\!\! &\frac{\sqrt{2}f_\pi}{3 Q^2}
	 \Bigg[c_0 \left\{1 + \frac{\omega^2}{5} +
 \frac{3\omega^4}{35} \right\}  + c_2 \left( \frac{\mu^2}{Q^2}
\right)^{\frac{\gamma_2}{2}} \frac{2\omega^2}{15} 
\left\{
1+\frac{(8+\gamma_2)(6+\gamma_2)}{8 (9+\gamma_2) } \omega^2
\right\} 
\nonumber\\
&&\hspace{3cm}\times\,
  6 N_2 B_2(\mu^2) + O(\omega^6)
\Bigg], \quad \mbox{for}\quad  0.4 \le |\omega| <  0.8\, ,
\end{eqnarray}
where the Wilson--coefficient $c_2\left(\alpha_s(\mu),\frac{Q}{\mu},
\frac{\partial}{\partial \gamma_j}\right)$ is defined by Eq.\
(\ref{Def-Coe-FulThe}) to NNLO accuracy and a consequent expansion in
$\alpha_s$ should be done to this accuracy. For $ | \omega | =0.8$, the higher-order
terms in $O(\omega^6)$ contribute 
at the 2\% and 20\% level for the $j=0$  and $j=2$ partial wave, respectively. 
These contributions can be reduced by a factor of
two [four] by taking the order $\omega^6$ [$\omega^8$] corrections into
account. For larger values of $ | \omega | $, the convergence of the Taylor
expansion at $\omega=0$ is rather slow for higher partial waves. For
instance, to approximate the third non-vanishing partial wave at
$ | \omega | =0.9$ to an accuracy of better than 10\%, one has to take into
account the first ten non-vanishing terms, i.e., up to $O(\omega^{24})$.

For ultra-large values of $ | \omega | $, let us say $|\omega| >0.95$, partial
waves with higher conformal spin start to contribute with increasing
$|\omega|$. However, as we have discussed, as long as we do not reach
the $|\omega| \to 1$ limit, there will be an exponential suppression for
higher values of $j$. Note that this limit can never be reached in any
experiment at an $e^+ e^-$ machine, where the mean value of the
virtuality of the untagged photon is set by the electron mass and there
are further kinematical restrictions arising from the detector geometry
and kinematical cuts. Just for illustration, we would like to mention
that for $ | \omega | =0.99$ and $ | \omega | =0.999$, the contributions of the 12th
and 38th partial waves are reduced by a factor $1/e\sim 0.37$ compared
to their contributions in the limit $ | \omega | \to  1$, while higher ones
start to be exponentially suppressed, since the $(j+1)
\ln(1+\sqrt{1-\omega^2}) > \sqrt{2}$ condition is fulfilled.

We now present the general result of the photon-to-pion transition form
factor for $|\omega| < 1$ in its expanded form to the NNLO approximation.
Notation analogous to that in Eqs.\ (\ref{eq:TjCS}) will be used and 
the Taylor expansions in $\omega^2$
for the contributing terms of the first five non-vanishing partial
waves are listed 
in Appendix \ref{App-OmegaRes}. 
At leading order the hypergeometrical functions can be expressed
in terms of elementary arctanh functions\footnote{The result can be
expressed in terms of $\ln$ functions by means of ${\rm
arctanh}(\omega)=1/2 \ln[(1+\omega)/(1-\omega)]$.}, e.g.,
\begin{eqnarray}
T_0^{(0)}(\omega)  &\!\!\! =\!\!\! &
\frac{3}{2\omega^2}
\left[1 -(1 - \omega^2)\frac{{\rm arctanh}(\omega)}{\omega}\right]
\, ,
\\
T_2^{(0)}(\omega) &\!\!\! =\!\!\! &
\frac{7}{24\omega^2}
\left[15 - 13\omega^2 - 
(5 - 6\omega^2 + \omega^4) \frac{3{\rm arctanh}(\omega)}{\omega}\right]
\, ,
\nonumber
\end{eqnarray}
and the expansion in $\omega$ is given in \req{eq:Tj0omegaExp}. The
radiative corrections for $\omega\not=0$ depend on the factorization
scheme even for the lowest partial waves. In comparison with Eqs.\
(\ref{eq:Tj1CS}) and (\ref{eq:Tj2CS}), here we will not include $\ln(2)$
terms in the factorization and renormalization log's, i.e. instead of $2
Q^2$ we rather employ the underlying scale $Q^2 \sim 1/z^2$ of the OPE
of two currents at the distance $z$. Thus, we have for the
$\overline{\rm MS}$ scheme the NLO result (\ref{Str-T1})
\begin{eqnarray}
T_j^{(1)}(\omega,Q/\mu_f) &\!\!\! = &\!\!\! C_F \left(T_{F,j}^{(1)}(\omega)
   + T^{(0)}_{j}(\omega)
 \ln \left( \frac{Q^2}{\mu_f^2} \right) v_j \right)\, ,
\label{eq:Tj1omega}
\end{eqnarray}
and analogous one for the CS scheme. Note that in comparison to this notation 
  the definitions of conformal
moments  in the $ | \omega | \to  1$ limit, Eqs.\ (\ref{eq:T1MS}) and (\ref{eq:Tj1FCS}),  differ by 
a  $\ln(2)$ proportional term:
\begin{eqnarray}
\lim_{\omega\to 1} T_{F,j}^{(1)}(\omega) =
T_{F,j}^{(1)} +\ln(2)\, T^{(0)}_{j}  v_j,\qquad
\lim_{\omega\to 1} T_{F,j}^{{\rm CS}(1)}(\omega) =
T_{F,j}^{{\rm CS}(1)} + \ln(2)\, T^{(0)}_{j} v_j\, .
\end{eqnarray}
Expanding the $T^{(0)}(\omega) {\rm LN}^i(\omega)$ terms in Eq.\
(\ref{Def-T1F}) provides after convolution with the corresponding
kernels the desired result in the $\overline{\rm MS}$ scheme
\req{eq:TjF1omegaExp}. In the considered order of $\omega^2$, it
coincides with the result of Ref.\ \cite{DieKroVog01}. The result in the
CS scheme can be easily derived by expanding Eqs.\ (\ref{Def-OveC}) in
order $\alpha_s$:
\begin{eqnarray}
\label{Def-TNLOomega}
T_{F,j}^{{\rm CS}(1)}(\omega) = 
\left[c^{(1)}- v_j\, s_j^{(1)}(\omega)\right] T_{j}^{(0)}(\omega), 
\end{eqnarray}
where $c^{(1)}$ and  $v_j$ are given by Eqs.\ (\ref{eq:cj1}) and
(\ref{Def-ConMomv}), respectively,
and  $s_j^{(i)}(\omega)$ is defined by the expansion
\begin{eqnarray}
\label{Def-sFun}
\frac{{_{2}F}_1\left({j+1+\epsilon, j+2+\epsilon \atop
2(j+2+\epsilon)}\Bigg|
\frac{2\omega}{1+\omega}\right)}{(1+\omega)^{j+1+\epsilon}}
=\left[
1+ s^{(1)}_j(\omega) \epsilon  + \frac{1}{2} s^{(2)}_j(\omega) \epsilon^2 +
O(\epsilon^3) \right]
 \frac{{_{2}F}_1\left({j+1, j+2 \atop
2(j+2)}\Bigg|\frac{2\omega}{1+\omega}\right)}{(1+\omega)^{j+1}} 
\, .
\end{eqnarray}
The corresponding expansions of $s_j^{(1)}$ and $s_j^{(2)}$ in $\omega^2$
are given for $j=2,4,6,8$ (because of current conservation $v_0=0$ and, 
consequently,  $j=0$ term does not 
contribute)  given in Eqs.
\req{eq:sj1omegaExp} and \req{eq:sj2omegaExp}, respectively. The NNLO
contribution in the CS scheme we write analogously to Eq.\
(\ref{eq:Tj2CS}) as
\begin{eqnarray}
\label{eq:Tj2CSomega}
T_j^{{\rm CS}(2)}(\omega) &\!\!\! = &\!\!\! C_F \left\{ 
   C_F \, T_{F,j}^{{\rm CS}(2)}(\omega)
  -\frac{\beta_0}{2} 
      T_{\beta,j}^{(2)}(\omega)
  -\left( C_F - \frac{1}{2} C_A \right) \, T_{G,j}^{{\rm CS}(2)}(\omega)
+ \ln \left( \frac{Q^2}{\mu_f^2} \right) T_j^{(0)}(\omega)
\right.\nonumber \\ & & \left.
   \qquad\times \left[
C_F \left( v_{F,j} +c_j^{(1)} v_j - v_j^2 s_j^{(1)} (\omega)
+ \frac{v_j^2}{2} \ln\left( \frac{Q^2}{\mu_f^2} \right)\right)
     -\frac{\beta_0}{2} \, v_{\beta,j}^{\rm CS, \Sigma}(\omega)
          \right. \right.   \\
& & \left.  
     \qquad  -\left( C_F - \frac{1}{2} C_A \right)\,  v_{G,j}  \Bigg]
 +\frac{\beta_0}{2} \ln \left( \frac{Q^2}{\mu_r^2} \right)
T_j^{{\rm CS}(1)}(\omega,Q/\mu_f)
 - \frac{\beta_0}{4} \ln^2 \left( \frac{Q^2}{\mu_f^2} \right)
      T_j^{(0)}(\omega)\, v_j
         \right\}\, .
\nonumber 
\end{eqnarray}
In comparison to the definitions in Eqs.\ (\ref{eq:Tj2FCS}),
(\ref{eq:Tj2GCS}), and (\ref{eq:Tj2betaMS}), one has again to take care
of $\ln(2)$ terms by means of the following correspondences:
\begin{eqnarray}
\label{Lim-omega-1}
&&\!\!\! \lim_{\omega\to 1} T_{F,j}^{{\rm CS}(2)}(\omega) =  T_{F,j}^{{\rm
CS}(2)}+\ln(2)\, T^{(0)}_{j} \left[
  v_{F,j} + c_j^{(1)} v_{j}  -
v_{j}^2 \left(s^{(1)}_j(\omega=1) + \frac{1}{2} \ln(2)  \right)
\right] \, ,
\nonumber\\
&&\!\!\! \lim_{\omega\to 1} T_{G,j}^{{\rm CS}(2)}(\omega) = T_{G,j}^{{\rm
CS}(2)}+\ln(2)\, T^{(0)}_{j} v_{G,j}\, ,
\\
&&\!\!\! \lim_{\omega\to 1} T_{\beta,j}^{(1)}(\omega) 
= T_{\beta,j}^{(1)} + \ln(2)\left[
T^{(0)}_{j} v^{\rm CS, \Sigma}_{\beta,j}-
T_{F,j}^{{\rm CS} (1)}  - \frac{1}{2} \ln(2)\,  T^{(0)}_j v_j \right]\, ,
\nonumber
\end{eqnarray}
and similarly for the quantities in the $\overline{\rm CS}$ scheme. Here
we have also employed the identity (\ref{ShiFunOme1}). For $T_{F,j}^{{\rm
CS}(2)}(\omega)$ and $T_{G,j}^{{\rm CS}(2)}(\omega)$, the expansion of
the Wilson--coefficients (\ref{Def-OveC}) gives
\begin{eqnarray}
T^{{\rm CS}(2)}_{F,j}(\omega) &\!\!\! = \!\!\!&
T^{(0)}_{j}(\omega)\left\{
c^{(2)}_{F,j} - \left(c^{(1)}_j v_j + v_{F,j} \right)
s_j^{(1)}(\omega)
+ \frac{v^2_j}{2} s_j^{(2)}(\omega)   \right\}\, ,
\label{eq:Tj2FCSomega}
\\
T^{{\rm CS}(2)}_{G,j}(\omega) &\!\!\!\! = \!\!\!\!&
T^{(0)}_{j}(\omega)\Big\{
c^{(2)}_{G,j} -  v_{G,j}\,  s_j^{(1)}(\omega)\Big\}
\, .
\label{eq:Tj2GCSomega}
\end{eqnarray}
The conformal moments proportional to $\beta_0$, are obtained from the
\MS\ result given by Eq.\ (\ref{Def-T2b}). The expansion in $\omega^2$
of the term $T_{\beta,j}^{(2)}$ is given by \req{eq:Tjbeta2omegaExp}. As
in the $ | \omega | =1$ case (see the discussion of expression
\req{eq:vbetasum}), we define the conformal moments of
$T^{(0)}(\omega)\otimes\left[v_\beta\right]_+$ by $T^{(0)}_j(\omega) \,
v_{\beta,j}^{\Sigma}(\omega)$. Note that the ``effective''conformal
moment $v_{\beta,j}^{\Sigma}$ now depends on $\omega$ and its expansion in
$\omega^2$ is given in \req{eq:vbetasumomegaExp}. Analogously to Eq.\
(\ref{Def-vCSbeta}), $v_{\beta,j}^{\rm CS,\Sigma} (\omega) $ is provided
by
\begin{eqnarray}
 v_{\beta,j}^{\rm CS,\Sigma} (\omega) =
v_{\beta,j}^{\Sigma}(\omega) +
\frac{T_{F,j}^{{\rm CS}(1)}(\omega)- 
T_{F,j}^{(1)}(\omega)}{T_j^{(0)}(\omega) }
\,  ,
\label{Def-vCSbetaOmega}
\end{eqnarray}
and the expansion can easily be obtained by means of Eqs.\
(\ref{Def-TNLOomega}), (\ref{eq:sj1omegaExp}), and 
(\ref{eq:TjF1omegaExp}).

The $\beta_0$ proportional NNLO terms in the $\overline{\rm CS}$ scheme 
are obtained by making the replacements
\begin{eqnarray}
\label{eq:T2omegaExpbarCS}
T^{(2)}_{\beta,j}(\omega) \to 
T^{\overline{\rm CS}(2)}_{\beta,j}(\omega) = 
T^{(0)}_{j}(\omega)\left\{
c^{(2)}_{\beta,j} -  v_{\beta,j}\, s_j^{(1)}(\omega)\right\}
\;\mbox{and}\;
v^{{\rm CS},\Sigma}_{\beta,j} \to
v^{\overline{\rm CS}}_{\beta,j} = v_{\beta,j}\,T^{(0)}_{j}(\omega) \, ,
\end{eqnarray}
while the other terms remain the same as in the CS scheme.

\section{Radiative corrections to the photon-to-pion transition form
factor}
\label{Sec-NumEst}
\setcounter{equation}{0}

This section is devoted to a model independent study of radiative
corrections to the pion-to-photon transition form factor in the case of one
quasi-real photon ($ | \omega | \to  1$) and in the small and intermediate
$ | \omega | $-regions. We also illustrate how the perturbative QCD approach to
exclusive processes can be tested in a novel way by a sum rule and how the
two lowest non-trivial conformal moments of the pion-distribution amplitude
could be extracted from experimental data in the intermediate 
$ | \omega | $-region.

In Section \ref{SubSec-NLO} we briefly review the features of the
radiative corrections to the first few conformal moments of the
hard-scattering amplitude in the \MS\ and CS schemes to NLO. We point
out that asymptotic formulas with respect to the conformal spin $j+1$
provide a very good approximation of these moments in question for a
rather low value of $j \ge 4$. As a byproduct, we propose a simple
method for reconstructing the amplitude from its conformal moments,
which is outlined in Appendix \ref{App-ReconHSA}.

In Section \ref{SubSec-NNLO} we present the numerical values of the NNLO
corrections to the first five non-vanishing conformal moments of the
hard--scattering amplitude in the CS and $\overline{\rm CS}$ schemes. We
point out their general features and discuss different possibilities for
treating the $\beta_0$ terms. In particular, we consider the lowest
conformal partial wave and compare its contribution to the
photon-to-pion transition form factor with experimental data. We study
the influence of radiative corrections to the sum rule and show that
higher-order corrections will interfere only slightly in the extraction
of the two lowest non-trivial conformal moments of the distribution
amplitude.

\subsection{Features of radiative corrections at NLO}
\label{SubSec-NLO}

\begin{table}[t]
\caption{
\label{Tab-NLO}
First moments of $T^{(0)}_{F,j}$, $T^{(1)}_{F,j}$,  
and $T^{{\rm CS}(1)}_{F,j}$ 
for $ | \omega | =1$ with $\mu_f^2= 2 Q^2$ and 
for $ | \omega | =0.8$ with $\mu_f^2= Q^2$.}
\begin{center}\begin{tabular}{|c|c|c|c|c|c|c|c|c|c|c|c|}
\hline
 & $j$ & 0 & 2 & 4 & 6 & 8 & 10 & 12 & 14 & 16 & 18 
\\\hline\hline 
$ | \omega | =1$ & $T^{(0)}$ 
   &1.5 & 0.58 & 0.37 & 0.27 & 0.21 & 0.17 & 0.15 & 0.13 & 0.11 & 0.10\\ 
$\mu_f^2= 2 Q^2$ 
     & $T^{(1)}_{F,j}$ 
   &-3.75 & 1.2 & 2.14 & 2.38 & 2.42 & 2.39 & 2.33 & 2.26 & 2.19 & 2.12 \\ 
& $T^{{\rm CS}(1)}_{F,j}$
    & -2.25 & 1.91 & 2.52 & 2.58 & 2.51 & 2.41 & 2.3 &  2.19 & 2.09 & 2 
\\\hline\hline
$ | \omega | =0.8$ & $T^{(0)}$       
   &1.19 & 0.15 & 0.03 & 0.01 & - & - & - & - & - & -\\ 
$\mu_f^2= Q^2$ & $T^{(1)}_{F,j}$ 
   &-2.02 & 0.14 & 0.1 & 0.03 & 0.01 & - & - & - & - & - \\ 
& $T^{{\rm CS}(1)}_{F,j}$
    & -1.78 & 0.16 & 0.1 & 0.03 & 0.01 & - & - &  - & - & -
\\\hline
\end{tabular}\end{center}
\end{table}
Let us first compare the NLO corrections in the $\overline{\rm MS}$ and
CS schemes for $ | \omega | \to  1$. The first ten non-vanishing moments
$T^{(1)}_{F,j}$ and $T^{{\rm CS} (1)}_{F,j}$, given by Eqs.\
(\ref{eq:T1MS}) and (\ref{eq:Tj1FCS}), are shown in Table \ref{Tab-NLO}.
From this table we realize that the main difference between the two schemes
is in the first two moments which differ by about 50\%, and in a
seemingly faster decrease of the moments in the CS scheme for large $j$.
Indeed, in the large $j$ asymptotics the leading terms are
\begin{eqnarray}
\label{App-T-NLO}
T^{(1)}_{F,j} &\!\!\!\! = \!\!\!\!&
\frac{2j+3}{(j+1)(j+2)}\left\{2 S_{1}(1 + j)^2 - \frac{9}{2} +
O\left((j+2)^{-1}\right)\right\},
\\
 T^{{\rm CS} (1)}_{F,j} &\!\!\!\! = \!\!\!\!&
 \frac{2j+3}{(j+1)(j+2)}\left\{
S_{1}(1 + j) \left[S_{1}(1 + j) + \frac{3}{2} + 4 \ln(2)\right] - 
\frac{9}{2} -
\zeta(2) - 3 \ln(2) + O\left((j+2)^{-1}\right)\right\}.
\nonumber
\end{eqnarray}
Taking into account the large $j$ asymptotics of the $S_{1}$ functions, 
given by
\begin{eqnarray}
\lim_{j\to\infty} S_{1}(1 + j) =
\ln(2 + j)+ \gamma_{\rm E}
\, ,
\label{log}
\end{eqnarray}
the ratio $T^{{\rm CS} (1)}_{F,j}/T^{(1)}_{F,j}$ slowly approaches 1/2.
The difference is caused by the (infinite) resummation of off-diagonal
terms in the $\overline{\rm MS}$ scheme. The asymptotic formulas
(\ref{App-T-NLO}) have a relative error of less than 2\% already for
$j\ge 4$. Thus, by knowing a few lowest moments and their asymptotics we
gain a complete insight into the radiative corrections for $|\omega|=1$.
In Appendix \ref{App-ReconHSA} we use this result to make an approximate
reconstruction of the hard-scattering amplitude in the momentum fraction
representation from its conformal moments. The consequence of the
logarithmical behaviour in Eq.\ (\ref{log}) is obviously an increase of
radiative corrections with growing conformal spin. It is shown in Table
\ref{t:Q2omega1} that already for $j=8$, radiative corrections are of
the size of $80\%$ for $\alpha_s/\pi \simeq 0.1$ (i.e., $\mu_r \approx
2$ GeV for one-loop $\alpha_s$ with $n_f=3$). From this point of view,
one might conclude that perturbation theory breaks down for rather large
values of $j$. Fortunately, higher conformal spin contributions are
suppressed by the non-perturbative input, see Eq.\
(\ref{Res-EndPoiBeh}), and so perturbative QCD remains applicable. In
the photon-to-pion transition form factor there might be also a
cancellation of the lowest partial wave with the remainder, which is due
to their relative minus sign. Of course, the net contribution of
radiative corrections depends on the model of the distribution amplitude
itself.

With decreasing $|\omega|$, higher conformal partial waves are starting
to be exponentially suppressed, and as we have shown in Section
\ref{SubSec-SmallOmega}, radiative corrections logarithmically enhance
this suppression. Also note that off-diagonal contributions to each
partial wave, which are relatively suppressed by powers of $\omega^2$
with respect to the diagonal ones, are becoming small. If we approach
the equal virtuality case, i.e., $\omega=0$, only a factorization scheme
independent constant, arising from the lowest partial wave, will
survive. Thus, by decreasing $|\omega|$ the differences between the
$\overline{\rm MS}$ and CS schemes must be washed out. In Table
\ref{Tab-NLO} we illustrate these effects for $ | \omega | =0.8$. For the two
lowest non-vanishing partial waves the differences between these two
schemes is reduced to about $\pm 14\%$ and for higher ones below 2\%.
Also in the CS scheme the contributions from the functions
$s^{(i)}(\omega) = O(\omega^2)$, cf.\ Eq.\ (\ref{Def-sFun}), are power
suppressed. So one expects from Eq.\ (\ref{Def-TNLOomega}) that
radiative corrections due to the normalization factors $c^{(i)}_j$ are
the essential ones, but with one exception. Since the coefficient
$c^{(1)}_2$ is relatively small compared to the anomalous dimension
$\gamma^{(1)}_2$, $O(\omega^2)$ corrections remain important for the
second partial wave in the intermediate $ | \omega | $ region.

\subsection{Predictions to NNLO accuracy}
\label{SubSec-NNLO}

\subsubsection{The quasi-real photon limit}

\begin{table}[t]
\caption{
\label{Tab-NNLOpred}
The first five non-vanishing Wilson--coefficients appearing in the 
perturbative expansions  of $T_j^{\rm CS}$ and $T_j^{\CSeq}$  
with respect to $\alpha_s/\pi$ at NNLO accuracy 
for $|\omega|=1$. The results are obtained employing $\mu_f^2=\mu_r^2=2 Q^2$.
}
\begin{center}\begin{tabular}{||c||c||c||c|c|c|c|||}
\hline
$j$ & 
$T_j^{(0)}$ & 
$\displaystyle \frac{C_F}{2}  T_{F,j}^{{\rm CS}(1)}$ & 
$\displaystyle \frac{C_F^2 }{4} T_{F,j}^{{\rm CS}(2)}$ & 
$\displaystyle - \frac{C_F(2C_F-C_A) }{8}  
              T_{G,j}^{{\rm CS}(2)}$ & 
$\displaystyle - \frac{C_F\beta_0}{8}   
          T_{\beta,j}^{(2)}$ &
$\displaystyle - \frac{C_F\beta_0}{8} T_{\beta,j}^{\CSeq(2)}$ 
\\\hline\hline 
0 &  1.5 &  -1.5 &  1.42 &  -0.04 & -12.23  
& -9.09
\\ \hline
2 &  0.58 &  1.27  &  -2.28 &   -0.53 & 8.58  
& 8.06
 \\ \hline
4 &  0.37 &  1.68  &  -0.46 &   -0.60 & 13.56
& 11.33
 \\ \hline
6 &  0.27 &  1.72  &  1.25 &   -0.58 & 15.17
& 12.06
 \\ \hline
8 &  0.21 &  1.67  &  2.54 &   -0.55 & 15.68
& 12.05
 \\ \hline
\end{tabular}\end{center}
\end{table}

We now turn to the discussion of NNLO effects in the CS and \CS\ schemes
starting with the limit $|\omega| \to 1$.
In Table \ref{Tab-NNLOpred} we present the numerical values of 
the Wilson--coefficients $T^{{\rm CS}(2)}_{F,j}$, $T^{{\rm CS}(2)}_{G,j}$, 
$T^{{\rm CS}(2)}_{\beta,j}=T^{(2)}_{\beta,j}$, and $T^{{\CSeq}(2)}_{\beta,j}$
corresponding to Eqs.\ (\ref{eq:Tj2FCS}), (\ref{eq:Tj2GCS}), 
(\ref{eq:Tj2betaMS}), and (\ref{eq:betaCSbar}), respectively. 
The values of $T^{{\rm CS}(2)}_{F,j}$, $T^{{\rm CS}(2)}_{G,j}$, 
and $T^{{\CSeq}(2)}_{\beta,j}$
have been obtained by means of
the NNLO result for the deep inelastic scattering structure function
$g_1$ \cite{ZijNee94}. 

Let us investigate in more detail
the contribution of the lowest partial wave 
to the transition form factor,
which is scheme dependent
for $\omega \neq 0$ 
and for $|\omega|=1$ reads:
\begin{itemize}
\item in the $\overline{\rm MS}$ scheme:
\begin{eqnarray}
\label{Res-ParWav0-MS}
F_{\gamma \pi}(Q) \!\!\!&=&\!\!\!
\frac{\sqrt{2} f_{\pi}}{2 Q^2}
\Bigg\{ 1 - \frac{5}{3} \, \frac{\alpha_s{(\mu_r)}}{\pi}  +
     \frac{\alpha_s^2{(\mu_r)}}{\pi^2}
   \Bigg[ \cdots -\frac{\beta_0}{2} \Bigg(-1.811 + \frac{5}{6} 
   \ln \left( \frac{2 Q^2}{\mu_r^2} \right)
\\
 &&\hspace{5.7cm}-\; 0.285
\ln \left( \frac{2 Q^2}{\mu_f^2} \right)\Bigg) \Bigg] +
O(\alpha_s^3) \Bigg\} \, ,
\nonumber
\end{eqnarray}
\item in the CS scheme:
\begin{eqnarray}
F_{\gamma \pi}(Q) \!\!\!&=&\!\!\!
\frac{\sqrt{2} f_{\pi}}{2 Q^2}
\Bigg\{ 1 - \frac{\alpha_s{(\mu_r)}}{\pi}  +
     \frac{\alpha_s^2{(\mu_r)}}{\pi^2}
   \Bigg[0.917 -\frac{\beta_0}{2} \Bigg(-1.811 + \frac{1}{2} 
   \ln \left( \frac{2 Q^2}{\mu_r^2} \right)
\\
 &&\hspace{5.7cm}+\; 0.048
\ln \left( \frac{2 Q^2}{\mu_f^2} \right)\Bigg)\Bigg] +
O(\alpha_s^3) \Bigg\} \, ,
\nonumber
\end{eqnarray}

\item in the $\overline{\rm CS}$ scheme:
\begin{eqnarray}
\label{0ParWavbarCS}
&&\hspace{-2.5cm}F_{\gamma \pi}(Q) =
\frac{\sqrt{2} f_{\pi}}{2 Q^2}
\Bigg\{ 1 - \frac{\alpha_s{(\mu_r)}}{\pi}  +
     \frac{\alpha_s^2{(\mu_r)}}{\pi^2}
   \Bigg[0.917 -\frac{\beta_0}{2} \Bigg(-1.347 + \frac{1}{2} 
   \ln \left( \frac{2 Q^2}{\mu_r^2} \right)\Bigg)\Bigg]
+O(\alpha_s^3) \Bigg\} \, .
\end{eqnarray}
\end{itemize}
For $\alpha_s(\mu_r^2)/\pi=0.1$, the ratio of the NLO to the LO
contribution is $-17\%$ in the $\overline{\rm MS}$ scheme, and $-10\%$
in the CS scheme. This difference arises from the fact that in the
$\overline{\rm MS}$ scheme off-diagonal terms of the hard-scattering
amplitude are resummed. In Eq.\ (\ref{Res-ParWav0-MS}) we see that the
$\ln (2 Q^2/\mu_f^2)$ term is rather small compared to the $\ln (2
Q^2/\mu_r^2)$ one. This is even more the case in the CS scheme, while in
the $\overline{\rm CS}$ scheme the $\ln (2 Q^2/\mu_f^2)$ term vanishes
completely, since all off-diagonal entries in the NLO evolution have
been removed. The sign alternating series of the $\beta_0$
non-proportional terms is due to the Sudakov effect, see Ref.\
\cite{MusRad97} for a detailed discussion.

Since factorization scale changing effects in the hard scattering
amplitude are for the lowest partial wave quite small and since they will be
compensated by the evolution of the non-perturbative part (see Section
\ref{SubSec-Evo}),  we set $\mu_f^2=2Q^2$ in the following and discuss the
scale setting of the residual $\mu_r$ dependence. First, let us equate
$\mu_r^2=2 Q^2$:
\begin{equation}
F_{\gamma \pi}(Q) =
\frac{\sqrt{2} f_{\pi}}{2 Q^2}
\left[
1 - \frac{\alpha_s{(2 Q^2)}}{\pi}  -
     \left\{ {7.23 \atop 5.14}  \right\} \frac{\alpha_s^2{(2 Q^2)}}{\pi^2}
   + O(\alpha_s^3)
\right] \quad \mbox{for}\quad \left\{ { \mbox{CS} \atop
\overline{\mbox{CS}} } \right.\; \mbox{-scheme}\, .
\end{equation}
Hence, for $\alpha_s(\mu_r^2)/\pi=0.1$, the ratio of the NNLO to the LO
contribution is $-7.2\%$ and $-5.1\%$, and the ratio of the NNLO to the
NLO contribution (the measure of the convergence of the perturbative QCD
expansion) is $\approx 70\%$ and $\approx 50\%$, in the CS and  
$\overline{\rm CS}$ schemes, respectively.

The main part of these rather large NNLO contributions arises from the
$\beta_0$-proportional term. Owing to the off-diagonal parts, it is
larger by about a factor of two in the CS scheme than in the 
$\overline{\rm CS}$ scheme. 
It is appealing to resum this large
contribution by the Brodsky--Lepage--McKenzie proposal
\cite{BroLepMac83}
(for application to exclusive processes see also \cite{BroJiPanRob97}),
in which all terms proportional to $\beta$ are
absorbed in the coupling by the scale setting $\mu_r=\mu_{\rm BLM}$:
\begin{equation}
F_{\gamma \pi}(Q) =
\frac{\sqrt{2} f_{\pi}}{2 Q^2}
\left\{ 1 - \, \frac{\alpha_s{(\mu_{\rm BLM})}}{\pi}  +
      0.92
     \frac{\alpha_s^2{(\mu_r)}}{\pi^2}
   + O(\alpha_s^3) \right\}
\, , 
\end{equation}
with
\begin{eqnarray}
 \mu_{\rm BLM}^2 = 2 Q^2 \left\{ {1/37.43  \atop 1/14.78
}\right\}\quad
\mbox{for} \quad \left\{ { {\rm CS} \atop \overline{\rm CS} }\right.
\mbox{-scheme}\, .
\end{eqnarray}
The ratio of the NNLO to the NLO coefficient is now only minus one and
reflects the Sudakov effect in the conformal theory. However, as we
realize, combining the COPE result with the $\overline{\rm MS}$ result
of the $\beta_0$-proportional piece induces a rather low scale. For
instance, for $2 Q^2 = 4\ \mbox{GeV}^2$ we have $\mu_{\rm BLM}^2 \sim
0.1\ \mbox{GeV}^2$ in the CS scheme and hence the non-perturbative
behaviour of the coupling is needed. If we completely remove the
off-diagonal terms, the BLM scale squared is enlarged by a factor of
$2.7$ and is now closer to that in the $\overline{\rm MS}$ scheme
\cite{MelNizPas01} given in Table \ref{t:BLMomega1}. What one is
actually doing here is to combine perturbative QCD with speculations
about the non-perturbative behaviour of the QCD coupling and so,
strictly speaking, one is leaving the perturbative ground on which the
whole analysis was based. However, one advantage of this proposal is
that the result predicted by conformal symmetry is recovered if we
consequently assume a hypothetical fixed-point of the $\beta$-function
during our considerations. What we in fact do by the freezing of the
coupling, is to assume that this non-perturbative fixed-point is at
$Q^2=0$.

\begin{figure}[t]
\unitlength1pt
\begin{center} 
\begin{picture}(400,200)(0,0)
\put(0,35){\rotate{$2 Q^2 F_{\gamma \pi}(\omega=\pm 1, Q)$\ [GeV]}}
\put(20,0){\insertfig{11}{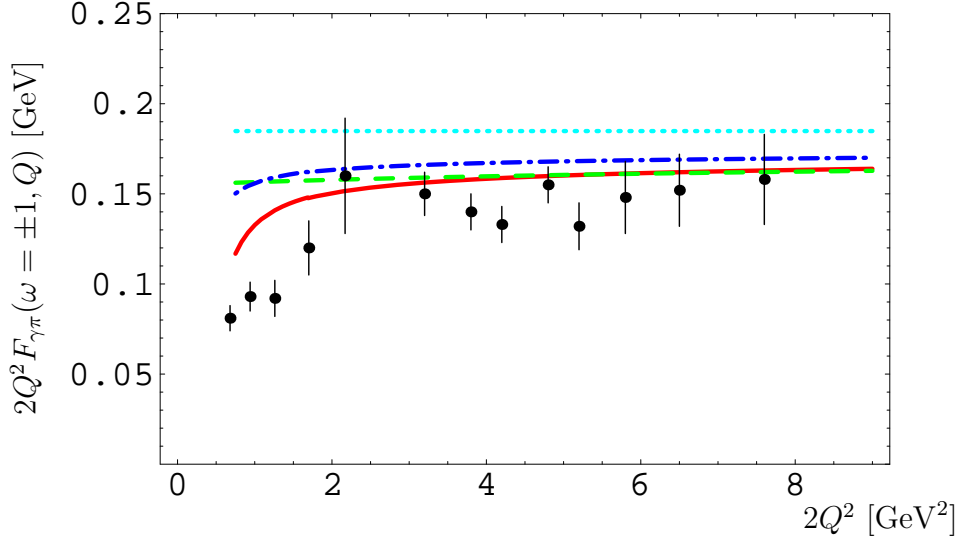}}
\put(300,-8){$2 Q^2\ [{\rm GeV}^2]$}
\end{picture}
\end{center}
\caption{
The contribution of the first partial wave to the scaled photon-pion
transition form factor $2 Q^2 F_{\gamma\pi}(\omega=\pm 1, Q)$ 
is shown in LO
(dotted), NLO (dash-dotted), and NNLO (solid and dashed) accuracy for the
$\overline{\rm CS}$ scheme. The renormalization scale has been set to
$\mu_r^2=2 Q^2$ (solid) and to the BLM scale $\mu_r^2=2
Q^2/14.78$ (dashed). The data are taken from Refs.\ \cite{Behetal91,Sav97}.
}
\label{Fig1}
\end{figure}
In Fig.\ \ref{Fig1} we compare the experimental data from the CLEO 
experiment with the
prediction arising from the lowest conformal moment, the only one that
survives in the asymptotic limit $Q^2\to \infty$. Without further
considerations, we assume, as  has also been done in the employed
method of data extraction (see Section 3 of Ref.\ \cite{Sav97}), that
the quasi-real photon limit has been reached. The prediction for
the asymptotically large $Q^2$ in this $ | \omega | \to  1$ limit is displayed as
a dotted line. As we have discussed, radiative corrections reduce the size of
this prediction for realistic values of $Q^2$. The dash-dotted line
represent the NLO and the solid line the NNLO prediction for the
standard $\overline{\rm MS}$ definition of the coupling in the
corresponding approximation with the normalization
$\alpha_s(\mu_r=M_Z)=0.118$. The dashed line is the prediction for the
BLM scale setting and accumulates non-perturbative effects by
freezing the coupling to $\alpha_s(\mu_r=0)=0.6$ by  adding an
effective gluon mass. Let us note that the scale setting ambiguities at 
$2 Q^2=1\ {\rm GeV}^2$ are of the order of $20\%$  if we
vary the renormalization scale $\mu_r^2$ from $0.5\ {\rm GeV}^2$ to
$2\ {\rm GeV}^2$. This ambiguity can be further reduced by going to 
higher orders in $\alpha_s$.

This comparison shows that theoretical uncertainties due to higher
order radiative corrections and scale setting ambiguities are much
smaller than the error of the experimental data. At larger value of $Q^2$
there is no significant contribution of higher partial waves. However,
there is a significant discrepancy of the results in Fig. 1 in the
region $0.5\ {\rm GeV}^2 < 2Q^2 < 2\ {\rm GeV}^2$, which may indicate
the presence of higher partial waves. Since evolution effects in this
kinematical region are rather strong, we could employ them to pin down
the size of higher partial waves. However, the question arises: 
Can we in this kinematical window rely on the leading twist result?

By considering the size of power suppressed contributions, we will now
argue that the answer to this question is positive. Since the (local)
matrix elements of any operator appearing in the OPE can only be built
with the momentum four-vectors $P$, Lorentz covariance immediately tells
us that power suppressed contributions are of even power in $Q$. The
only dimensional parameters that can appear are the mass of the pion
$m_\pi\sim 0.14\ {\rm GeV}$ and the QCD scale parameter $\Lambda\sim
0.2\ {\rm GeV} $. Assuming that multi--partonic correlation functions
will not have a strong numerical enhancement, we expect that the
contributions proportional to $m_\pi^2$ will provide a relative
correction of the order of $0.02\ {\rm GeV}^2/2 Q^2$. The size of the
remaining non-perturbative corrections arises from ambiguities in
summing the perturbative series and can be estimated in the framework of
renormalons. In the conformal scheme, we might again borrow the results
from the analysis of the coefficient of the structure function $g_1$,
which gives for the lowest moment an uncertainty of a similar size, see
Ref.\ \cite{HerMauManSteSch96} and references therein. Thus, we might
conclude that the non-perturbative effects to the contribution of the
zero partial wave are smaller than $10\%$ at a scale of $2 Q^2 \sim 1\
{\rm GeV}^2$. Certainly, this rather optimistic speculation should be
confronted with other methods used. The lowest conformal moments of
twist-four quark-gluon-quark operators have been obtained by means of
QCD sum rules \cite{BraFil90Bal98,Kho97}. 
Here it has been found that, relative to $m_\pi^2$, 
a certain matrix element is numerically enhanced by a factor of ten.
This certainly would strike our point of view and indicate that
the so-called Wandzura-Wilczek approximation, in which higher
multi-parton correlations are neglected, fails. Renormalon induced
corrections have been studied in a model-dependent way in Ref.\
\cite{GodKiv97}, where their relative size has been estimated to be $0.2
\, {\rm GeV}^2/2 Q^2$. Note, however, that in this analysis excitations
of higher conformal partial waves have been included. Making it short,
we stress that the estimate of higher twist-contributions has to be made
in a consistent framework that is set by the scheme in which one
started. Combining estimates from different approaches is a popular but
rather awkward procedure.

\begin{table}
\caption{
\label{t:Q2omega1}
The ratio of NLO to LO and NNLO to NLO radiative corrections
in units of $\alpha_s/\pi$ for $|\omega|=1$
and $\mu_r^2=\mu_f^2=2 Q^2$ in the $\overline{\rm MS}$, CS, and
$\overline{\rm CS}$ schemes.
}
\begin{center}\begin{tabular}{||c||c|c||c|c|c||}
\hline
$j$ & 
$\displaystyle  \frac{T_j^{(1)}}{2 \, T_j^{(0)}}$ &
$\displaystyle  \frac{T_j^{(2)}}{2 \, T_j^{(1)}}$ &
$\displaystyle  \frac{T_j^{{\rm CS}(1)}}{2 \, T_j^{(0)}}$ &
$\displaystyle  \frac{T_j^{{\rm CS}(2)}}{2 \, T_j^{{\rm CS}(1)}}$ &
$\displaystyle  \frac{T_j^{\CSeq(2)}}{2 \, T_j^{\CSeq(1)}}$ 
\\[0.2cm]\hline\hline 
0 &  -1.67 &  - & -1  &  7.23 & 5.14 
\\ \hline
2 &  1.37 &  -  & 2.18  & 4.54   & 4.13
 \\ \hline
4 &  3.88 &  -  & 4.58 &  7.44  & 6.11
 \\ \hline
6 &  5.92 &  -  & 6.42  & 9.21  & 7.39
 \\ \hline
8 &  7.64 &  -  & 7.93  & 10.56  & 8.39
 \\ \hline
\end{tabular}\end{center}
\end{table}

\begin{table}
\caption{
\label{t:BLMomega1}
Ratio $a_{\rm BLM} =\mu_{\rm BLM}^2/2 Q^2$ of  the BLM scale squared  to
$2 Q^2$ and the ratio of the NNLO to the NLO coefficient  
in  units of $\alpha_s/\pi$
for $\mu_r^2=\mu_{\rm BLM}^2$, $\mu_f^2=2 Q^2$, and $ | \omega | =1$. 
}
\begin{center}\begin{tabular}{||c||c|c|c||c|c|c||}
\hline
 $j$ &
$a_{\rm BLM}^{\overline{\rm  MS}}$&
$a_{\rm BLM}^{\rm CS}$&
$a_{\rm BLM}^{\CSeq}$&
$\displaystyle  \frac{T_j^{(2)}}{2 \, T_j^{(1)}}$ &
$\displaystyle  \frac{T_j^{{\rm CS}(2)}}{2 \, T_j^{{\rm CS}(1)}} = 
\displaystyle  \frac{T_j^{\CSeq(2)}}{2 \, T_j^{\CSeq(1)}}$ 
\\ \hline \hline
0 & 1/8.79 & 1/37.43 & 1/14.78 & - & -0.92
 \\ \hline
2 & 1/120.08 & 1/20.1 & 1/16.76 & - & -2.22
 \\ \hline
4 & 1/68.73 & 1/36.17 & 1/20.05 & - & -0.63
 \\ \hline
6 & 1/70.3 & 1/50.41 & 1/22.52 & - & 0.39
 \\ \hline
8  & 1/75.29 & 1/64.4 & 1/24.54 & - & 1.19
\\ \hline
\end{tabular}\end{center}
\end{table}

In this process with a quasi-real photon higher partial waves are
summed. Even if the $ | \omega | \to  1$ limit is not reached in the
experiment, a rather large number of terms will contribute. Without any
knowledge about the shape of the distribution amplitude, it is a rather
vague assumption to truncate this series by hand to extract the values
of the lowest partial waves from the normalization of the
pion-to-pion transition form factor. Figure\ \ref{Fig1} clearly shows that
the dominant contribution, at least for $2Q^2 > 2\ {\rm GeV}^2$, arises
from the lowest partial wave and the remainder is small. The fact that
the contributions of higher partial waves cancel each other is not excluded
and it remains a claim that the asymptotic shape of the distribution
amplitude is established by experimental data. In principle,
one can gain more information on the remainder if one also employs the
evolution of the distribution amplitude. However, even if rather high
precision data are available, the deconvolution problem is 
not easy to solve. As we have already mentioned, at NLO the perturbative
correction will increase with growing conformal spin. The same tendency
can be read off from Table \ref{t:Q2omega1} also in NNLO, where the
$\beta_0$ proportional term is the dominant one. This is also reflected
by the decrease of the BLM scale as shown in Table \ref{t:BLMomega1},
where we can also see that the remaining corrections at NNLO are
moderate. Note that the BLM scale is rather low for $ 2 \le j$ in the
$\overline{\rm MS}$ and CS schemes, which is due to off-diagonal terms.

\subsubsection{What can we learn from the small and intermediate
$ | \omega | $ regions ?}

\begin{table}[h]
\caption{
\label{t:Q2omega}
Same as in Table \ref{t:Q2omega1} for $|\omega|=0.8$ and 
$\mu_r^2=\mu_f^2=Q^2$.
}
\begin{center}\begin{tabular}{||c||c|c||c|c|c||}
\hline
$j$ & 
$\displaystyle  \frac{T_j^{(1)}}{2 \, T_j^{(0)}}$ &
$\displaystyle  \frac{T_j^{(2)}}{2 \, T_j^{(1)}}$ &
$\displaystyle  \frac{T_j^{{\rm CS}(1)}}{2 \, T_j^{(0)}}$ &
$\displaystyle  \frac{T_j^{{\rm CS}(2)}}{2 \, T_j^{{\rm CS}(1)}}$ &
$\displaystyle  \frac{T_j^{\CSeq(2)}}{2 \, T_j^{\CSeq(1)}}$ 
\\[0.2cm]\hline\hline 
0 &  -1.14 &  - & -1  &  4.21 & 3.58  
\\ \hline
2 &  0.6 &  -  & 0.7  & 1.3  & 0.8 
 \\ \hline
4 &  2.3 &  -  & 2.4 & 4.6  & 3.8
 \\ \hline
6 &  3.7 &  -  & 3.7  & 5.8  & 5.1
 \\ \hline
\end{tabular}\end{center}
\end{table}

\begin{table}[h]
\caption{
\label{t:BLMomega}
Analogous to Table \ref{t:BLMomega1} for the ratio
$a_{\rm  BLM} = \mu_{\rm BLM}^2/ Q^2$, where  $\mu_f^2=Q^2$
and $ | \omega | =0.8$.
}
\begin{center}\begin{tabular}{||c||c|c|c||c|c|c||}
\hline
 $j$ &
$a_{\rm BLM}^{\overline{\rm MS}}$&
$a_{\rm BLM}^{\rm CS}$&
$a_{\rm BLM}^{\CSeq}$&
$\displaystyle  \frac{T_j^{(2)}}{2 \, T_j^{(1)}}$ &
$\displaystyle  \frac{T_j^{{\rm CS}(2)}}{2 \, T_j^{{\rm CS}(1)}} = 
\displaystyle  \frac{T_j^{\CSeq(2)}}{2 \, T_j^{\CSeq(1)}}$ 
\\ \hline \hline
0 & 1/7.4 & 1/9.7 & 1/7.4 & - & -0.92
 \\ \hline
2 &1/155 & 1/80 & 1/62 & - & -8.54
 \\ \hline
4 & 1/40 & 1/38 & 1/28 & - & -3.63
 \\ \hline
6 & 1/37 & 1/37 & 1/27 & - & -2.3
 \\ \hline
\end{tabular}\end{center}
\end{table}

As it has been clearly spelt out in Ref.\ \cite{DieKroVog01} and
explained in a more general way in Section \ref{SubSec-SmallOmega}, the
small $ | \omega | $ region is suitable for a novel test of the perturbative
QCD approach to the class of exclusive light-cone dominated processes.
As we noted in Section \ref{SubSec-NLO}, for decreasing $|\omega|$ the
differences between different schemes will decrease too. This is
illustrated for $ | \omega | =0.8$ in Tables \ref{t:Q2omega} and
\ref{t:BLMomega}. For the lowest [second] partial wave we have about a
$-40\%$ $[10\%]$ effect at NNLO compared to NLO or a $-4\%$ $[< 1\%]$
correction compared to LO for $\mu_r^2=\mu_f^2=Q^2$ and $\alpha_s/\pi
\simeq 0.1$. Altogether, we find $\sim -15\%$ reduction of the LO
prediction for the lowest partial wave and an increase of about $7\%$
and $33\%$ for the second and fourth one, respectively. The main part of
the NNLO corrections arises from the $\beta_0$ proportional term. Its
absorption in the running coupling via the BLM scale setting prescription
again requires the knowledge of the non-perturbative behaviour of
$\alpha_s$. Table \ref{t:BLMomega} shows that then a sign change occurs at
NNLO, where the BLM scale for the second partial wave is quite low and
its remaining NNLO correction is rather large.

\begin{figure}[t]
\unitlength1pt
\begin{center}
\begin{picture}(600,200)(0,0)
\put(0,25){\rotate{$2 Q^2 F_{\gamma \pi}(\omega,Q)$\ [GeV]}}
\put(20,0){\insertfig{8}{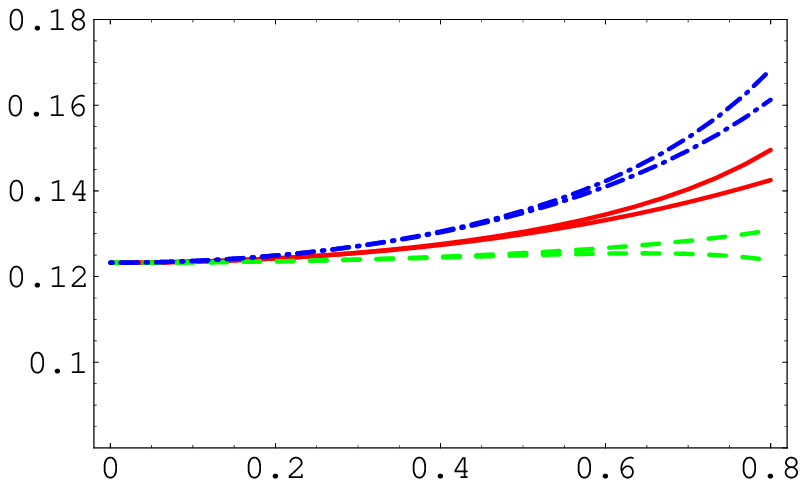}}
\put(260,0){\insertfig{8}{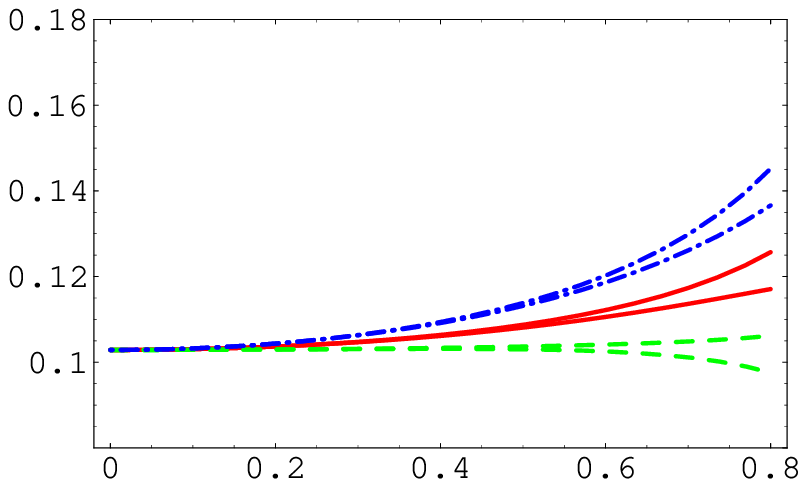}}
\put(230,-5){$  \omega  $}
\put(470,-5){$  \omega  $}
\put(140,-15){(a)}
\put(380,-15){(b)}
\end{picture}
\end{center}
\caption{
The $\omega$--dependence of the scaled photon-to-pion transition form
factor $2 Q^2 F_{\gamma \pi}(\omega,Q)$ at $Q^2 = 2\ {\rm GeV}^2$ in LO (a) and
in $\overline{\rm CS}$ at NNLO (b) for three different values of $B_2
=\{0,-0.5,0.5\}$ is shown as solid, dashed and dash-dotted lines,
respectively. The spread of the corresponding lines display the
sensitivity of the predictions with respect to the parameter $B_4$,
which is equated to $-0.25$ and $0.25$ for the lower and upper curve,
respectively. 
}
\label{Fig23}
\end{figure}

Certainly, the resummation of the $\beta$ proportional corrections is
associated with a new input that is not well known. Thus, in the
following discussion concerning the extraction of non-perturbative
conformal moments of the distribution amplitude we prefer the naive
scale setting prescription $\mu_r^2=\mu_f^2=Q^2$. In panels (a) and (b)
of Fig.\ \ref{Fig23} we display the $\omega$--dependence for the scaled
photon-to-pion transition form factor evaluated in the $\overline{\rm
CS}$ scheme at LO and NNLO, respectively. One clearly sees that the
prediction is almost independent of $\omega$ for $ | \omega |  \le 0.2$ and
only a negligible dependence arises for $ 0.2 < | \omega | < 0.4$. Radiative
corrections will only shift this prediction downward. Note that this
shift will slightly increase, if we go to higher orders of $\alpha_s$.
For the lowest partial wave they can be taken from the calculation of
the radiative corrections to the Bjorken sum rule, which are evaluated
in the third-loop approximation \cite{GorLar86LarVer91} and roughly
estimated at four loops \cite{BroKat02}. Consequently, confronting these
predictions with experimental measurements would provide either a novel
test of perturbative QCD or an insight into the size of higher-twist
contributions. To enhance statistics, one can even integrate over the
small $ | \omega | $ region: 
\begin{eqnarray}
\frac{Q^2}{\omega^{\rm cut}} \int_0^{\omega^{\rm cut}} d\omega
 F_{\gamma \pi}(\omega,Q) =
 \frac{\sqrt{2}f_\pi}{3}\Bigg\{1\!\!\! &-&\!\!\!\frac{\alpha_s(Q)}{\pi} -
3.583 \frac{\alpha_s^2(Q)}{\pi^2}-
20.215\frac{\alpha_s^3(Q)}{\pi^3}
\\
\!\!\! &-&\!\!\! (\sim 200) \frac{\alpha_s^4(Q^2)}{\pi^4}  + 
O(\alpha_s^5) + O\left(m_\pi^2/Q^2,\Lambda^2/Q^2\right)  \Bigg\},
\nonumber
\end{eqnarray}
where $\omega^{\rm cut} < 0.4$ and $n_f=3$. 

If we vary the unknown $B_2$ and $B_4$ parameters in the range that is
suggested by non-perturbative estimates, i.e., $-1/2 \le B_2 \le 1/2$
and $-1/4 \le B_4 \le 1/4$, it can be realized that in the intermediate
$ | \omega | $ region the perturbative QCD predictions start to be dependent
on the value of $B_2$ and for larger $ | \omega | $ even on that of $B_4$, while
higher partial waves can be safely neglected. Here it is important that
radiative corrections do not spoil a possible extraction. They rather
shift the curves in the whole region and slightly enhance the spread of
the curves associated with $B_4$. This is caused by the fact that the
radiative corrections to the lowest two partial waves are negative and
approximately of the same size, while they are positive in all other
cases. Since the gap between the curves for both extreme cases,
where $B_2 =\pm 0.5$ is about 30\% or even larger, we might expect
that these curves can be distinguished by a measurement. This would
also test the reliability of non-perturbative methods. Certainly, a
precise extraction of $B_2$ or even $B_4$ also requires high precision
measurements with high statistics. Assuming that such data are
available, the extraction of the non-perturbative parameter itself is
rather simple. First, a measurement in the small $ | \omega | $ region should
be confronted with the parameter free prediction that depends only on
the running of the coupling. As we argued, we do not expect higher twist
corrections to be essential. If that were experimentally
established, a simple two parameter dependent fit
\begin{eqnarray}
F_{\gamma \pi}(\omega,Q)=
f_{\pi}  \left[ T_0(\omega,Q) +  
\frac{18}{7} T_2(\omega,Q) B_2(Q) + \frac{45}{11}\, 
T_4(\omega,Q) B_4(Q)   \right]
\end{eqnarray}
could be used to extract the two conformal moments $B_2(Q)$ and
$B_4(Q)$ as long as the statistics is high enough. Moreover, a consistency
check is provided by the $Q$-dependence of these parameters. Note that
in the $\overline{\rm CS}$ scheme the mixing between different
partial waves is caused by evolution at NNLO, while in the
$\overline{\rm MS}$-scheme the mixing appears already at NLO in both the
hard-scattering amplitude and the evolution of $B_i$. In the CS scheme the
mixing appears first in the evolution to NLO accuracy.

Finally, we want to comment on the size of evolution effects which are
caused by the off-diagonal entries in the anomalous dimension matrix. We
numerically observe that the conformal symmetry breaking terms at NLO,
(compare the \MS\ scheme results with the CS ones in Table \ref{Tab-NLO},
as well as at NNLO, see Table \ref{Tab-NNLOpred}) can provide an
enhancement of the corresponding corrections up to 50\%. A similar
relative effect one would naively expect from the evolution due to the
unknown mixing in the \CS\ scheme arising at NNLO. However, since it
does not appear at the input scale $Q_0$ due to the initial condition
(\ref{IntConB}), this mixing effect is in fact small \cite{Mue98}. For
instance, at NLO in the \MS\ and CS schemes this mixing effect in the
lowest partial wave goes up to -1.3\% and 2.3\%, respectively, for the
evolution from $Q_0= 0.5$ GeV to $Q= 20$ GeV and $ | \omega | =1$. Note,
however, that in the \MS\ scheme cancellation appears in the off-diagonal
terms between the $C_F^2$ and $C_F \beta_0$ proportional parts 
and that the contribution from only the $C_F^2$ proportional term might
be of the order of 2\%. This number should be compared with the
correction in the hard-scattering induced by the off-diagonal terms,
which is of about 7\%. As we discussed above, these corrections are
reduced for $|\omega| < 1$. Since the ratio of diagonal entries in NNLO
to those in NLO is smaller than $3 \alpha_s/\pi$ 
for the first five even diagonal
terms, we might argue that the ratio of off-diagonal entries in the \MS\
scheme is of the same size. Assuming so, one would expect that unknown
mixing effects, induced by $\Delta^{\overline{\rm CS}(1)}_{jk}$ due to
evolution are reduced by a factor 1/3 or even more with respect to
those at NLO. Thus, we expect from the numbers given above at NLO that
the contribution of $\Delta^{\overline{\rm CS}(1)}_{jk}$ is smaller than
0.5\% in the \MS\ scheme. In the \CS\ scheme, the contribution of the
log term in Eq.\ (\ref{Def-B-barCS}) can be estimated by comparison with
the NLO effect in the CS scheme. Here one finds a relative contribution
smaller than $\alpha_s(Q) \ln[Q/Q_0]/\pi$. For the example discussed
above that would produce a mixing effect smaller than 1\%. Therefore,
for higher values of the input scale, e.g., $Q_0 \sim 1$ GeV, we expect
a rather tiny mixing in the \CS\ scheme.

\section{Conclusions}
\setcounter{equation}{0}

We have employed conformal symmetry in the perturbative sector to
evaluate the NNLO corrections for the pion production through two-photon
fusion. The requirement of a manifestly conformal invariant result partly
removes the ambiguities arising from the factorization. However, the
ambiguities are retained in the scheme dependence of the forward
Wilson-coefficients and anomalous dimensions and in the treatment of the
conformal symmetry breaking induced by the trace anomaly,
proportional to the $\beta$-function. The latter ambiguity has been
studied here in two alternative schemes: (i) combining the conformal
predictions with the \MS\  result and (ii) improving the partial wave
decomposition of the conformal invariant theory by the renormalization
group equation. The second possibility minimizes the mixing of partial
waves and gives us an almost good quantum number, namely, the conformal
spin. For decreasing values of $|\omega|$, the differences between these
schemes are removed, since off-diagonal terms are suppressed by powers in
$\omega^2$.

As it has been known before, for $ | \omega | =1$ NLO corrections can be
considered to be small for only the two lowest partial waves, since the NLO
corrections logarithmically increase with the conformal spin. This
behaviour is analogous to the large $j$ behaviour of the
Wilson-coefficients in DIS arising from soft gluon configurations. The
effect is manifested by $\ln(1-x)$ - terms that are associated with
factorization log's and, consequently, are
absent in the lowest partial wave in the conformal schemes. 
Other $\ln(1-x)$ terms are related to
the Sudakov effect and are manifested in a sign alternating series for
$\beta=0$. However, the numerical study has shown that the NNLO
corrections are dominated by the $\beta_0$ proportional term, as
expected. In general, this term is rather large compared with the NLO
coefficient and thus the BLM scale is rather low, which drives the
coupling in the non-perturbative region.

We have compared the NNLO predictions with the existing data for the
quasi-real photon case and have found that for $2Q^2 > 2 {\rm GeV}^2$
the contribution from the lowest partial wave is compatible with the
data. The deviation below this scale is induced by non-perturbative
effects or by the contribution of higher partial waves and it requires a
deeper insight in power suppressed contributions. Although, there is no
doubt in the literature that the CLEO measurement can be analyzed in
this limit, we should state here that partial waves with sufficiently
large conformal spin will be exponentially suppressed. This can
affect the analysis only if the matrix elements of conformal operators with
rather large conformal spin will contribute. On the other hand, we know
that the net effect of all partial waves with the conformal spin $j+1
\ge 3$ is small. Obviously, this does not necessarily mean that the
matrix elements themselves are small and therefore we cannot say that
the asymptotic form of the pion distribution amplitude is experimentally
established. Also, having a look at non-perturbative results from other
methods given in the literature, a strong statement that the asymptotic
form is suggested by those estimates cannot in fact be made.

These problems we have spelled out, can be separately studied apart from
the $ | \omega | \to  1$ limit. Indeed, in the small $ | \omega | $ region,
perturbative QCD predicts a sum rule that has the same status as the
Bjorken sum rule in deep inelastic scattering, evaluated at order
$\alpha_s^3$. A first test of this sum rule might be possible with
existing $e^+ e^-$ machines and would offer us a first insight into the
size of power suppressed contributions for exclusive processes from
experimental data. We expect that such contributions will turn out to be
small. If this should be established experimentally, one might attack
the extraction of the first and second lowest conformal moments of the
distribution amplitude. This is an important task, since it would open a
window to testing the reliability of non-perturbative methods applied to
exclusive quantities. Having in mind that the collinear factorization,
applied here to the photon-to-pion transition form factor, is also
adopted for the analysis of exclusive B-physics, it is timely to
confront such methods with experimental measurements.

Let us finally give a short outlook for the application of the
conformal approach to other processes. After a simple replacement of the
decay constant and matrix elements the NNLO result obtained can be used
for the analysis of $\eta$ production, i.e., its flavour octet
component. Moreover, the formalism can be extended in a straightforward
manner to the $\eta^\prime$-to-photon transition form factor. Guided by
the large $j+1$ asymptotics of the conformal moments, it is also possible
to reconstruct the hard-scattering amplitude in DVCS. The reliability
of this technique can be tested at NLO and partly also in NNLO, i.e.,
for $\beta$ proportional terms. We also want to add that one can go one
order further in $\alpha_s$ in the approximation of the first few
conformal moments of the hard-scattering amplitude, since we can borrow
the forward Wilson--coefficients from the non-singlet sector of
the deep-inelastic structure function $F_3$, evaluated in ${\rm N}^3$LO 
\cite{RetVer00}.

\vspace{1cm}
\noindent{\bf Acknowledgments }
We would like to thank M. Diehl and P. Kroll for useful discussions. 
One of us (B.M.) acknowledges the support by the Alexander von Humboldt
Foundation. This work was partially supported by the Ministry of Science
and Technology of the Republic of Croatia under Contract No. 0098002.

\renewcommand{\theequation}{\Alph{section}.\arabic{equation}}%

\begin{appendix}

\section{Structure of the hard-scattering amplitude in the \MS\ scheme}
\label{App-MSstr}
\setcounter{equation}{0}

Here we prove that the ${\rm LN}(\omega,x)$ terms, appearing in Eqs.\
(\ref{Def-T1F}- \ref{Def-T2G}), are related to the factorization log's.
We do not distinguish between the renormalization and the factorization
scales, since the difference appears only in the $\omega$-independent
term $\ln(\mu_r/\mu_f)$. The hard-scattering amplitude is given by the
sum over all Feynman diagram ${\cal F}$ contributions
\begin{eqnarray}
\frac{1}{Q^2} T(\omega,x,Q/\mu) = \sum_{\cal F} T_{\cal F}(\omega,x,Q/\mu),
\end{eqnarray}
where we have rescaled the individual contributions by $Q^2$ to have  
dimensionless amplitudes. Each of these contributions is given as a
product of propagators and vertices integrated over the virtual loop
momenta. The two photon vertices are connected by a chain of quark
propagators $S$ and quark-gluon-quark vertices $V$: 
\begin{eqnarray}
\slash\!\!\!\!S([1-2 x]P/2 + q - l) V_1\,\,
\slash\!\!\!\!S([1-2 x]P/2 + q - l + k_1) 
\dots \slash\!\!\!\!S([1-2 x]P/2 + q - l + \dots + k_n),
\end{eqnarray}
where $P = q_1 + q_2$ and the large momentum $q= (q_1 - q_2)/2$ flows
only into this chain. Momentum conservation requires that $l+l^\prime =
\sum_{i=1}^n k_i$, where $l$ and $l^\prime$ is the sum of virtual
momenta flowing into the first and flowing out of the second photon
vertex, respectively. Interchanging the two photon vertices will give
the crossed contributions with $ | \omega | \to  -\omega$. Obviously, there are
further propagators that depend only on the virtual momenta and $x P$ or
$(1-x)P$, but not on $q$. Introducing the Feynman-Schwinger
representation for the propagators, integrating over the virtual momenta
and making use of the on-shell condition $P^2=0$, give us the following
representation for the regularized contribution: 
\begin{eqnarray}
 T_{\cal F}(\omega,x,Q/\mu) = 
 \int_{0}^1 dz_1 \int_{0}^{z_1} dz_2 \dots \int_{0}^{z_{m-1}} dz_m 
\frac{\mu^{2\epsilon}  {\cal T}_{\cal F}(\underline{z}|\epsilon)}
{Q^{2\epsilon} \left[x\, \omega\,  {\cal B}_{1}(\underline{z}) + 
(1-x)\, \omega {\cal B}_{2}(\underline{z}) +1\right]^{1+\epsilon} }  \, .
\end{eqnarray}
Here $\epsilon$ is the dimensional regularization parameter and the
functions ${\cal B}_i(\underline{z})$ depend on the Feynman--Schwinger
variables $\underline{z}= \left\{z_1, \dots, z_m\right\}$ with $n \le
m$. We introduce the new variable \begin{eqnarray} y = \frac{1}{2}-
\frac{x}{2} {\cal B}_1(\underline{z}) - \frac{1-x}{2} {\cal
B}_2(\underline{z}) \, 
\end{eqnarray}
and write 
\begin{eqnarray}
 T_{\cal F}(\omega,x,Q/\mu) = \int_0^1 dy\, 
\frac{\mu^{2\epsilon} }
{Q^{2\epsilon}\left[1-(2 y-1)\omega  \right]^{1+\epsilon}} \,
V_{\cal F}(y,x|\epsilon)\, ,
\end{eqnarray}
where the unrenormalized convolution kernel is defined by
\begin{eqnarray}
V_{\cal F}(y,x|\epsilon) = 
\int_{0}^1 dz_1 \int_{0}^{z_1} dz_2 \dots \int_{0}^{z_{m-1}} dz_m  
\delta\left(y -\frac{1}{2}+
\frac{x}{2} {\cal B}_1(\underline{z})+\frac{1-x}{2} 
{\cal B}_2(\underline{z})\right) 
  {\cal T}_{\cal F}(\underline{z}|\epsilon) \, .
\end{eqnarray} 
The renormalization procedure provides the factorization log's
$\ln(Q^2/\mu^2)$, which always appear in combination with ${\rm
LN}(\omega,x)$ terms. The factorization theorem tells us that after
resummation of all Feynman diagrams the corresponding convolution
kernel is given by the evolution one. Obviously, the log independent terms
can also be represented as convolution. The support 
of all these kernels is known and follows from the restrictions of the
${\cal B}_i(\underline{z})$, which are obtained from their definition 
and the topology of Feynman graphs 
(see, for instance \cite{Nak71,MueRobGeyDitHor94}).

\section{Consistency check with the forward-limit results}
\label{App-ConChe}
\setcounter{equation}{0}

In this section we present a consistency check between
the results for the non-singlet coefficient function
of the DIS polarized structure function $g_1$  \cite{ZijNee94}
and the hard-scattering amplitude of 
the pion transition form factor \cite{BelSch98,MelNizPas01}. 
The former quantity is known to NNLO, while
for the latter one
discussed in Sect. 3.1
the calculation has been performed 
up to $\beta_0$-proportional NNLO terms.
Both results have been obtained in the
$\overline{\mbox{MS}}$ scheme. 
Making use of the fact that both quantities, 
the photon-to-pion transition form factor
and the structure function $g_1$, 
are defined by the two-photon amplitudes
belonging to a general class of the scattering amplitudes
for the two-photon process at light-like distances,
we are able to transform 
the results for the photon-to-pion transition form factor 
to the results for $g_1$.

The general scattering amplitude for the two-photon processes 
is given by the time-ordered product of two electromagnetic currents
sandwiched between the in- and out- hadronic states
with momenta $P_1$ and $P_2$, respectively.
Using the notation $q=(q_1+q_2)/2$ 
($q_1$ and $q_2$ are incoming and outgoing photon momenta), 
$P=P_1+P_2$, and $\Delta=P_2-P_1$,
the following generalized Bjorken region can be defined 
\cite{MueRobGeyDitHor94,Mue97a}:
\begin{equation}
\nu = P\cdot q \to \infty
\qquad \mbox{and} \qquad 
Q^2=-q^2 \to \infty \, ,
\end{equation}
with the scaling variables
\begin{equation}
\omega=\frac{\nu}{Q^2}
\qquad \mbox{and} \qquad 
\eta=\frac{\Delta\cdot q}{\nu}
\,.
\end{equation}
In the forward case, corresponding to DIS, 
$1/\omega$ can be identified with the Bjorken variable
$x_{Bj}$ and $\eta$ vanishes, 
while for the two-photon production of a hadron 
$\eta=1$.
The relations between the non-forward ER-BL kernels
and forward DGLAP kernels were extensively studied
and derived in Ref. \cite{MueRobGeyDitHor94},
while in Ref. \cite{Mue97a} the consistency between
the transition form factor and $g_1$ results 
has been reported up to NLO. 
Here we explain in more detail the technical side of 
these consistency checks and extend them to $\beta_0$-proportional 
NNLO terms.
 
The building blocks of the 
hard-scattering amplitude for the
photon-to-pion transition form factor $T(\omega,x,Q,\mu_f)$ 
given in Eqs. \req{eq:Texp}, \req{eq:T0omega} and 
(\ref{Str-T1}-\ref{Def-C2b})
can generally be written as
\begin{eqnarray}
{\cal A}_1^{(\gamma^* \pi)}(\omega,x) & = &
\frac{1}{1-\omega (2 y -1) -i \varepsilon} \otimes
           \delta(y-x)  \, ,
\nonumber \\[0.25cm]
{\cal A}_{2,n}^{(\gamma^* \pi)}(\omega,x) & = &
\frac{\ln^n(1-\omega (2 y -1) -i \varepsilon)}{1-\omega (2 y -1) -i 
\varepsilon} \otimes
 [ \VF(y,x) ]_+ \, ,
\label{eq:A12n}
\end{eqnarray}
where $n=0,1,2$. 
Note that we have reintroduced 
the $i \varepsilon$ term, originally present 
in the definition of Feynman propagators.
For the kernels that appear in Eqs.  
(\ref{Str-T1}-\ref{Def-C2b})
we use  a generic symbol $\VF(y,x)$. 
Furthermore, for the kernels of interest, given in 
(\ref{eq:g}-\ref{Def-dotKer}) and \req{Def-EvoKer-Beta},
the function $\VF$ is of the general form
\begin{equation}
\VF(x,y)=\theta(y-x) \FF(x,y) +
\left\{x \to \bar{x} \atop y \to \bar{y}  \right\} \, .
\end{equation}

We have to extend our restricted non-forward kinematics to the
whole kinematical region. 
The extension of the BL kernels $\VF$
to the whole $x,y$ region ($-\infty < x,y < \infty$) 
\cite{MueRobGeyDitHor94}
is accompanied by the change of the $\theta$ function  as 
\begin{equation}
\VF^{ext}(x,y) =
\VF(x,y)_{ \left| \theta(y-x) \rightarrow 
\theta\left(1 - \frac{x}{y}\right) \,
\theta \left(\frac{x}{y} \right) \, \mbox{\scriptsize sign}(y) \right.}\, .
\label{eq:Vext}
\end{equation}
Furthermore, the dependence on $\eta$ has to be restored
and one performs the following change of variables:
\begin{equation}
x  \rightarrow \frac{1 + t/\eta}{2} \, , \quad
y  \rightarrow \frac{1 + t'/\eta}{2} \, , \quad
\omega \rightarrow \eta \omega \, .
\end{equation}
The definition of the distribution amplitudes,
with which the hard-scattering amplitude is convoluted, 
as well as the definition of its generalized counterpart,
introduce the restriction  $-1<t<1$. 
After examining the $\theta$ functions in the kernels \req{eq:Vext} 
and taking into account $|\eta|\le 1$, one obtains $-1<t'<1$.
The building blocks of the generalized two-photon
scattering amplitude are thus obtained and they are
of the form
\begin{eqnarray}
{\cal A}_1(\omega,\eta,t) & = &\frac{1}{1-\omega t -i \varepsilon} \; ,
\nonumber \\[0.25cm]
{\cal A}_{2,n}(\omega,\eta,t)  & = &\int_{-1}^{1}  \; dt' \; \; 
\frac{\ln^n(1-\omega t' -i \varepsilon)}{1-\omega t' -i \varepsilon} 
 \;  \; \;\frac{1}{2 \eta} 
\left[ \VF^{ext}\left(\frac{1+t'/\eta}{2},\frac{1+t/\eta}{2}\right) 
\right]_+ \; .
\label{eq:A12nGEN}
\end{eqnarray}
It is easy to see that  relations \req{eq:A12n}
indeed represent the $\eta =1$ limit of \req{eq:A12nGEN}.

The forward limit (i.e. the forward Compton amplitude) corresponds to
$\eta \rightarrow 0$ and, due to the optical theorem, the non-singlet
coefficient functions of the DIS polarized structure function $g_1$
(contributing to the total cross-section) are determined by taking the
imaginary part of the forward amplitude. Taking into account the
definition of $g_1$ and its coefficient functions (see \cite{ZijNee94}),
one arrives at the following recipe\footnote{ Factor $1/(2 \pi)$ comes
from the dispersion relation, and the additional factor of $2$ from the
definition of $g_1$. The origin of the factor $\omega$ in Eq. \req{eq:Ag1}
lies in the fact that the transition form factor is scaled by $Q^2$,
while the forward Compton amplitude is scaled by $P\cdot q$.} for the
building blocks of the non-singlet coefficient functions $C_q^{NS}$: 
\begin{equation}
{\cal A}_i^{(g_1)}(z) = \frac{\omega}{\pi} \, \mbox{Im}
\left[\,  \lim_{\eta \to 0} \, {\cal A}_i (\omega,\eta,t)\,  \right]_{
| \; \omega=1/z, \, t=1 } \, .
\label{eq:Ag1}
\end{equation}
The $\eta \to 0$ limit of the extended ER-BL kernels \cite{MueRobGeyDitHor94}
results in the corresponding DGLAP kernels $\PF$ of the general
form
\begin{equation}
\PF(z) = \theta(z) \theta(1-z) \; \pF(z)  \quad \mbox{with} \quad
\left[ \PF \left( z \right) \right]_+ =
\PF(z) - \delta(1-z) \int_0^1 dz' \PF(z') \, .
\end{equation}
It is straightforward to derive
\begin{equation}
\lim_{\eta \to 0} \; \frac{1}{2 \eta} 
\left[ \VF^{ext}\left(\frac{1+t'/\eta}{2},\frac{1+t/\eta}{2}\right)
\right]_+
=
\; \mbox{sign}(t) \, \frac{1}{t} \left[ \PF \left( \frac{t'}{t} \right) 
\right]_+
\, ,
\end{equation}
with $\pF$ given by
\begin{equation}
\frac{1}{t} \;\;  \pF\left( \frac{t'}{t} \right) \; = \; 
 \lim_{\eta \to 0} \; \frac{1}{2 \eta} 
\left[ \FF \left(\frac{1+t'/\eta}{2},\frac{1+t/\eta}{2}\right) 
-\FF \left(\frac{1-t'/\eta}{2},\frac{1-t/\eta}{2}\right)
\right]_+ \, . 
\label{eq:LIM}
\end{equation}
The imaginary part of expressions
\req{eq:A12nGEN} is obtained
by making use of 
\begin{equation}
\mbox{Im} \frac{1}{1-t' \omega - i \varepsilon}
= \pi \delta(1-t' \omega) \, ,
\end{equation}
and for more complicated functions, containing
$\ln^n(1-t' \omega -i \varepsilon)$ ($n=1,2$), 
we derive the following decompositions 
\begin{eqnarray}
\frac{\ln(1-\omega t' -i \varepsilon)}{1-\omega t' -i \varepsilon} 
& = & \frac{1}{(1-s)_+} \otimes \frac{1}{1-s t' \omega -i \varepsilon}
\, ,
\nonumber \\[0.25cm]
\frac{\ln^2(1-\omega t' -i \varepsilon)}{1-\omega t' -i \varepsilon} 
&=& 2 \left( \frac{\ln(1-s)-\ln s}{1-s} \right)_+ \otimes 
\frac{1}{1-s t' \omega -i \varepsilon}
\, .
\end{eqnarray}
Alternatively, for the imaginary parts of expressions,
containing logarithms, one can refer to \cite{GelShi64}.
Finally, we present the results relevant to the NNLO calculation:
\begin{eqnarray}
{\cal A}_1^{(g_1, {\rm NS})}(z) & = & \delta(1-z) 
\, , \nonumber \\[0.25cm]
{\cal A}_{2,0}^{(g_1,{\rm NS})}(z)  &=&  \left[ \PF(z) \right]_+
\, , \nonumber \\[0.25cm]
{\cal A}_{2,1}^{(g_1,{\rm NS})}(z) & = & 
\theta\left(\frac{z}{z'}\right) \theta\left(1-\frac{z}{z'}\right)
\frac{1}{\left(z'-z\right)_+}
\otimes \; \left[ \PF(z') \right]_+
\, , \nonumber \\[0.25cm]
{\cal A}_{2,2}^{(g_1,{\rm NS})}(z) & = & 
\theta\left(\frac{z}{z'}\right) \theta\left(1-\frac{z}{z'}\right)
\; 2 \left( \frac{\ln(z'-z)-\ln z}{z'-z}\right)_+
\otimes \; \left[ \PF(z') \right]_+
\, .
\label{eq:A12nG1}
\end{eqnarray}

Hence, the building blocks for the hard-scattering amplitude
of the photon-to-pion transition form factor, 
${\cal A}_i^{(\gamma^* \pi)}(\omega,x)$
given in \req{eq:A12n},
can be brought into correspondence with the building blocks of 
the non-singlet coefficient function
of the polarized structure function $g_1$,
${\cal A}_i^{(g_1, {\rm NS})}(z)$ displayed in \req{eq:A12nG1}.
For various ER-BL kernels $\VF(x,y)$, the corresponding
DGLAP kernels $\PF(z)$ are obtained by taking
the limit \req{eq:LIM} with $t=1$ (and $t' \to z$) taken into account. 
In Table \ref{t:transition} we list some selected results.
We mention here that the integration of two ``+''
forms results again in the ``+'' form,  
but the contributing terms should be appropriately  
rearranged.
\renewcommand{\arraystretch}{2}
\begin{table}
\caption{Selected forward counterparts of the non-forward quantities. }
\begin{center}
\begin{tabular}{l|c}
\hline \hline
$\delta(x-y)$ &  $\delta(1-z)$ 
\\
$v^a(x,y)$ &  
    $ (1-z)
     \, \theta(z) \theta(1-z) $
\\
$v^b(x,y)$ & 
    $\displaystyle \frac{2 z}{1-z}
     \, \theta(z) \theta(1-z) $
\\
$v(x,y)$ & 
    $\displaystyle \frac{1+z^2}{1-z}
     \, \theta(z) \theta(1-z) $
\\
$g(x,y)$ & 
    $ \displaystyle -\frac{2 \ln(1-z)}{1-z}
     \, \theta(z) \theta(1-z) $
\\
$\dot{v}^a(x,y)$ & 
    $ (1-z) (\ln z+1)
     \, \theta(z) \theta(1-z) $
\\
$\dot{v}(x,y)$ & 
     $ \displaystyle 
        \left[ (1-z)+ \frac{1+z^2}{1-z} \ln z \right]
     \, \theta(z) \theta(1-z) $
\\
$\ddot{v}(x,y)$ & 
     $ \displaystyle 
       \left[ 2 (1-z) \ln z+ \frac{1+z^2}{1-z} \ln^2 z \right]
     \, \theta(z) \theta(1-z) $
\\
$\dot{g}(x,y) $ & 
     $ \displaystyle 
    \left[ -\frac{\pi^2}{3(1-z)}- \frac{\ln^2(1-z)}{1-z} \ln z
           + \frac{2 \mbox{Li}_2(1-z)}{1-z} \right]
     \, \theta(z) \theta(1-z) $
\\[0.2cm] \hline
\multicolumn{2}{l}{$
\begin{array}{lcl}
\displaystyle
\frac{\ln(1-\omega(2 y-1))}{1-\omega(2 y -1)} \otimes v(y,x)
&\rightarrow& \displaystyle
\left\{ -(1-z) +  \frac{3}{2(1-z)} 
+ \frac{2}{1-z} \ln(1-z)
+ \frac{1+z^2}{1-z} \Big[\ln(1-z)-\ln z\Big] \right\}\\
& &
\times \, \theta(z) \theta(1-z) 
\end{array}
$}
\\[0.2cm]
\hline \hline
\end{tabular}
\end{center}
\label{t:transition}
\end{table}
\renewcommand{\arraystretch}{1}

Following the procedure explained above, we finally obtain
the forward counterparts of the elements of the hard-scattering
amplitude for the photon-to-pion transition form factor:
\begin{eqnarray}
T^{(0)}(\omega,x)
& \rightarrow & \delta(1-z) 
\, ,
\label{eq:T0lim} \\[0.2cm]
T_F^{(1)}(\omega,x)
& \rightarrow & 
\left[ -(1-z)+\frac{3}{2(1-z)}-\frac{3z}{(1-z)}
+ \frac{1+z^2}{1-z} \Big(\ln(1-z)-\ln z\Big) \right]_+
\nonumber \\ & &
-\frac{3}{2} \delta(1-z)
\, ,
\label{eq:T1lim}\\[0.2cm]
T_{\beta}^{(2)}(\omega,x)
& \rightarrow & 
\left[ -\frac{31}{12}(1-z)+\frac{19}{4(1-z)}
-\frac{209(1+z^2)}{36(1-z)} 
\right. \nonumber \\ & & \left.
+ \left( \frac{3}{2} (1-z) -\frac{3}{2(1-z)} 
+ \frac{19(1+z^2)}{6(1-z)} \right) \ln(1-z)
\right. \nonumber \\ & & \left.
+ \left( -\frac{1}{4} (1-z) + \frac{5}{2} z 
- \frac{19(1+z^2)}{4(1-z)} \right) \ln(z)
\right. \nonumber \\ & & \left.
+\frac{1+z^2}{1-z}
\left(-\frac{5}{4} \ln^2 z
- \frac{1}{2} \ln^2(1-z) + 2 \ln z \ln (1-z)
+ \mbox{Li}_2(1-z)+\frac{\pi^2}{3} 
\right)
\right]_+
 \nonumber \\ & & 
-3 \delta(1-z) 
\, .
\label{eq:T2betalim}
\end{eqnarray}
Here, the expressions on the r.h.s.\ represent the scale independent LO,
NLO and $\beta_0$-proportional NNLO terms of the non-singlet coefficient
function ($C_q^{NS}$) of $g_1$. Similar expressions can be written for
the terms proportional to $\ln^n (Q^2/\mu^2)$. Following the notation of
\cite{ZijNee94}, the $\theta(z) \theta(1-z)$ factors are not shown in
Eqs.\ (\ref{eq:T0lim}-\ref{eq:T2betalim}). We note that the limit of
$T^{(i)}(\omega,1-x)=T^{(i)}(-\omega,x)$ corresponds to the antiquark
case ($C_{\bar{q}}^{NS}$).

Our results (\ref{eq:T0lim}-\ref{eq:T2betalim})%
\footnote{
The representation of the coefficient functions
in a form
$
A \delta(1-z) + \left[ F(z) \right]_+
$,
as given in
Eqs.\ (\ref{eq:T0lim}-\ref{eq:T2betalim})
and naturally emerging in our calculation,
is convenient
for the determination of the Mellin moments
$
c_j =\int_0^1 z^{j} c(z)
$
since the $j=0$ Mellin moment corresponds
to the term proportional to $\delta(1-z)$.
}
are numerically in agreement with the Mellin moments 
and up to a typo also with the analytical expression of the
$n_f$-proportional term displayed in the appendix of Ref.
\cite{ZijNee94}. Namely, in Eq.\ (A.2) the term $1/3 (1+11 z) \ln z$
should read $1/3 (1-11 z) \ln z$. 

\section{Determination of conformal moments}
\label{App-ConfMom}
\setcounter{equation}{0}

In this section we present a method for computing 
moments with respect to conformal partial waves with the index $k$.
We introduce the notation
\begin{equation}
\left< F(x) \right>_k \equiv
\int_0^1 \; dx \; F(x) \; \frac{x (1-x)}{N_k} \; C_k^{3/2}(2 x -1)
\, ,
\label{eq:Fxk}
\end{equation}
while $N_k$ is defined in \req{Rel-DA-ConMom}.
It follows trivially that
$
\left< F(1-x) \right>_k =
(-1)^k \, \left< F(x) \right>_k 
$. 
In the calculation of the photon-to-pion transition form factor
for the special case $ | \omega | =1$ we encounter
the functions $F(x)$ of the forms $f(x)/x$ and $f(x)/(1-x)$, 
with $f(x) \in \{ 1$, $\ln^n(x) \ln^m(1-x)$, 
$\mbox{Li}_2(x)$, $\mbox{Li}_3(x)$, $\mbox{S}_{1 2}(x)$
$\}$. 

It is convenient to use the following expression
for the Gegenbauer polynomials:
\begin{eqnarray}
\frac{x (1-x)}{N_k} \; C_k^{3/2}(2 x -1) &=&
(-1)^k \; \frac{2 (2 k + 3)}{(k+1)!} \;
\frac{d^k}{dx^k} \left[ x (1-x) \right]^{k+1} \nonumber \\
  &=&
(-1)^k \; \frac{2 (2 k + 3)}{(k+1)} \;
\sum_{i=0}^{k+1} (-1)^i \;
\left( k+1 \atop i \right) 
\left( k+i+1 \atop i+1 \right)
\; x^{i+1}
\, . \quad
\label{eq:expCk}
\end{eqnarray}
The evaluation of the conformal moments, i.e.,
in our case the evaluation of the expressions 
\begin{equation}
\left< \frac{f(x)}{x} \right>_k =
(-1)^k \; \frac{2 (2 k + 3)}{(k+1)} \;
\sum_{i=0}^{k+1} (-1)^i \;
\left( k+1 \atop i \right) 
\left( k+i+1 \atop i+1 \right)
\; \int_0^1 \; x^{i} \; f(x)
\, ,
\label{eq:fxxk}
\end{equation}
and
\begin{equation}
\left< \frac{f(x)}{1-x} \right>_k =
\frac{2 (2 k + 3)}{(k+1)} \;
\sum_{i=0}^{k+1} (-1)^i \;
\left( k+1 \atop i \right) 
\left( k+i+1 \atop i+1 \right)
\; \int_0^1 \; x^{i} \; f(1-x)
\, ,
\label{eq:fx1xk}
\end{equation}
consists then in calculating
the Mellin moments and
performing the summation.
The Mellin moments for the functions we encounter
in our calculation are well known
(see, for example \cite{DevDuk84,BluKur99}), 
and most of the nontrivial sums we are left with
can be found in \cite{Ver99}. The sums that usually appear are 
\begin{equation}
\begin{array}{ll}
\displaystyle \mbox{S}_m(n)=\sum_{i=1}^n \; \frac{1}{i^m} \, ,
&
\displaystyle \mbox{S}_{m,j_1,\ldots,j_p}(n)=\sum_{i=1}^n \; \frac{1}{i^m} 
\, \mbox{S}_{j_1,\cdots,j_p}(i) \, ,
\\[0.5cm]
\displaystyle \mbox{S}_{-m}(n)=\sum_{i=1}^n \; \frac{(-1)^i}{i^m}
\, ,
\qquad
&
\displaystyle 
\mbox{S}_{-m,j_1,\ldots,j_p}(n)=\sum_{i=1}^n \; \frac{(-1)^i}{i^m} 
\, \mbox{S}_{j_1,\cdots,j_p}(i)
\, .
\end{array}
\label{eq:Sm}
\end{equation}
The  functions $S_m(z)$ are expressed by $\psi(z)= d\ln\Gamma(z)/dz$ ones:
\begin{eqnarray}
\psi(z) = -\gamma_{\rm E} + S_1(z-1)\, , \qquad 
\frac{d^m}{dz^m}\psi(z) = m! (-1)^{(m+1)} \left[\zeta(m+1) -
 S_{m+1}(z-1)\right]\, .
\end{eqnarray}

For $\left< \mbox{Li}_3(x)/(1-x) \right>_k$
and $\left< \mbox{S}_{1 2}(x)/(1-x) \right>_k$,
the corresponding sums are missing in \cite{Ver99},
and to obtain them we turn to expressing the relevant functions
as convolutions of appropriate functions with
the known diagonal kernels.
Generally, the conformal moments of a kernel 
$[\tilde{v}]_+$ are defined by
\begin{equation}
\tilde{v}_{lk} \equiv \left< \, [\tilde{v} (x,y)]_+ \, \right>_{lk} =
\int_0^1 dx \, \int_0^1 dy \;   
  C_{l}^{3/2}(2 x-1)  \; [\tilde{v}(x,y)]_+ \;
 \frac{y(1-y)}{N_k} \; C_{k}^{3/2}(2 y -1)
 \, ,
\label{eq:tvlk}
\end{equation}
and for the kernels appearing in this calculation 
$\tilde{v}_{lk}=0$ for $l<k$ and $l-k$ odd.
The conformal moments of the convolution
$F(x)=G(y) \otimes [\tilde{v}(y,x)]_+$ then take the form 
\begin{equation}
\left< F(x) \right>_k =
\left< \, G(y) \otimes [\tilde{v}(y,x)]_+ \, \right>_k =
\sum_{l \ge k} \left< G(y) \right>_l \; \tilde{v}_{lk} 
\, .
\label{eq:FxkconvND}
\end{equation}
As before we use the simplified notation
for the diagonal moments
\begin{equation}
\tilde{v}_{kk} \equiv \tilde{v}_{k}
\, ,
\label{eq:simplv}
\end{equation}
i.e.,
for the diagonal conformal moments we retain just one 
index and 
the relation \req{eq:FxkconvND}
simplifies to 
\begin{equation}
\left< F(x) \right>_k =
\left< \, G(y) \otimes [\tilde{v}(y,x)]_+ \, \right>_k =
\left< G(y) \right>_k \tilde{v}_{k}
\; .
\label{eq:FxkconvD}
\end{equation}
Hence, in order to determine
$\left< \mbox{Li}_3(x)/(1-x) \right>_k$ and
$\left< \mbox{S}_{1 2}(x)/(1-x) \right>_k$,
we make use of the identities
\begin{eqnarray}
\frac{\mbox{Li}_2(1-y)}{1-y} \; \otimes \; [ v^a(y,x) ]_+
 & = &
- \frac{\mbox{Li}_3(1-x)}{x} + \frac{\zeta(3)}{x}
-\frac{\ln(x)}{1-x} + \frac{\ln(x) \ln(1-x)}{2 (1-x)}
 \nonumber \\ & & 
+\,  \frac{1}{2}\,  \left( \frac{\mbox{Li}_2(x)}{1-x} - 
                     \frac{\zeta(2)}{1-x} \right)
+ \left( \frac{\mbox{Li}_2(1-x)}{x} - 
         \frac{\zeta(2)}{x} \right)
\, , \quad
\label{eq:convLi3}
\end{eqnarray}
and
\begin{eqnarray}
\frac{\mbox{Li}_2(y)}{1-y} \; \otimes \; [ v(y,x) ]_+
 & = &
- \frac{\mbox{S}_{1 2}(x)}{1-x} 
+  \frac{\zeta(3)}{1-x} 
+\frac{\ln(1-x)}{x} 
+ \frac{\mbox{Li}_2(x)}{2 (1-x)} 
\nonumber \\ & &
+ \, \zeta(2)\,  \left( \frac{\ln (1-x)}{1-x} 
+  \frac{1}{1-x} \right) 
\, .
\label{eq:convS12}
\end{eqnarray}
The kernels $v^a$ and $v$ are defined in \req{Def-va}, \req{Def-vb} 
and \req{Def-V-LO},
while the corresponding moments 
can be read off from Eqs.\ (\ref{eq:vnab}) (for $\epsilon=0$) and
(\ref{Def-ConMomv}), respectively.

Finally, in Table \ref{t:confmom} we summarize the conformal
moments of the functions relevant to our calculation.

\renewcommand{\arraystretch}{2}
\begin{table}
\caption{The conformal moments of some relevant functions.}
\begin{tabular}{l|l}
\multicolumn{2}{c}{} \\ \hline \hline
$\displaystyle \left< \frac{1}{1-x} \right>_k$ &
$\displaystyle \frac{2 (2 k+3)}{(k+1)(k+2)}
 = \frac{1}{2 N_k}$ \\[0.3cm] \hline
$\displaystyle \left< \frac{\ln (1-x)}{1-x} \right>_k$ &
$\displaystyle \frac{1}{2 N_k}
\Big[ -\mbox{S}_1(k+2)-\mbox{S}_1(k) \Big]$ \\
$\displaystyle \left< \frac{\ln^2 (1-x)}{1-x} \right>_k$ &
$\displaystyle \frac{1}{2 N_k}
\left\{ \Big[ -\mbox{S}_1(k+2)-\mbox{S}_1(k) \Big]^2
+ \Big[ \mbox{S}_2(k+2)-\mbox{S}_2(k) \Big]  \right\}$ \\
$\displaystyle \left< \frac{\ln^3 (1-x)}{1-x} \right>_k$ &
$\displaystyle \frac{1}{2 N_k}
\left\{ \Big[ -\mbox{S}_1(k+2)-\mbox{S}_1(k) \Big]^3
+ 2 \, \Big[ -\mbox{S}_3(k+2)-\mbox{S}_3(k) \Big]
 \right. $ \\[-0.5cm]
 &
$ \displaystyle \qquad \: \: \left.
 + 3 \, \Big[-\mbox{S}_1(k+2)-\mbox{S}_1(k) \Big] \;
 \Big[ \mbox{S}_2(k+2)-\mbox{S}_2(k) \Big] \right\} $ \\[0.3cm] \hline
$\displaystyle \left< \frac{\ln (1-x)}{x} \right>_k$ &
$\displaystyle \frac{1}{2 N_k}
\left[ -\frac{1}{(k+1)(k+2)} \right]$ \\
$\displaystyle \left< \frac{\ln^2 (1-x)}{x} \right>_k$ &
$\displaystyle \frac{1}{2 N_k}
\left\{ \frac{-2}{(k+1)(k+2)}\Big[-\mbox{S}_1(k+2)-\mbox{S}_1(k)+1\Big] 
\right\}$ \\[0.3cm] \hline
$\displaystyle \left< \frac{\ln(x) \ln (1-x)}{1-x} \right>_k$ &
$\displaystyle \frac{1}{2 N_k}
\left[ -\zeta(2) - \mbox{S}_{-2}(k+2)-\mbox{S}_{-2}(k)
+ \mbox{S}_{2}(k+2)-\mbox{S}_{2}(k)
  \right.$ \\[-0.5cm] & $ \qquad \: \: \left. \displaystyle
-\left( 1 - (-1)^k \right) \frac{1}{(k+1)(k+2)} 
\right]$ \\[0.3cm] \hline
$\displaystyle \left< \frac{\mbox{Li}_2 (x)}{1-x} \right>_k$ &
$\displaystyle \frac{1}{2 N_k}
\left[ \zeta(2)- \mbox{S}_{2}(k+2)+\mbox{S}_{2}(k)
+\frac{1}{(k+1)(k+2)} \right]$ \\
$\displaystyle \left< \frac{\mbox{Li}_2 (1-x)}{1-x} \right>_k$ &
$\displaystyle \frac{1}{2 N_k}
\left[ \zeta(2) + \mbox{S}_{-2}(k+2)+\mbox{S}_{-2}(k)
-\frac{(-1)^k}{(k+1)(k+2)} \right]$ \\[0.3cm] \hline
$\displaystyle \left< \frac{\mbox{Li}_3 (x)}{1-x} \right>_k$ &
$\displaystyle \frac{1}{2 N_k}
\left\{ \zeta(3)
- \frac{(-1)^k}{(k+1)(k+2)}
\Big[\zeta(2)+\mbox{S}_{-2}(k+2)+\mbox{S}_{-2}(k) \Big] 
\right\}$ \\[0.3cm] \hline
$\displaystyle \left< \frac{\mbox{S}_{1 2} (x)}{1-x} \right>_k$ &
$\displaystyle \frac{1}{2 N_k}
\left\{ \zeta(3)
- \frac{1}{(k+1)(k+2)}
\Big[-\mbox{S}_{1}(k+2)-\mbox{S}_{1}(k)-\frac{1}{(k+1)(k+2)}\Big]
  \right. $ \\[-0.2cm] 
 & $\displaystyle  \left. \qquad \: \:
+ \, 
\Big[-\mbox{S}_{1}(k+2)-\mbox{S}_{1}(k)\Big]
\Big[\mbox{S}_{2}(k+2)-\mbox{S}_{2}(k)\Big]
\right\} $  \\ \hline  \hline
\end{tabular}
\label{t:confmom}
\end{table}
\renewcommand{\arraystretch}{1}

As a byproduct of this calculation, 
we list the following nontrivial sums:
\begin{eqnarray}
\sum_{i=0}^{n} (-1)^j \;
\left( n \atop j \right) 
\left( n+j \atop j+1 \right)
S_{1,2}(j+1) &=&
\frac{1}{n+1} 
\left[ -\frac{1}{n (n+1)}
- (-1)^n \Big( S_{-2}(n+1) + S_{-2}(n-1) \Big) \right]
   \nonumber \\[0.2cm]
\sum_{i=0}^{n} (-1)^j \;
\left( n \atop j \right) 
\left( n+j \atop j+1 \right)
\frac{S_{1,2}(j+1)}{j+1} &=&
\frac{1}{n+1} 
\left[ \frac{(-1)^n}{n (n+1)}
 \Big( S_{-2}(n+1) + S_{-2}(n-1) \Big) \right]
   \nonumber \\[0.2cm]
\sum_{i=0}^{n} (-1)^j \;
\left( n \atop j \right) 
\left( n+j \atop j+1 \right)
\frac{S_{3}(j+1)}{j+1} &=&
\frac{1}{n+1} 
\left[ -\frac{1}{n^2 (n+1)^2}
 + \Big( S_{1}(n+1) + S_{1}(n-1) \Big) 
   \right. \nonumber \\ & & \left. \quad \times
 \left(  \frac{1}{n (n+1)} + \frac{1}{n^2 (n+1)^2}\right)\right]
  \, ,
\label{eq:sums}
\end{eqnarray}
which complement the collection of sums found in \cite{Ver99}.

\section{Taylor expansions in $\omega$}
\label{App-OmegaRes}
\setcounter{equation}{0}

We now present the results for the five lowest even partial waves,
which are expanded in $\omega^2$ to the first seven non-vanishing terms.
For brevity, we will not denote the neglected terms.

The LO result can be simply expanded by means of its representation in
terms of hypergeometrical functions. Employing the identity
\begin{eqnarray*}
\frac{1}{(1+\omega)^{j+1+\epsilon}}
{_{2}F}_1\left({j+1+\epsilon, j+2+\epsilon  \atop
 2(j+2+\epsilon)}\Bigg|\frac{2\omega}{1+\omega}\right)
= {_{2}F}_1\left({j/2+\epsilon/2+1/2, j/2+\epsilon/2+1  \atop
 j+\epsilon+5/2}\Bigg|\omega^2\right)
\end{eqnarray*}
and representing the hypergeometrical functions as power series in $\omega$, 
after a few simple manipulation with $\Gamma$ functions we find:
\begin{eqnarray}
T_j^{(0)}(\omega) = \frac{3 + 2 j}{4} 
\sum_{n=0}^\infty \frac{ \sqrt{\pi}\,  \Gamma(1 + j + 2 n)}
  {n! \Gamma(5/2 + j + n)} \left(\frac{\omega}{2}\right)^{2 n +j} \, .
\end{eqnarray}
The first few moments read
\begin{eqnarray}
\label{eq:Tj0omegaExp}
T_0^{(0)}&\!\!\! \simeq\!\!\! &
1 + 0.2{\omega }^2 + 0.0857{\omega }^4 + 0.0476\,{\omega }^6 + 
   0.0303 {\omega }^8 + 0.0210 {\omega }^{10} + 0.0154 {\omega }^{12}
  \, ,
\nonumber\\
T_2^{(0)}&\!\!\! \simeq \!\!\! &
\frac{2 {\omega }^2}{15} \left(
1 + 0.6667{\omega }^2 + 0.4545{\omega }^4 + 0.3263{\omega }^6 + 
   0.2448{\omega }^8 + 0.1900{\omega }^{10} + 0.1517{\omega }^{12}
 \right)
  \, ,
\nonumber\\
T_4^{(0)} &\!\!\!\simeq \!\!\! &
\frac{8{\omega }^4}{315} 
 \left(
1+ 1.1538{\omega }^2 + 1.0769{\omega }^4 + 0.9502{\omega }^6 + 
   0.8252{\omega }^8 + 0.7152{\omega }^{10} + 0.6219{\omega }^{12}
\right)
  \, ,
\\
T_6^{(0)} &\!\!\!\simeq \!\!\! &
\frac{16{\omega }^6}{3003} \left(
1 + 1.6471{\omega }^2 + 1.9505{\omega }^4 + 2.0433{\omega }^6 + 
   2.0211{\omega }^8 + 1.9403{\omega }^{10} + 1.8325{\omega }^{12}
\right)
  \, ,
\nonumber\\
T_8^{(0)} &\!\!\!\simeq \!\!\! &
\frac{128 {\omega }^8}{109395} \left(
1 + 2.1429{\omega }^2 + 3.0745{\omega }^4 + 3.7304{\omega }^6 + 
   4.1449{\omega }^8 + 4.3736{\omega }^{10} + 4.4677{\omega }^{12}
\right)
  \, .
\nonumber
\end{eqnarray}
The relative error of these approximations is
for $j=\{0,2,4,6,8\}$ about
$\{0.1\%, 0.7\%, 2\%, 4.4\%,8\%\}$ for $ | \omega | =0.8$ and increases to
$\{0.6\%, 4\%, 10\%, 19\%,30\%\}$ for $ | \omega | =0.9$ .

The expansion of the $s^{i}_j(\omega)$ functions 
from Eq.\ (\ref{Def-sFun}) can be found in an analogous way:
\begin{eqnarray}
\label{eq:sj-gen}
s^{(i)}_j(\omega) \!\!\! &=&\!\!\! 
\frac{\sum_{n=0}^\infty  {\cal S}^{(i)}(j,n) \frac{\Gamma(1 + j + 2 n)}
  {n! \Gamma(5/2 + j + n)} \left(\frac{\omega}{2}\right)^{2 n}}{
\sum_{n=0}^\infty \frac{ \Gamma(1 + j + 2 n)}
  {n! \Gamma(5/2 + j + n)} \left(\frac{\omega}{2}\right)^{2 n}}
\\
{\cal S}^{(1)}(j,n)\!\!\! &=&\!\!\! S_1(j+2n)  -S_1(3/2 +j+n)- S_1(j)
+S_1(3/2 +j)\, ,
\\
{\cal S}^{(2)}(j,n) \!\!\! &=&\!\!\! \left[{\cal S}^{(1)}(j,n)\right]^2
 - S_2(j+2n)  +S_2(3/2 +j+n) + S_2(j) - S_2(3/2 +j)\, . 
\nonumber
\end{eqnarray}
The approximation of $s^{1}_j(\omega)$  reads
\begin{eqnarray}
\label{Def-s1}
s^{(1)}_2(\omega) &\!\!\! \simeq \!\!\!&
\frac{13 {\omega }^2}{54}
\left(1 + 0.3642 {\omega }^2 + 0.1973 {\omega }^4 + 0.1266 {\omega }^6 + 
  0.0894 {\omega }^8 + 0.0670 {\omega }^{10} \right) 
  \, ,
\label{eq:sj1omegaExp}
\nonumber\\
s^{(1)}_4(\omega) &\!\!\!\simeq \!\!\!&
\frac{83{\omega }^2}{338}\left(
1 + 0.3694 {\omega }^2 + 0.2025 {\omega }^4 + 0.1313 {\omega }^6 + 
  0.0935 {\omega }^8 + 0.0707 {\omega }^{10} \right) 
  \, ,
\\
s^{(1)}_6(\omega) &\!\!\! \simeq\!\!\!&
\frac{143\omega^2}{578}\left(
1 + 0.3716 {\omega }^2 + 0.2048 {\omega }^4 + 0.1334 {\omega }^6 + 
  0.0953 {\omega }^8 + 0.0723 {\omega }^{10} \right) 
  \, ,
\nonumber \\
s^{(1)}_8(\omega) &\!\!\! \simeq\!\!\!&
\frac{73 {\omega }^2}{294}\left(
1 + 0.3727{\omega }^2 + 0.2059{\omega }^4 + 0.1344{\omega }^6 + 
   0.0963{\omega }^8 + 0.0732{\omega }^{10}
\right)
  \, .
\nonumber
\label{dd}
\end{eqnarray}
Note that the prefactor of these series is given by $\left(1 -
\frac{3}{{\left( 5 + 2j \right) }^2} \right){\omega }^2/4 \sim {\omega
}^2/4$ and that for $j=\{2,4,6,8\}$ and $ | \omega | =0.8$ the relative error 
of these approximations is about
$\{1.2\%,2.5\%,4.4\%,6.7\%\}$ and increases to
$\{ 6\%, 10\%, 14\%,19\%\}$ for $ | \omega | =0.9$. 
The approximation of $s^{2}_j(\omega)$ is given by
\begin{eqnarray}
s^{(2)}_2(\omega) &\!\!\! \simeq \!\!\!&
\frac{{\omega }^2}{243} \left(
1 + 14.7002 {\omega }^2 + 10.7004 {\omega }^4 + 7.7684 {\omega }^6 +
5.8690 {\omega }^8 + 4.5967 {\omega }^{10}
 \right)
  \, ,
\nonumber\\
s^{(2)}_4(\omega) &\!\!\! \simeq \!\!\!&
\frac{3 {\omega }^2}{2197}
\left(1 + 44.8125 {\omega }^2 + 33.1180 {\omega }^4 + 24.3070 {\omega }^6 +
 18.5325 {\omega }^8 + 14.6322 {\omega }^{10} \right)
  \, ,
\\
s^{(2)}_6(\omega) &\!\!\! \simeq \!\!\!&
\frac{3 \omega^2}{4913} \left(
1 + 100.913 {\omega }^2 + 75.0187{\omega }^4 + 55.3152{\omega }^6 + 42.3458
{\omega }^8 + 33.5573{\omega }^{10}
\right)
  \, ,
\nonumber\\
s^{(2)}_8(\omega) &\!\!\! \simeq \!\!\!&
  \frac{{\omega }^2}{3087}
\left(
1 + 191.007{\omega }^2 + 142.411{\omega }^4 + 105.261{\omega }^6 + 
   80.7569{\omega }^8 + 64.1255{\omega }^{10}
\right)
  \, .
\nonumber
\label{eq:sj2omegaExp}
\end{eqnarray}
Here we mention that the analytical expansion reads 
$3 {\omega }^2/(5 + 2 j)^3 + {\omega }^4
(1+ O(1/(5+2j)^2)/16 $. Thus, the ${\omega }^2$ term 
is numerically suppressed. In the $ | \omega | \to  1$ limit,
the functions $s_j^{(1,2)}$ take the values:
\begin{eqnarray}
\label{ShiFunOme1}
s^{(1)}_j(\omega=1) \!\!\! & \equiv & \!\!\!
2 S_1(2j+3)-S_1(j+1)-S_1(j+2)-\ln(2)\, ,
\\
s^{(2)}_j(\omega=1) \!\!\! & \equiv & \!\!\!
\left[s^{(1)}_j(\omega=1)\right]^2 - 4 S_2(2j+3)+S_2(j+2)+S_2(j+1)+2 \zeta
(2)\, .
\nonumber
\end{eqnarray}

The quantities in the \MS\ scheme are evaluated from
Eqs.\ (\ref{Str-T1}--\ref{Def-C2b}). For $\mu_f^2=Q^2$
the NLO contribution $T_{j}^{(1)} (\omega,\mu_f^2/Q^2=1)$  reads
\begin{eqnarray}
\label{eq:TjF1omegaExp}
T_{0}^{(1)}&\!\!\! \simeq\!\!\! & 
  -2 \left(
1 + 0.3333{\omega }^2 + 0.1873{\omega }^4 + 0.1245{\omega }^6 + 
  0.0904{\omega }^8 + 0.0694{\omega }^{10} + 0.0554{\omega }^{12}\!
 \right)
\, ,
\\
T_{2}^{(1)} &\!\!\! \simeq\!\!\! &\frac{{\omega }^2}{9}
\left(
1 + 1.2815 {\omega }^2 + 1.0293 {\omega }^4 + 0.7770 {\omega }^6 + 
  0.5844 {\omega }^8 + 0.4436 {\omega }^{10} + 0.3407{\omega }^{12}
\right)
\, ,
\nonumber\\
T_{4}^{(1)} &\!\!\! \simeq \!\!\!&
\frac{4184{\omega }^4}{42525}
\left(
1 + 1.4076 \omega ^2 + 1.4571 \omega^4 + 1.3683 \omega^6 +
 1.2370 \omega^8 + 1.1013 \omega^{10} + 0.9754 \omega^{12}
\right)
\, ,
\nonumber\\
T_{6}^{(1)} &\!\!\! \simeq \!\!\!&
\frac{96182{\omega }^6}{2837835}
\left(
1 + 1.8560 {\omega }^2 + 2.3723 {\omega }^4 + 2.6207 {\omega }^6 +
2.6949 {\omega }^8 + 
  2.6644 {\omega }^{10} + 2.5747 {\omega }^{12}
\right)
\, ,
\nonumber\\
T_{8}^{(1)} &\!\!\! \simeq \!\!\!&
\frac{568352 {\omega }^8}{57432375}
\left(
1 + 2.3324{\omega }^2 + 3.5542{\omega }^4 + 4.5116{\omega }^6 +
5.1921{\omega }^8 + 
  5.6346{\omega }^{10} + 5.8894{\omega }^{12}
\right)
 \, ,
\nonumber
\end{eqnarray}
while the conformal moments of the factorization log proportional term, i.e.,
$C_F v_j \, T^{(0)}_{j}(\omega) $, are obtained by
multiplying the results (\ref{eq:Tj0omegaExp}) with the values of
$C_F v_j$, given in Table \ref{Tab-EigValEvoKer}. At NNLO, only the 
$\beta_0$ proportional term has been evaluated:
\begin{eqnarray}
\label{eq:Tjbeta2omegaExp}
T^{(2)}_{\beta,0} \!\!\! &\simeq& \!\!\!
 -3 \left( 
1 + 0.3505{\omega }^2 + 0.1785{\omega }^4 + 0.1098{\omega }^6 + 
   0.0750{\omega }^8 + 0.0548{\omega }^{10} + 0.0420{\omega }^{12}
\right)
\, ,
\\
T^{(2)}_{\beta,2} \!\!\! &\simeq& \!\!\!
\frac{4369{\omega }^2}{8640}
\left( 1+ 0.9159{\omega }^2 + 0.7298{\omega }^4 + 0.5740{\omega }^6 + 
   0.4566{\omega }^8 + 0.3691{\omega }^{10} + 0.3031{\omega }^{12}\right)
\, ,
\nonumber\\
T^{(2)}_{\beta,4}  \!\!\! &\simeq& \!\!\!
\frac{2356859{\omega }^4}{8505000}
\left(1+ 1.3588{\omega }^2 + 1.4063{\omega }^4 + 1.3333{\omega }^6 + 
   1.2209{\omega }^8 + 1.1022{\omega }^{10} + 0.9899{\omega }^{12} \right)
\, ,
\nonumber\\
T^{(2)}_{\beta,6}  \!\!\! &\simeq& \!\!\!
 \frac{20352710029{\omega }^6}{222486264000}
\left(  1+ 1.8408{\omega }^2 + 2.3592{\omega }^4 + 2.6233{\omega }^6 + 
   2.7190{\omega }^8 + 2.7108{\omega }^{10} + 2.6415{\omega }^{12}\right)
\, ,
\nonumber\\
T^{(2)}_{\beta,8}  \!\!\! &\simeq& \!\!\!
  \frac{363260060687{\omega }^8}{13676945782500}
\left(1 + 2.3308{\omega }^2 + 3.5655{\omega }^4 + 4.5516{\omega }^6 + 
   5.2715{\omega }^8 + 5.7582{\omega }^{10} + 6.0576{\omega }^{12} \right)
\, .
\nonumber
\end{eqnarray}
To restore the factorization log, one needs
\begin{eqnarray}
\label{eq:vbetasumomegaExp}
v_{\beta,0}^{\Sigma} \!\!\! &\simeq& \!\!\!
-\frac{\omega^2}{6}
\left(1 + 0.5657\,{\omega }^2 + 0.3830 {\omega }^4 + 
  0.2837 {\omega }^6 + 0.2220 {\omega }^8 + 
  0.1804{\omega }^{10}\right)
\, ,
\\
v_{\beta,2}^{\Sigma} \!\!\! &\simeq& \!\!\!
-\frac{83{\omega }^2}{216}
\left(
1. + 0.0488{\omega }^2 + 0.0308{\omega }^4 + 
  0.0225{\omega }^6 + 0.0177{\omega }^8 + 
  0.0146{\omega }^{10} + 0.0123{\omega }^{12}
\right)
\, ,
\nonumber\\
v_{\beta,4}^{\Sigma} \!\!\! &\simeq& \!\!\!
-\frac{7783{\omega }^4}{70875}
\left(
1. + 0.0261{\omega }^2 + 0.0172{\omega }^4 + 
  0.0130{\omega }^6 + 0.0105{\omega }^8 + 
  0.0088{\omega }^{10} + 0.0076{\omega }^{12}
\right)
\, ,
\nonumber\\
v_{\beta,6}^{\Sigma} \!\!\! &\simeq& \!\!\!
-\frac{3745727 {\omega }^6}{132432300}
\left(
1. + 0.0174{\omega }^2 + 0.0118{\omega }^4 + 
  0.0091{\omega }^6 + 0.0075{\omega }^8 + 
  0.0064{\omega }^{10} + 0.0056{\omega }^{12}
\right)
\, ,
\nonumber\\
v_{\beta,8}^{\Sigma} \!\!\! &\simeq& \!\!\!
-\frac{76991788{\omega }^8}{10854718875}
\left(1 + 0.0130{\omega }^2 + 0.0089{\omega }^4 + 
  0.0070{\omega }^6 + 0.0058{\omega }^8 + 
  0.0050{\omega }^{10} + 0.0044\,{\omega }^{12}
\right)
\, .
\nonumber
\end{eqnarray}

In the CS scheme, the NLO result
$T_{j}^{{\rm CS}(1)} (\omega,\mu_f/Q^2=1)$ 
we find from Eq.\ (\ref{Def-TNLOomega}):
\begin{eqnarray}
\label{eq:Tj1omegaExpCS}
T_{0}^{{\rm CS}(1)}&\!\!\! \simeq\!\!\! & 
  -2 \left(
1 + 0.2 {\omega }^2 + 0.0857{\omega }^4 + 0.0476{\omega }^6 + 
   0.0303{\omega }^8 + 0.0210{\omega }^{10} + 0.0154{\omega }^{12}
\right)
 \, ,
\\
T_{2}^{{\rm CS}(1)} &\!\!\! \simeq\!\!\! &\frac{{\omega }^2}{9}
\left(
 1 + 1.4691{\omega }^2 + 1.2818{\omega }^4 + 1.0443{\omega }^6 + 
   0.8466{\omega }^8 + 0.6933{\omega }^{10} + 0.5752{\omega }^{12}
\right)
 \, ,
\nonumber\\
T_{4}^{{\rm CS}(1)} &\!\!\! \simeq \!\!\!&
\frac{4184{\omega }^4}{42525}
\left(
1 + 1.4102{\omega }^2 + 1.4674{\omega }^4 + 1.3875{\omega }^6 + 
   1.2644{\omega }^8 + 1.1354{\omega }^{10} + 1.0148{\omega }^{12}
\right)
 \, ,
\nonumber\\
T_{6}^{{\rm CS}(1)} &\!\!\! \simeq \!\!\!&
\frac{96182{\omega }^6}{2837835}
\left(
1 + 1.8373{\omega }^2 + 2.3344{\omega }^4 + 2.5697{\omega }^6 + 
   2.6372{\omega }^8 + 2.6050{\omega }^{10} + 2.5171{\omega }^{12}
\right)
 \, ,
\nonumber\\
T_{8}^{{\rm CS}(1)} &\!\!\! \simeq \!\!\!&
\frac{568352 {\omega }^8}{57432375}
\left(
  1 + 2.3052{\omega }^2 + 3.4829{\omega }^4 + 4.3926{\omega }^6 + 
   5.0299{\omega }^8 + 5.4373{\omega }^{10} + 5.6655{\omega }^{12}
\right)
 \, .
\nonumber
\end{eqnarray}
while the factorization log proportional term is the same as in the \MS\
scheme. The NNLO correction
$T_{j}^{{\rm CS}(2)} (\omega,\mu_f/Q^2=1,\mu_r/Q^2=1)$ for
$\beta_0=0$ reads
\begin{eqnarray}
\label{eq:Tj2omegaExpCS}
T_{0}^{{\rm CS}(2)}&\!\!\! \simeq\!\!\! &
3.6667
\left(\!
1+ 0.2 {\omega }^2 + 0.0857{\omega }^4 + 0.0476{\omega }^6 + 
   0.0303{\omega }^8 + 0.021{\omega }^{10} + 0.0154{\omega }^{12}
\!\right)_{|\beta_0=0}
\, ,
\\
T_{2}^{{\rm CS}(2)} &\!\!\! \simeq\!\!\! &
-3.2331{\omega }^2 
\left(\!
1 + 0.6769\,{\omega }^2 + 0.4557{\omega }^4 + 0.3224{\omega }^6 + 
   0.2386{\omega }^8 + 0.1832{\omega }^{10} + 0.1448{\omega }^{12}
\!\right)_{|\beta_0=0}
\, ,
\nonumber\\
T_{4}^{{\rm CS}(2)} &\!\!\! \simeq \!\!\!&
-0.9338 {\omega }^4
\left(\!
1+ 1.0967{\omega }^2 + 0.9764{\omega }^4 + 0.8272{\omega }^6 + 
   0.6939{\omega }^8 + 0.5838{\omega }^{10} + 0.4949{\omega }^{12}
\!\right)_{|\beta_0=0}
\, ,
\nonumber\\
T_{6}^{{\rm CS}(2)} &\!\!\! \simeq \!\!\!&
-0.21 {\omega }^6
\left(\!
 1 + 1.5133{\omega }^2 + 1.6617{\omega }^4 + 1.6285{\omega }^6 + 
   1.5184{\omega }^8 + 1.3829{\omega }^{10} + 1.2459{\omega }^{12}
\!\right)_{|\beta_0=0}
\, ,
\nonumber\\
T_{8}^{{\rm CS}(2)} &\!\!\! \simeq \!\!\!&
 -0.0427{\omega }^8
\left(\!
 1 + 1.905{\omega }^2 + 2.4505{\omega }^4 + 2.6858{\omega }^6 + 
   2.7134{\omega }^8 + 2.6181{\omega }^{10} + 2.4577{\omega }^{12}
  \!\right)_{|\beta_0=0}
\, .
\nonumber
\end{eqnarray}
The $Q^2$ independent and $\beta_0$-proportional term in the CS scheme
is the same as in the \MS\ one, given in Eq.\ (\ref{eq:Tj2omegaExpCS}).
The factorization and renormalization log proportional terms, appearing
in Eq.\ (\ref{eq:Tj2CSomega}) can be easily restored by means of 
the results from Table
\ref{Tab-EigValEvoKer}, Eqs.\ (\ref{eq:Tj0omegaExp},\ref{Def-s1}) as
well as Eqs.\ (\ref{Def-vCSbetaOmega},
\ref{eq:TjF1omegaExp},\ref{eq:vbetasumomegaExp},\ref{eq:Tj1omegaExpCS}).

The difference between the CS and $\overline{\rm CS}$ schemes
arises only from the $\beta_0$-proportional terms 
(\ref{eq:T2omegaExpbarCS}).
The $\beta_0$-proportional NNLO term
$T_{\beta,j}^{\overline{\rm CS}(2)} (\omega)$ 
reads
\begin{eqnarray}
\label{eq:Tj2omegaExpbarCSbeta}
T_{\beta,0}^{\overline{\rm CS}(2)}&\!\!\! \simeq\!\!\! &
-3\left(\!
1+ 0.2 {\omega }^2 + 0.0857{\omega }^4 + 0.0476{\omega }^6 + 
   0.0303{\omega }^8 + 0.0210{\omega }^{10} + 0.0154{\omega }^{12}
\!\right)
\, ,
\\
T_{\beta,2}^{\overline{\rm CS}(2)} &\!\!\! \simeq\!\!\! &
0.50567{\omega }^2\left(\!
1. + 0.8496{\omega }^2 + 0.6431{\omega }^4 + 0.4900{\omega }^6 + 
   0.3820{\omega }^8 + 0.3048{\omega }^{10} + 0.2482{\omega }^{12}
\!\right)
\, ,
\nonumber\\
T_{\beta,4}^{\overline{\rm CS}(2)} &\!\!\! \simeq \!\!\!&
0.277114{\omega }^4\left(\!
1 + 1.2512{\omega }^2 + 1.2252{\omega }^4 + 1.1162{\omega }^6 + 
   0.9919{\omega }^8 + 0.8747{\omega }^{10} + 0.7710{\omega }^{12}
\!\right)
\, ,
\nonumber\\
T_{\beta,6}^{\overline{\rm CS}(2)} &\!\!\! \simeq \!\!\!&
0.0914785 {\omega }^6\left(\!
  1. + 1.7236{\omega }^2 + 2.1049{\omega }^4 + 2.2550{\omega }^6 + 
   2.2689{\omega }^8 + 2.2076{\omega }^{10} + 2.1078{\omega }^{12}\!\right)
\, ,
\nonumber\\
T_{\beta,8}^{\overline{\rm CS}(2)} &\!\!\! \simeq \!\!\!&
0.02656{\omega }^8\left(\!
 1 + 2.2092{\omega }^2 + 3.2413{\omega }^4 + 4.0009{\omega }^6 + 
   4.5064,{\omega }^8 + 4.8081{\omega }^{10} + 4.9570{\omega }^{12}
   \!\right)
\, .
\nonumber
\end{eqnarray}
The restoration of the factorization log in the $\beta_0$ sector
requires the knowledge of $v_{\beta,j}$, given in Table
\ref{Tab-EigValEvoKer}.

\begin{table}[t]
\caption{
The first four non-vanishing and even eigenvalues of the evolution
kernel (\ref{eq:Vexp}) to NNLO  accuracy
(for color decomposition see (\ref{eq:gammaj})). 
The values of $v_{j}^{(2)}$ are taken from
the non-singlet result for the deep inelastic structure function $F_3$
\cite{RetVer00}.
}
\begin{center}\begin{tabular}{||c||c||c|c||c||}
\hline
$j$ & 
$\displaystyle  C_F v_j$ &
$\displaystyle  C_F^2 v_{F,j}+\frac{C_F}{2N_c} v_{G,j}  $ &
$\displaystyle   v_{\beta,j}$ &
$\displaystyle   v_{j}^{(2)} = -1/2 \gamma_j^{(2)}$
\\[0.2cm]\hline\hline
2 &  -2.77778 &   3.41307  &-2.88194  & 
$-155.614 + 24.5592\, n_f + 0.220250\, n_f^2  $
 \\ \hline
4 &  -4.04444 &  7.15867  & -4.32389 & 
$-215.118 + 34.7698\,n_f + 0.295776\,n_f^2  $
 \\ \hline
6 &  -4.89048 &  9.82554 & -5.30857  & 
$-254.562 + 41.3602\,n_f + 0.342420\,n_f^2 $ 
 \\ \hline
8 &   -5.52910 &  11.86905\,\,  & -6.06196 &   
$-284.650 + 46.3238\,n_f + 0.375806\,n_f^2$
 \\ \hline
\end{tabular}\end{center}
\label{Tab-EigValEvoKer}
\end{table}

\section{Reconstruction of the hard-scattering amplitude
in the momentum fraction representation}
\label{App-ReconHSA}
\setcounter{equation}{0}

Let us now discuss the reconstruction of the hard-scattering amplitude
in the momentum fraction representation from the conformal moments. This
technical problem is of immense importance for the discussion of 
two-photon processes in which the operator product expansion is not
convergent. The solution is known in forward kinematics and is given by
the Mellin transformation of moments. In non-forward kinematics, the
problem is solved  in principle \cite{BelMue98a}, however, one has
to evaluate rather cumbersome integrals. Here we propose a simple
approximative solution which is based on the asymptotic behaviour,
presented in Section 4.1, and it is applicable to deeply virtual
Compton scattering. In the following we demonstrate this method at NLO
order and its generalization to higher orders is straightforward.

First, we consider the quality of the asymptotic formulas (\ref{App-T-NLO}).
Since the neglected terms are of the order $1/(j+2)$, we expect that the
approximations (\ref{App-T-NLO}) have an accuracy of the level of
10\% for $j=10$. Surprisingly, the accuracy is already below 1\%
in both cases, which indicates that the $1/(j+2)$ term is small. Thus, we
completely neglect such terms and improve the approximate formula by adding
subleading terms of the form
\begin{eqnarray}
\frac{\alpha  + \beta\, S_{1}(j+1)}{(j+1)(j+2)} 
 + \frac{ \gamma + \delta\, S_{1}(j+1)}{(j+1)^2(j+2)^2} + \cdots,
\nonumber
\end{eqnarray}
where the coefficients $\alpha, \dots, \delta$ are determined from a fit of
the lowest moments. In this way, we obtain an approximation that is
better than 1\% for all moments. Now we can 
recover an approximate expression for the hard-scattering amplitude as
a convolution by the following recipe:
\begin{itemize}
\item
Substitute the LO Wilson--coefficients by the corresponding hard-scattering
amplitude:\\
\vspace{-0.4cm}
\begin{eqnarray*}
\displaystyle \frac{2j+3}{(j+1)(j+2)}\to \frac{1}{2(1-x)}.
\end{eqnarray*}
\vspace{-1cm}
\item
Restore kernels:\\
\vspace{-0.4cm}
\begin{eqnarray*}
{\rm cons} \to {\rm cons}\, {\rm \bf I}, \quad
\frac{1}{(j+1)(j+2)} \to v_a(x,y), \quad
S_{1}(1 + j)\to -\frac{1}{2} [v_b(x,y)]_+ +1\, {\rm \bf I}.
\end{eqnarray*}
\item
Consider the multiplication of the above given conformal moments,
which corresponds to convolution in the momentum fraction space.
\end{itemize}
Here we have introduced a shorthand notation for the identity 
${\rm \bf I} \equiv
 \delta(x-y)$.
In the
$\overline{\rm MS}$ scheme, using Eq.\ (\ref{eq:T1MS}) and the
recipe given above we restore the exact expression (\ref{eq:T1o1}) 
in the momentum
fraction space. For
the CS scheme, using the improved form (\ref{App-T-NLO}), we get a good
approximation of Eq.\ (\ref{eq:Tj1FCS}), for the lowest moments also,  
by taking
\begin{eqnarray}
\alpha = \frac{83}{10} + 8 \zeta(2) - 26 \ln(2) \approx 3.438,\ 
  \gamma= -\frac{73}{5} - 12 \zeta(2) + 48 \ln (2)
\approx -1.068,\  \beta=\delta= 0
\end{eqnarray}
and the hard-scattering part reads
\begin{eqnarray}
T^{{\rm CS}(1)}_{F}(x) &\!\!\!\! \approx \!\!\!\!&
 T^{(0)} \otimes \left\{ \frac{1}{2} [v_b]_+\otimes \left(
\frac{1}{2} [v_b]_+ - 6.273\, {\rm \bf I} \right)
-2.952\, {\rm\bf I} + 3.438\, v_a -1.068\, v_a\otimes  v_a \right\}(x)
\nonumber\\
&\!\!\!\! \approx \!\!\!\!&
\frac{0.25 \ln^2(1 - x)- 2.136\ln(1 - x) - 8.224}{2(1 - x)}  + 
  \frac{\ln(1 - x)\left[0.25 \ln(1 - x) -6.642\right]}{2 x}
\nonumber\\
&& +  1.068 \frac{{\rm Li}_2(x) - {\rm Li}_2(1)}{2(1 - x)}.
\end{eqnarray}
After analytical continuation in $x$ this result corresponds to the NLO
correction of the deeply virtual Compton scattering for the quark-quark 
channel in the parity odd sector.

\end{appendix}



\end{document}